\documentclass[prb,twocolumn,amsmath,amssymb,superscriptaddress,longbibliography]{revtex4-1}

\usepackage{amsfonts}
\usepackage{amsmath}
\usepackage{amssymb}
\usepackage{graphicx}
\usepackage[T1]{fontenc}
\usepackage[utf8]{inputenc}
\usepackage{hyperref}
\usepackage{xcolor}
\usepackage{color}
\usepackage{bm}
\usepackage{times}
\usepackage{comment}

\usepackage[section]{placeins}
\usepackage[T1]{fontenc}
\usepackage[utf8]{inputenc}
\usepackage{hyperref}
\usepackage{xcolor}
\usepackage{color}
\usepackage{bm}
\usepackage{times}
\usepackage{comment}
\usepackage{footnote}

\newcommand{\q}[1]{``#1''}

\begin{document}
\title{Refrustration and competing orders in the prototypical {\bf
    Dy$_2$Ti$_2$O$_7$} spin ice material}

\author{P. Henelius$^*$} \affiliation{Department of Theoretical
  Physics, Royal Institute of Technology, SE-106 91 Stockholm, Sweden}
\author{T. Lin}\affiliation{Department of Physics and Astronomy,
  University of Waterloo, Waterloo, Ontario, N2L 3G1, Canada}
\author{M. Enjalran}\affiliation{Department of Physics, Southern
  Connecticut State University, 501 Crescent Street, New Haven,
  Connecticut 06515-1355, United States} \affiliation{Connecticut
  State Colleges and Universities Center for Nanotechnology, Southern
  Connecticut State University, New Haven, Connecticut 06515-1355,
  United States} \author{Z. Hao}\affiliation{Department of Physics and
  Astronomy, University of Waterloo, Waterloo, Ontario, N2L 3G1,
  Canada} \author{J. G. Rau}\affiliation{Department of Physics and
  Astronomy, University of Waterloo, Waterloo, Ontario, N2L 3G1,
  Canada} \author{J. Altosaar}\affiliation{Department of Physics and
  Astronomy, University of Waterloo, Waterloo, Ontario, N2L 3G1,
  Canada} \affiliation{Department of Physics, Princeton University,
  Princeton, New Jersey 08544, United States}
\author{F. Flicker}\affiliation{ Department of Physics, University of
  California, Berkeley, California 94720 USA}
\author{T. Yavors'kii}\affiliation{Applied Mathematics Research
  Centre, Coventry University, Coventry, CV1 5FB, United Kingdom}
\author{M. J. P.  Gingras}\affiliation{Department of Physics and
  Astronomy, University of Waterloo, Waterloo, Ontario, N2L 3G1,
  Canada} \affiliation{Canadian Institute for Advanced Research, 180
  Dundas St. W., Toronto, Ontario, M5G 1Z8, Canada}
\affiliation{Perimeter Institute for Theoretical Physics, 31 Caroline
  St. N., Waterloo, Ontario, N2L 2Y5, Canada}

\begin{abstract} Spin ices, frustrated magnetic materials analogous to
  common water ice, have emerged over the past fifteen years as
  exemplars of high frustration in three dimensions.  Recent
  experimental developments aimed at interrogating anew the
  low-temperature properties of these systems, in particular whether
  the predicted transition to long-range order occurs, behoove
  researchers to scrutinize our current dipolar spin ice model
  description of these materials.  In this work we do so by combining
  extensive Monte Carlo simulations and mean-field theory calculations
  to analyze data from previous magnetization, elastic neutron
  scattering and specific heat measurements on the paradigmatic
  Dy$_2$Ti$_2$O$_7$ spin ice material.  In the present work, we also
  reconsider the possible importance of the nuclear specific heat,
  $C_{\rm nuc}$, in Dy$_2$Ti$_2$O$_7$. We find that $C_{\rm nuc}$ is
  not entirely negligible below a temperature $\sim 0.5$ K and must
  therefore be taken into account in a quantitative analysis of the
  calorimetric data of this compound below that temperature.  We find
  that in this material, small effective spin-spin exchange
  interactions compete with the magnetostatic dipolar interaction
  responsible for the main spin ice phenomenology.  This causes an
  unexpected ``refrustration'' of the long-range order that would be
  expected from the incompletely self-screened dipolar interaction and
  which positions the material at the boundary between two competing
  classical long-range ordered ground states.  This allows for the
  manifestation of new physical low-temperature phenomena in
  Dy$_2$Ti$_2$O$_7$, as exposed by recent specific heat measurements.
  We show that among the four most likely causes for the observed
  upturn of the specific heat at low temperature -- an
  exchange-induced transition to long-range order, quantum non-Ising
  (transverse) terms in the effective spin Hamiltonian, the nuclear
  hyperfine contribution and random disorder -- only the last appears
  to be reasonably able to explain the calorimetric data.

\end{abstract}

\maketitle

\section{Introduction}
\label{sec:Intro}

Highly frustrated magnetism (HFM) arises when the leading effective
spin-spin interactions are in strong competition among
themselves\cite{Springer_frust_book,Gardner10,Balents_Nature}. High
frustration dramatically weakens a material's tendency towards
conventional long-range magnetic order. This opens up an avenue to the
discovery of spin liquids, intriguing states of matter where the
magnetic degrees of freedom are disordered by quantum mechanical
fluctuations even at absolute zero
temperature\cite{Anders87,Balents_Nature,Gingras14}.  It is convenient
to follow Anderson's perspective\cite{Anders87} and divide models of
HFM into two classes. In the first, the lattice architecture
frustrates the predominant antiferromagnetic (AF) nearest-neighbor
interactions and the formation of a unique collinear long-range
ordered N{\'e}el state. Popular examples of such highly frustrated
lattice structures include the kagome and pyrochlore lattices in two
and three dimensions,
respectively\cite{Springer_frust_book,Balents_Nature,Gingras14}.  In
the second class, one finds a number of interactions of roughly the
same scale that compete to control the development of distinct
magnetic correlations.  A well-known example of this second class is
the $J_1$-$J_2$ model on the square
lattice\cite{Anders87,Chandra,Melzi,Bombardi}.  More recently, it has
been suggested that frustration based on competing interactions, as
opposed to the common geometrical antiferromagnetic nearest-neighbor
frustration, is at play in the kagome-based antiferromagnetic
Kappellasite material~\cite{Kapel}.

The first class of HFM systems has attracted the most attention from
experimentalists because, unlike the second class, it does not rely on
a fine-tuning accident of nature (\emph{e.g.} $J_2 \sim J_1/2$ in the
$J_1$-$J_2$ model\cite{Chandra}) to be at a strong frustration point
to induce large quantum spin fluctuations.  In this class, there is no
unique long-range ordered state selected at the mean-field level and
the spectrum of soft modes is dispersionless throughout the whole
Brillouin zone\cite{Reimers91,Moessner98,Gingras_CJP}.  Consequently,
these systems are very soft even at the classical level and show
limited propensity towards ordering\cite{Villain79}, suggesting that
magnetic materials involving such lattices are attractive candidates
in the search for a spin liquid state\cite{Balents_Nature,Gingras14}.
High frustration does not only enable large thermal and quantum
fluctuations. It also allows for random disorder, in the form of
off-stoichiometry or inter-site mixing for example, to have dramatic
effects on the low-temperature properties of a
system\cite{Villain79,Sen}.
 
Spin ices, in which the magnetic moments obey an energetic ``ice
rule'' similar to that governing the proton positions in common water
ice\cite{Gingras11,Bramwell_Science}, have traditionally been viewed
as belonging to the first category of HFM
systems\cite{Gingras11,Gingras_CJP,Bramwell_Science, Gardner10}.  The
two textbook examples of spin ice materials are the rare-earth
pyrochlore oxides Ho$_2$Ti$_2$O$_7$ (Ref.~\onlinecite{Harris}) and
Dy$_2$Ti$_2$O$_7$ (Ref.~\onlinecite{Ramirez}) in which Ho$^{3+}$ and
Dy$^{3+}$ are the magnetic ions.  The key signature of spin ices --
the formation of an exponentially large number of nearly degenerate
low-energy states (called the spin-ice manifold) at sufficiently low
temperature $T$ -- does not require fine-tuning\cite{Gingras_CJP}.
Indeed, the spin-ice phenomenon is robust and consistent with the two
dominant spin-spin interactions: the nearest-neighbor exchange $J_1$
and the long-range dipolar interactions of strength $D$.  However, in
this $J_1$-$D$ dipolar spin ice model (DSM)
\cite{Siddharthan,denHertog}, the imperfect screening of the
magnetostatic dipolar $1/r^3$ tail\cite{Gingras_CJP,Isakov_SS} is
theoretically expected to induce a transition to long-range order at
$T\approx 0.13 D$ as found in Monte Carlo
simulations\cite{Melko01,Melko04}.  Experiments\cite{Harris98, Harris,
  Ramirez, Fuka02}, on the other hand, have so far not found evidence
for a transition to long-range order.  It has generally been assumed
that this is the result of the dynamical arrest and freezing out of
gapped spin-flip excitations (``monopoles'')\cite{Castelnovo_Nature}
at low temperatures\cite{Castel10,Paulsen14}, as made clear by recent
experiments aimed at studying the low-temperature dynamical and
thermodynamical properties of spin ices\cite{Revell13, Poma13,
  Paulsen14, Sala14}.  It is therefore surprising and noteworthy that
one of these experiments on Dy$_2$Ti$_2$O$_7$
(Ref.~\onlinecite{Poma13}) finds an \emph{increase} in the magnetic
specific heat $C(T)$ below a temperature $T^*\sim 0.5$ K, as opposed
to a rapid plunge of $C(T)$ to zero around that temperature as found
in all previous specific heat measurements on this
compound\cite{Ramirez,Zhou_chem_press, Klemke11, Higa02}.  The rise of
$C(T)$ observed in Ref.~[\onlinecite{Poma13}] results in a concurrent
continuous decrease of the thermal magnetic entropy with no signature
of a residual Pauling entropy plateau over any extended temperature
window.  Moreover, this observed rise in $C(T)$ is not in accord with
the predictions of the original DSM\cite{denHertog, Melko01, Melko04}
or its refinement\cite{Taba06,Yavo08} that includes long-range dipolar
interactions as well as exchange interactions up to third nearest
neighbor.  Given the successes of this refined DSM\cite{Taba06,Yavo08}
in rationalizing a number of experimental aspects of
Dy$_2$Ti$_2$O$_7$, the results of Ref.~[\onlinecite{Poma13}] beg the
question that motivated the present study: ``What is the microscopic
origin of the observed rise in $C(T)$ for $T\lesssim 0.5$ K?''.

This is an important question because the $C(T)$ upturn could suggest
some heretofore unexposed physics going on deep in the low-temperature
spin ice regime of Dy$_2$Ti$_2$O$_7$.  For example, this upturn could
signal the emergence of quantum effects and suggest that the
previously assumed strictly classical\cite{Gingras11}
Dy$_2$Ti$_2$O$_7$ spin ice may be in fact entering a quantum spin ice
state at sufficiently low temperature, thus offering itself as an
unexpected realization of a quantum spin liquid\cite{Gingras14}.

To address the above question, we step all the way back to reassess
the premises defining the spin-spin couplings of the
DSM\cite{denHertog,Taba06,Yavo08} and discuss what we consider the
most cogent way to proceed.  In the Dy$_2$Ti$_2$O$_7$ and
Ho$_2$Ti$_2$O$_7$ dipolar spin ice materials, the spin ice regime is a
collective paramagnetic state\cite{Villain79}, a classical spin liquid
of sorts\cite{Balents_Nature}.  In that liquid regime, the thermal
evolution of most thermodynamic quantities is smooth and without sharp
features (e.g. specific heat).  This is the reason why the temperature
and magnetic field dependence of several quantities need to be
simultaneously fitted to parametrize the spin-spin coupling of the DSM
beyond the nearest-neighbor exchange. This problem did not arise in
the formulation of the simplest original DSM~\cite{denHertog} which
contains only one independent (nearest-neighbor) exchange coupling.
This exchange could be determined by fitting independently the
temperature at which the specific heat peaks or the height of that
peak since the dipolar strength $D$ is a priori known from the
saturation magnetization of the material and the lattice spacing. Such
a fitting of multiple thermodynamic quantities is the procedure that
was followed in Ref.~[\onlinecite{Yavo08}].  An important conclusion
of that study was that the so-determined Hamiltonian would most likely
display a transition to long-range order near 100 mK. This prediction
would seem to be in significant disagreement with the recent
experiment of Pomaranski {\it et al.}\cite{Poma13} if the upturn of
the specific heat at $T\sim 0.5$K were to be interpreted as a
harbinger of an impending transition to long-range order.

In the present work we follow an approach that differs in two ways
from the work of Ref.~[\onlinecite{Yavo08}].  First of all, we
consider an experiment on Dy$_2$Ti$_2$O$_7$
\cite{Ruff,Higa_112,Sato06} performed with a magnetic field applied
along a direction such that one can invoke symmetry considerations and
settings that allow us to determine two symmetry-distinct subleading
(third nearest-neighbor) exchange parameters that would be difficult
to determine from measurements that do not exploit such an astute
experimental symmetry set-up. Secondly, as spin-spin correlations, and
therefore their reciprocal space description as probed by neutron
scattering, are the observables most directly linked to the details of
the spin Hamiltonian, we scrutinize the reciprocal space details of a
neutron scattering intensity map previously obtained on
Dy$_2$Ti$_2$O$_7$ spin ice\cite{Fennell04}.

By comparing detailed experimental information on the spin-spin
correlations contained in neutron scattering results on
Dy$_2$Ti$_2$O$_7$ with that of Monte Carlo simulations, we show below
that the best set of couplings parametrizing its spin Hamiltonian
positions Dy$_2$Ti$_2$O$_7$ near the phase boundary between two
competing classical long-range ordered spin ice states.  We find that
the recovery of the Pauling entropy is a more intricate process than
previously thought, with the boundary region rich in unusual phenomena
such as a nearly energetically degenerate stacking of ordered spin
planes.  Through a remarkable coincidence of nature, the competing
distance-dependent exchange and dipolar interactions in
Dy$_2$Ti$_2$O$_7$ ``refrustrate'' this material at low temperature
\footnote{By ``refrustrate'' we mean that exchange interactions beyond
  nearest-neighbor compete with the meager tendency of the long-range
  part of the dipolar interactions to lift the degeneracy of the
  ice-rule states\cite{Gingras_CJP,Isakov_SS} with an associated
  transition at a temperature $T_c\sim 0.13D$\cite{Melko01,Melko04}
  and further depress the ordering transition to an even lower
  temperature.}  and, therefore, make it a new example of the second
class of HFM systems discussed above.

Such a competition between various classical ground states may in
principle allow for an enhanced level of quantum fluctuations and
contribute to driving dipolar spin ices, traditionally viewed as
strictly classical Ising systems\cite{Gingras11}, into a U(1) quantum
spin liquid state\cite{Gingras14,McClarty}.  This is reminiscent of
the $J_1$-$J_2$ model at the $J_2$=$J_1/2$ point where two classical
ground states are also degenerate\cite{Chandra}.  Because the critical
transition temperature for the exchange parameters determined from the
analysis of the neutron scattering data is far below the upturn seen
in the specific heat data\cite{Poma13}, we are unable to describe the
$C(T)$ data of Ref.~[\onlinecite{Poma13}] for $T\lesssim 0.7$ K with
those parameters, even after accounting for the non-negligible nuclear
contribution to the total specific heat below that temperature.  This
leads us to reconsider how strong the heretofore largely ignored
quantum (non-Ising) terms in the Hamiltonian may be for
Dy$_2$Ti$_2$O$_7$ (see, however, Ref.~[\onlinecite{McClarty}] for a
study that does consider possible quantum effects).  Motivated by
recent works having found that various forms of random disorder are
present in magnetic rare-earth pyrochlore oxides\cite{Sala14,Ross12,
  Cava15}, we also explore what effect random disorder may have on the
low-temperature properties of this material. We reaffirm that the
quantum terms in Dy$_2$Ti$_2$O$_7$ should be very small indeed and
unlikely to be responsible for the development of a low-temperature
quantum coherence (e.g. coherent hopping of spinons\cite{Hao_hopping},
i.e. spin ice monopoles\cite{Castelnovo_Nature}) that would be
signaled by a rise in $C(T)$, similarly to that recently reported in a
quantum Monte Carlo simulation study of a quantum spin ice
model\cite{Kato14}.  On the other hand, we show that random disorder,
in the form of intersite disorder\cite{Ross12, Cava15} (\emph{e.g.}
``stuffing,''), could potentially explain the rise of $C(T)$ without
dramatically affecting the neutron scattering intensity pattern over
the same temperature range.

The rest of the paper is organized as follows. In Section
\ref{sec:Model_Methods} we define the dipolar spin ice model that we
study in this paper along with the Monte Carlo simulation that we
employ to analyze the various experimental data that we consider.
Section \ref{sec:Results} contains the essential results of our work.
Subsection \ref{sec:112} presents the analysis of the $[112]$ magnetic
field magnetization measurements used to constrain the exchange
parameters defining our dipolar spin ice model.  Subsection
\ref{sec:phasediagram} discusses the long-range ordered phases that
the dipolar spin ice model displays within the constrained spin-spin
coupling constants determined in Subsection \ref{sec:112}.  Subsection
\ref{sec:neutronscatt} reports the analysis of previously published
neutron scattering data on Dy$_2$Ti$_2$O$_7$ that allows us to
position this compound in the phase diagram determined in Subsection
\ref{sec:phasediagram}.  In Subsection \ref{sec:specheat}, we analyze
the recent specific heat data in relation to the phase diagram
presented in Subsection \ref{sec:phasediagram}.  In Subsection
\ref{sec:specheatneutroncomp}, we make a detailed comparison of the
Monte Carlo results for the specific heat and neutron scattering
intensity obtained in the distinct parts of the phase diagram and show
that we are unable to reconcile the experimental specific heat
measurements of Ref.~[\onlinecite{Poma13}] with the spin-spin
couplings identified in Subsection \ref{sec:phasediagram} that
describe the main features of the neutron scattering data.  In
Subsection \ref{sec:quantumdisorder}, we argue that quantum
(non-Ising) exchange couplings that one might want to consider in the
dipolar spin ice model\cite{McClarty} are likely to be much too small
to explain the upturn in the specific heat below a temperature of
$T^*\sim0.5$~K or so in Dy$_2$Ti$_2$O$_7$.  On the other hand, the
same subsection explores a toy-model of random disorder (in the form
of stuffed spins) that could possibly rationalize the specific heat
upturn.  Finally, we close the paper with a brief conclusion in
Section \ref{sec:conclusion}.  With the aim of providing a streamlined
reading of the key results presented in the main text, we have
packaged all the technical details of the simulations, data analysis
and other calculations supporting the key results discussed in the
main text in a series of appendices.

\section{Model and  Monte Carlo Method}
\label{sec:Model_Methods}
\subsection{Model}
\label{sec:model}

\begin{figure}[h!]
  \resizebox{\hsize}{!}{\includegraphics[clip=true]{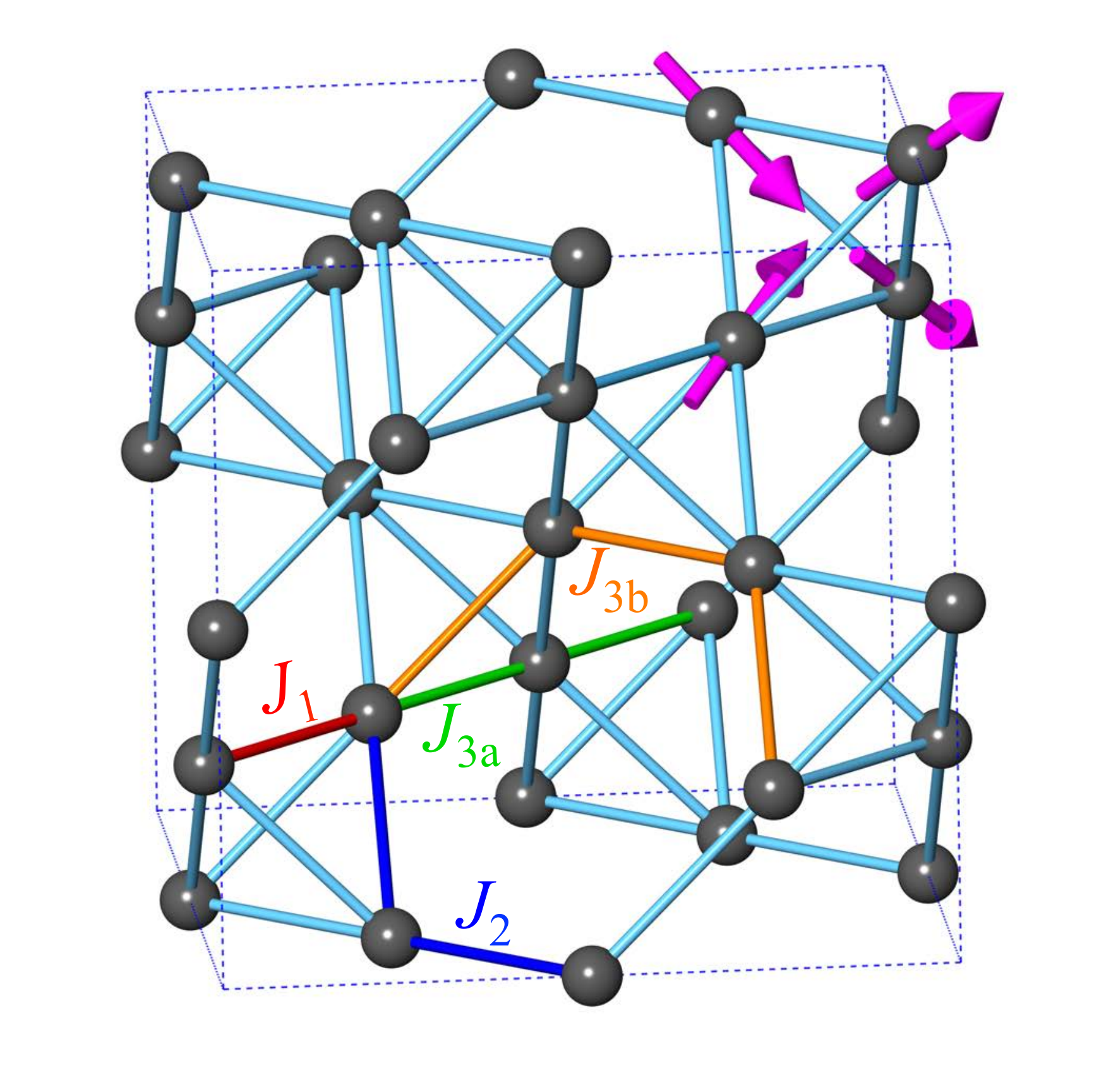}}
  \caption{Set of interacting neighbors on the pyrochlore lattice. The
    first ($J_1$), second ($J_2$) and two distinct third ($J_{3a}$ and
    $J_{3b}$) nearest-neighbor pathways are indicated by red, blue,
    green and orange connections, respectively, on the pyrochlore
    lattice of corner-sharing tetrahedra. A two-in/two-out state of
    two spins pointing into the center of the tetrahedron and two
    spins pointing out from the tetrahedron is shown in the upper
    right hand corner of the lattice. An ice-rule obeying state is
    characterized by all tetrahedra being in a two-in/two-out spin
    configuration, but with no other constraint on the orientation of
    the spins.}
\label{pyro3d}
\end{figure}

In Dy$_2$Ti$_2$O$_7$, the magnetic Dy$^{3+}$ rare-earth moments occupy
the sites of the pyrochlore lattice shown in Fig.~\ref{pyro3d}.  This
structure can be described as a face-centred cubic (FCC) space lattice
with a primitive basis that consists of a tetrahedron\cite{Gardner10}
(four sites).  The microscopic Dy$^{3+}$--Dy$^{3+}$ interionic
magnetic and superexchange couplings are small ($[10^{-2}-10^{-1}]$ K)
compared to the energy splitting ($\Delta \sim 300$ K) between the
crystal-field ground state doublet and the first excited
doublet\cite{Gardner10}.  One can thus project these interactions into
a reduced Hilbert space comprised solely of the single-ion
crystal-field ground states\cite{Gingras11}, ignoring for the moment
unusually strong high rank multipolar Dy-Dy
interactions\cite{santini2009multipolar,Rau15}. Due to the specific
spectral decomposition of the crystal-field ground state
doublet\cite{Gardner10,Gingras11}, the effective (projected)
Hamiltonian can then be expressed as a classical Ising
model\cite{Gingras11}.  We return to this fundamental assumption in
Subsection \ref{sec:quantumdisorder}.  The ``spins'' in that Ising
model interact via distance-dependent ``exchange'' couplings,
$J_{ij}$, between ions $i$ and $j$, as well as through the long-range
magnetostatic dipole-dipole interaction. The Hamiltonian for this
generalized dipolar spin ice model
\cite{denHertog,Taba06,Yavo08,Melko01, Ruff} (g-DSM) in an external
magnetic field $\bm H$ reads:
\begin{align} {\cal H} = &Dr^3_{\rm nn} \sum_{i >j} \frac{ {\bm S}_i
    \cdot {\bm S}_j}{|{\bm r}_{ij}|^3}-\frac{3( {\bm S}_i \cdot {\bm
      r}_{ij})({\bm S}_j
    \cdot {\bm r}_{ij})}{|{\bm r}_{ij}|^5} \nonumber \\
  &+\sum_{i>j} J_{ij} {\bm S}_i\cdot {\bm S}_j - g\mu_B\langle
  J^z\rangle \sum_i {\bm S}\cdot {\bm H} .
\label{eq:ham} 
\end{align} 
The scale of the dipolar interaction at the nearest-neighbor distance,
$r_{\rm nn}$, is given by $D = \mu_0 (g \mu_B \langle J^z \rangle)^2 /
4 \pi r_{\rm nn}^3=1.3224 $ K (Ref.~[\onlinecite{Yavo08}]) using
$\langle J^z \rangle=7.40$ (Ref.~[\onlinecite{Yavo08}]), $r_{\rm
  nn}=3.58$ \AA ~(Ref.~[\onlinecite{Fuka02}]) and the Land\'{e} $g$
factor is $g=4/3$.  Here ${\bm S}_i$ is a classical unit vector
representing the Dy$^{3+}$ magnetic moment at site $i$ which is
constrained by the crystal field Ising anisotropy to point along the
local $\hat z_i$ cubic $[111]$ direction. We thus have ${\bm S}_i =
\sigma_i \hat z_i$ with $\sigma_i=\pm 1$ as the Ising variable
\footnote{ Here, we have chosen to write the expected effective Ising
  exchange interactions for Dy$^{3+}$ ions\cite{Rau15} as ${\bm S}_i
  \cdot {\bm S}_j$ in order to keep with the notation used in several
  previous works.  }. For sites $i$ and $j$ that belong to different
sublattices, $\hat z_i \cdot \hat z_j =-1/3$ while $\hat z_i \cdot
\hat z_j =1$ if they belong to the same sublattice. For most of our
paper, we consider first ($J_1$), second ($J_2$) and two distinct
third-nearest-neighbor interactions ($J_{3a}$ and $J_{3b}$), but take
$J_4$ into account to check the validity of our conclusions.  The
intersite connectivity defined by the $J_1$, $J_2$, $J_{3a}$ and
$J_{3b}$ couplings is illustrated in Fig.~\ref{pyro3d}.

\subsection{Monte Carlo method}
\label{sec:method}

For the most part of this paper, we employed Monte Carlo simulations
to investigate the thermodynamic properties of the model defined by
Eq.~(\ref{eq:ham}).  In order to help interpret the Monte Carlo
results, we also used in Appendix \ref{app:phasediagram} conventional
mean-field theory, formulated in reciprocal space, as described in
Ref.~[\onlinecite{Enjalran_mft}].
 
The Monte Carlo calculations were performed with periodic boundary
conditions and using the Ewald summation technique to handle the
long-range dipolar interaction\cite{Melko04,Enjalran_mft}. The systems
simulated consisted of $L^3$ cubic unit cells each with $16$ sites.
Loop updates\cite{Melko01,Melko04} were used when the single-spin-flip
acceptance rate fell below $1\%$.  In order to ease the investigation
of the phase diagram discussed in Subsection \ref{sec:phasediagram},
loop Monte Carlo simulations were further supplemented by the parallel
tempering method\cite{Marinari92} using $72$ distinct temperatures
distributed between $0.05$ K and $0.7$ K.

The neutron scattering structure factor to be compared with
experiments was calculated according to
\begin{equation}
  S(\boldsymbol{q})=\frac{[f(|\boldsymbol{q}|)]^2}{N}\sum_{ij}\langle s_is_j\rangle
  ({\hat z}_i^{\perp}\cdot {\hat z}_j^{\perp})e^{i\boldsymbol{q}\cdot \boldsymbol{r}_{ij}},
  \label{eq:sq}
\end{equation}
where ${\hat z}_i^{\perp}$ is the component of the local Ising axis
perpendicular to the wave vector $\boldsymbol{q}$ and
$f(|\boldsymbol{q}|)$ is the magnetic form factor of Dy$^{3+}$
(Ref.~[\onlinecite{Brown93}]). Points of particular interest are
$\boldsymbol{q}$=$(0\: 0\: 3)$ and $\boldsymbol{q}$=$(\frac{3}{2}\:\:
\frac{3}{2}\:\:\frac{3}{2})$ with $[f(|0\: 0\: 3|)]^2=0.8224$ and
$[f(|\frac{3}{2}\:\: \frac{3}{2}\:\: \frac{3}{2}|)]^2=0.8627$.

In Subsection \ref{sec:quantumdisorder}, we explore the effects of
local effective magnetic degrees of freedom that may be generated by
some form of random disorder such as stuffing (magnetic ions on the B
site of the pyrochlore lattice, oxygen vacancies, or both).  To do so,
we consider a minimal \emph{effective} impurity model that consists of
impurity magnetic moments on the B sites of the pyrochlore lattice,
termed ``stuffed'' spins\cite{Ross12}
\textsuperscript{,}\footnote{Recent x-ray synchrotron
  work\cite{Cava15} has shown that a small percentage of rare-earth
  (RE) ions occupying (``stuffing'') the B-site nominally occupied by
  the tetravalent transition metal ion (e.g. Ti$^{4+}$) in the
  RE$_2$Ti$_2$O$_7$ pyrochlore oxides is quite common, and found to
  occur, for example, in Ho$_2$Ti$_2$O$_7$, Er$_2$Ti$_2$O$_7$ and
  Yb$_2$Ti$_2$O$_7$.  } .

As a proof of principle, we assume for simplicity that the impurity
spins are Ising-like, with a local moment ${\bm L}_{\alpha}$ pointing
along the line defined by the centers of the B-site tetrahedra on the
pyrochlore lattice\cite{Gardner10}.  The stuffed spin ${\bm
  L}_{\alpha}$ is coupled to its six nearest neighbors ${\bm S}_i$ on
the A sites with an effective exchange interaction
\begin{equation}
  H_{\Omega}= -\Omega\sum_{\langle \alpha, i\rangle} {\bm
    L}_{\alpha}\cdot {\bm S}_i=
\frac{\Omega}{3}\sigma_{\alpha}\sigma_i,
  \label{eq:Omega}
\end{equation}
where $\Omega$ is the coupling constant. Here the index $\alpha$ runs
over all randomly stuffed spins, which are chosen randomly to occupy a
fraction $p$ of the B sites, and $\sigma=\pm 1$.

\section{Results}
\label{sec:Results}

\subsection{[112] magnetic field experiment analysis}
\label{sec:112}

\begin{figure}[h!]
\resizebox{\hsize}{!}{\includegraphics[clip=true]{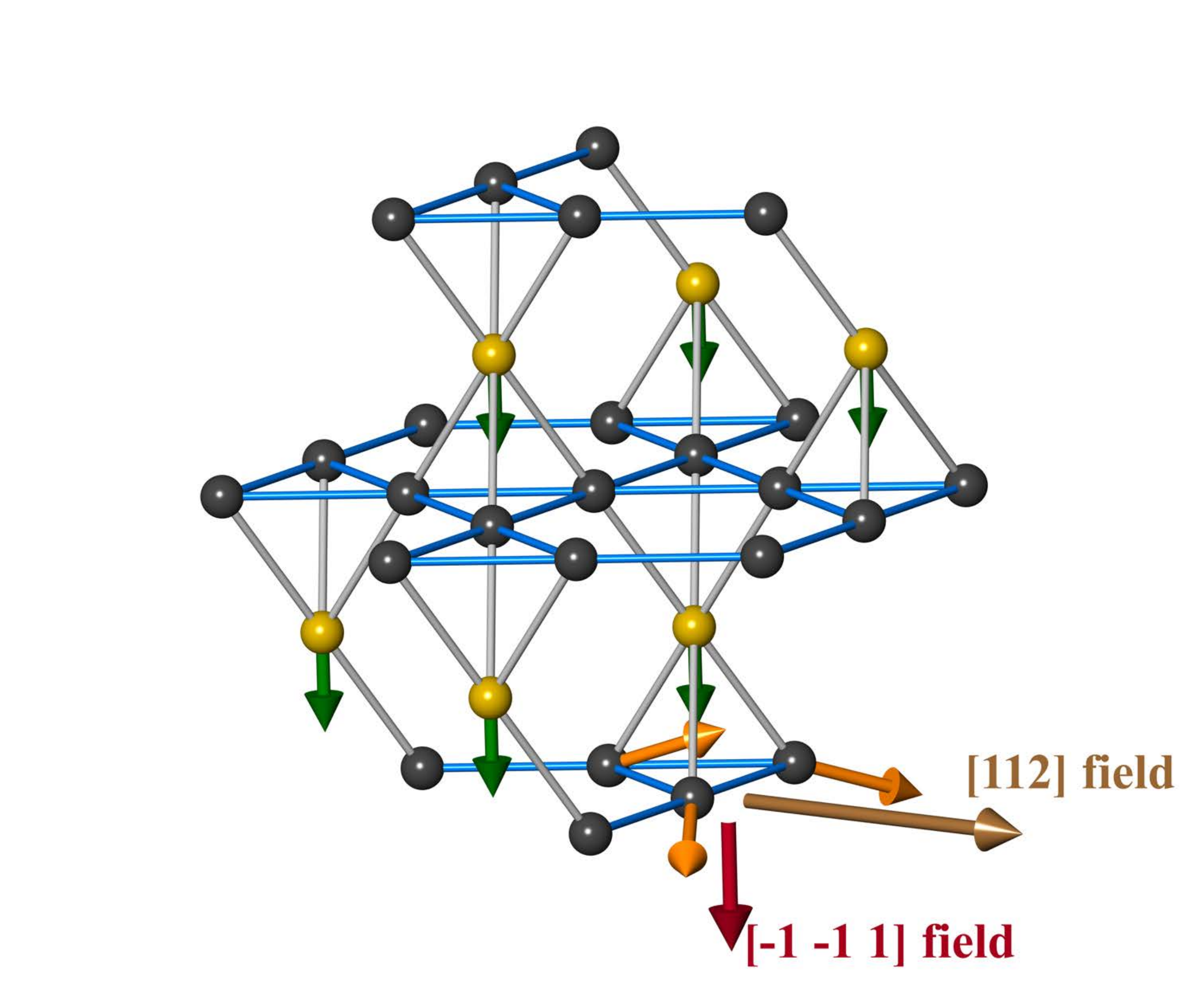}}
\caption{Schematic illustration of the $[112]$ field magnetization
  measurement with the structure of the pyrochlore lattice viewed as
  alternating triangular and kagome planes. An applied field in the
  $[112]$ direction is shown as a brown arrow. An applied field along
  $[\bar {1} \bar {1} 1]$ is shown in red. This field points along the
  local Ising axis of the green spin vectors on the triangular planes
  (yellow sites).  }
\label{112_3D}
\end{figure}

In this subsection, we begin to revisit the set of $J_{ij}$ values
describing Dy$_2$Ti$_2$O$_7$ and, unlike in other g-DSM works
\cite{Ruff,Taba06,Yavo08}, we do not \emph{a priori} assume that
$J_{3a}=J_{3b}$.  To do so, we first analyze magnetization data
measured in a magnetic field near the $[ 1 1 2]$ direction, with a
small magnetic field component along $[ \bar{1} \bar{1} 1]$
(Ref.~[\onlinecite{Sato06}]).  Fig.~\ref{112_3D} illustrates the
geometry associated with this experiment.  The analysis of this
experiment allows us to establish constraints among the $J_{ij}$
parameters and thus reduce the dimensionality of the model parameter
space that needs to be explored to describe Dy$_2$Ti$_2$O$_7$.

This ``$[1 1 2]$ experiment'' is rather remarkable in that it provides
us with direct access to the third-nearest-neighbor interactions,
$J_{3a}$ and $J_{3b}$.  In this experiment, the large $[1 1 2]$
component of the applied field saturates the magnetization on three of
the four face-centred cubic (FCC) sublattices that constitute the
pyrochlore lattice.  At the same time, a small $[ \bar{1} \bar{1} 1]$
magnetic field component can be tuned so that the vector sum of the
external applied plus the combined exchange and dipolar internal
fields lead to a decoupling of the remaining FCC sublattice from the
three fully polarized sublattices\cite{Ruff}.

In Fig.~\ref{112_3D}, these decoupled magnetic moments reside on the
triangular lattice indicated by the yellow sites with a downward green
arrow.  Since the nearest-neighbor distance on the FCC sublattice
corresponds to the third-nearest-neighbor distance on the pyrochlore
lattice, an analysis of the susceptibility for the field component
along $[ \bar{1} \bar{1} 1]$ yields a relation between $J_{3a}$ and
$J_{3b}$ that is almost linear and can be approximately described by
(all in kelvin),
\begin{equation}
  J_{3b}=\left\{ \begin{array}
      {r@{\quad:\quad}l}
      -0.667 J_{3a} +0.03 & J_{3a}<0.1\\
      -0.842 J_{3a} +0.0474 &  J_{3a}>0.1.
\end{array} \right.
 \label{eq:j3}
\end{equation}
Furthermore, the applied experimental critical field in the
[$\bar{1}\bar{1}$1] direction that decouples the corresponding FCC
sublattice from the internal ice rules, enforced by the combined
exchange plus dipolar field, yields another linear relation between
$J_1$ and $J_2$,
\begin{equation}
  J_2=-\frac{1}{2} J_1 + 1.555.
 \label{eq:j2}
\end{equation}
A derivation of these relations is given in Appendix \ref{app:112}.
In the following three subsections, we make use of the constraints
defined by Eqs. (\ref{eq:j3}) and (\ref{eq:j2}) to analyze the ground
state, spin-spin correlations and associated neutron scattering
function and thermodynamic properties of the model in the resulting
$J_1$-$J_{3a}$ parameter space.

\subsection{Phase diagram}
\label{sec:phasediagram}

\begin{figure}[h!]
\resizebox{\hsize}{!}{\includegraphics[clip=true]{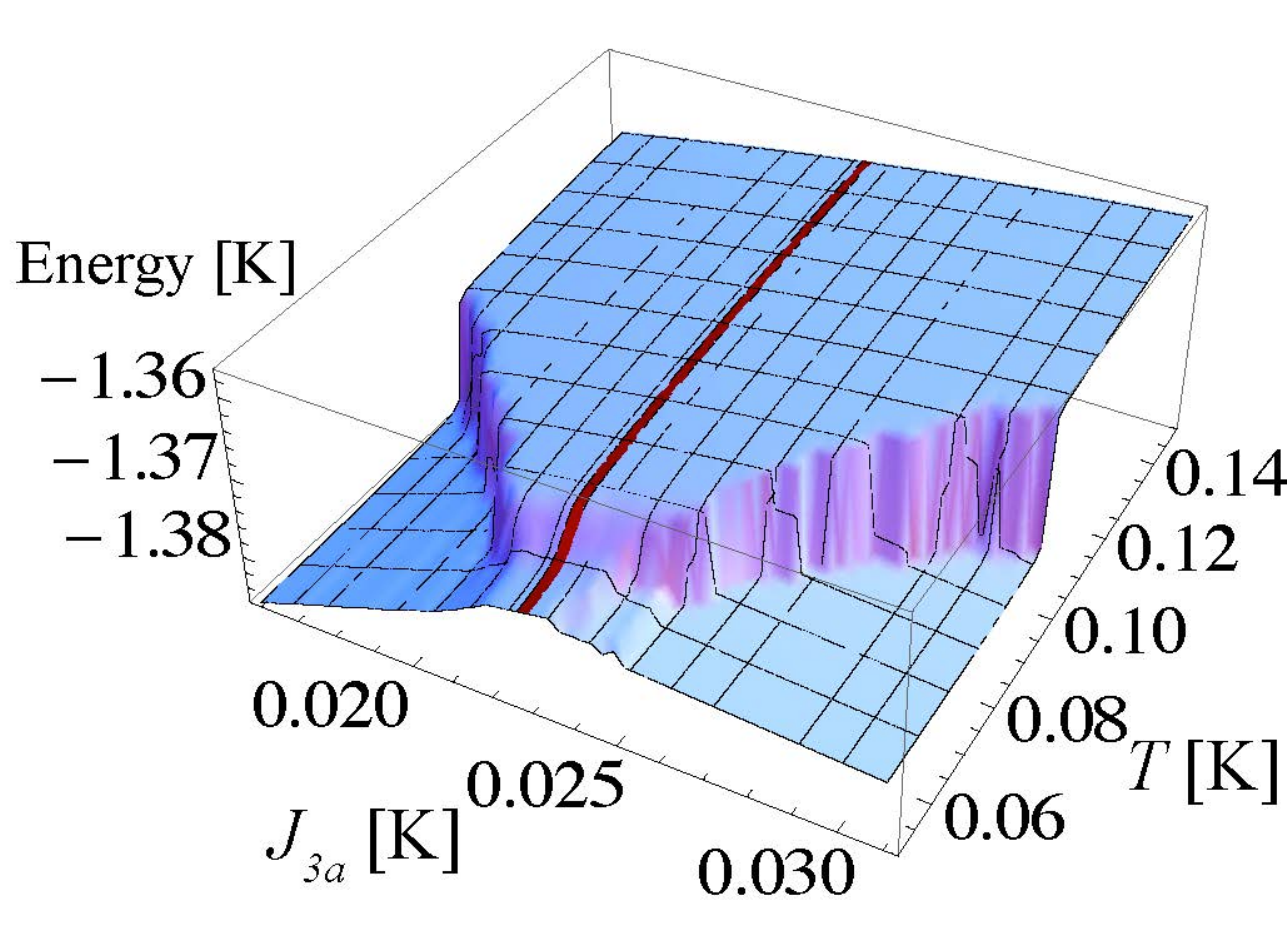}}
\caption{The average thermal energy as a function of $T$ and $J_{3a}$
  near the boundary between the single- and double-chain states for
  $J_1=3.41$ K. The phase boundary is located at $J_{3a} \approx
  0.0228$ K and is marked by a red band. The phase transition is
  clearly first order away from the boundary, while the critical
  temperature is suppressed with the transition appearing to be close
  to continuous at the boundary. }
\label{energy_boundary}
\end{figure}

We now proceed to explore the phases and ordering wave vectors in the
constrained $J_1\mbox{-}J_{3a}$ parameter space where $J_2$ and
$J_{3b}$ have been eliminated via Eqs.~(\ref{eq:j3}) and
(\ref{eq:j2}).  From extensive experimental and theoretical work, we
know that $J_1 = 3.3\pm 0.2$ K, and $|J_3| < 0.2$ K
(Refs.~[\onlinecite{Bramwell_Science, Yavo08}]). Furthermore, work on
the simple DSM with $J_2=J_{3a}=J_{3b}=0$ has
shown\cite{Gingras_CJP,Melko01} that, despite its highly frustrated
nature, (reciprocal) ${\bm q}$-space mean-field
theory\cite{Enjalran_mft} could be used to identify the candidate
ordered state that ultimately develops if non-local loop dynamics are
used to maintain equilibrium down to the ordering temperature
\cite{Melko01,Melko04}.  By combining a mean-field theory survey of
ordering wave vectors at the mean-field critical temperature with a
direct analysis of the actual ordered states that appear in the Monte
Carlo calculations (see Appendix \ref{app:phasediagram} for details of
the calculations), we find that there are two long-ranged ordered
ground states in the parameter range appropriate for Dy$_2$Ti$_2$O$_7$
(and as constrained by the $[112]$ experimental results above).  From
the Monte Carlo results, the propagation vector of each ground state
is $(1\ 1\ 0 )$ and $(\frac{1}{2}\ \frac{1}{2}\ 0)$. The first state
was previously identified for the dipolar spin ice model with
$J_2=J_{3a}=J_{3b}=0$ (Ref.~[\onlinecite{Melko01}]). In this state,
parallel chains of spins order antiferromagnetically when viewed along
a cubic $\langle100\rangle$ axis (see Fig.~\ref{chains}a). We call
this the ``single-chain state''.  In the second state, pairs of
adjacent spin chains are aligned, but each pair is antiparallel with
the adjacent pairs and we refer to this as the ``double-chain state''
(see Fig.~\ref{chains}b).  The double-chain state is particularly
interesting since the energy difference between different stackings in
the $z$-direction of the $(001)$ plane of spins, shown in
Fig.~\ref{chains}c, is only $O(10^0)$~mK, and this state is thus
quasi-degenerate.  Without the long-ranged dipolar interaction,
different stackings have exactly the same (degenerate) energy. This is
yet another manifestation of a self-screening effect emerging in spin
ice\cite{denHertog, Gingras_CJP, Isakov_SS}.  This stacking degeneracy
signals the \q{refrustration} of the DSM\cite{denHertog,Melko01}
alluded to in the title. It arises from the mutual competition of the
(dimensionless) perturbative energy scales ($J_2/J_{\rm eff}$,
$J_{3a}/J_{\rm eff}$, $J_{3b}/J_{\rm eff}$; $J_{\rm eff} \equiv
[5D-J_1]/3$) (Refs.~[\onlinecite{Bramwell_Science,denHertog}]) which
mask the true ground-state order parameter at
$\boldsymbol{q}=(\frac{1}{2}\ \frac{1}{2}\ 0)$. For example, the main
intensity peak in $S(\boldsymbol{q})$ at a parameter point within the
double chain region at temperatures above the transition temperature
appears at $\boldsymbol{q}=(\frac{1}{2}\ \frac{1}{2}\ \frac{1}{2})$,
rather than at the ground state order wave vector
$\boldsymbol{q}=(\frac{1}{2}\ \frac{1}{2}\ 0)$.  These results, which
are reminiscent of the phenomenology at play in the three-dimensional
ANNNI model\cite{Selke}, are further discussed in Appendix
\ref{app:phasediagram}.

To determine the phase boundary, we are prompted by the observation
from the Monte Carlo simulations that the two ordered states are
formed by ferromagnetically-ordered spin chains. By considering these
spin chains as the fundamental units of the system, we determine the
following equation for the phase boundary,
\begin{equation}
  J_{3a}+\frac{J_{2}}{3} + 0.02D=0. 
\label{eq:boundary}
\end{equation} 
In the boundary region, phase competition increases the energy while
\emph{decreasing} the critical ordering temperature towards either
ground state. This is explicitly demonstrated in
Fig.~\ref{energy_boundary}, where the average energy is displayed as a
function of temperature and $J_{3a}$ as one crosses the boundary at
$J_{3a}\cong 0.023$ K for $J_1=3.41$ K. While there is a clear
first-order transition away from the boundary, it is much smoother at
the boundary.  The precise determination of the order of the phase
transitions is beyond the scope of the present work.  Having
determined the key candidate long-range ordered phases for
Dy$_2$Ti$_2$O$_7$, as well as the location of the phase boundary
separating the competing ground states, we now proceed to position
this material within this parameter space.

\subsection{Neutron scattering data analysis}
\label{sec:neutronscatt}

\begin{figure*}[ht]
  \begin{center}
 \includegraphics[width=18cm]{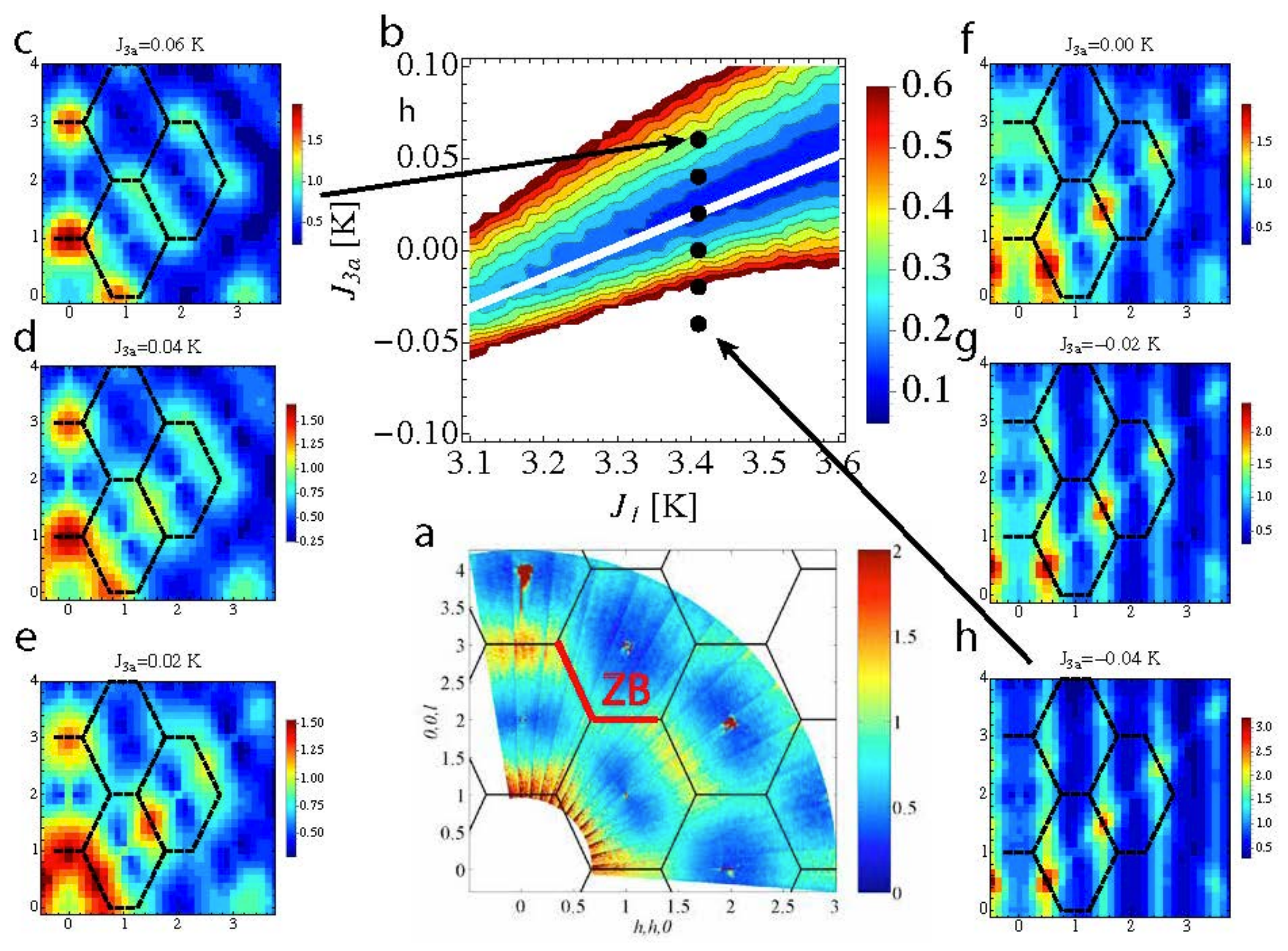}
 \caption{ Neutron structure factor and zone boundary scattering in
   the $(hhl)$ plane across the phase boundary. {\bf a,} Experimental
   neutron scattering data at $0.3$ K (Ref.~[\onlinecite{Fennell04}])
   with a section of the zone-boundary, ZB, highlighted. {\bf b,}
   Deviation from constant scattering along the boundary segment ZB in
   the $J_1\mbox{-}J_{3a}$ plane calculated through Monte Carlo
   simulations using Eq.~(\ref{eq:ZBS}). The phase boundary according
   to Eq.~\ref{eq:boundary} is shown as a white line.  Note that the
   most ``${\bm q}$-space intensity-flat'' ZBS (dark blue) is centered
   along the phase boundary. The corresponding value for the
   experimental data is $\sigma_{\rm ZBS}=0.11$, in remarkably good
   agreement with the Monte Carlo result along the phase boundary.
   {\bf c-h,} Monte Carlo results for $S(\boldsymbol{q})$ in the
   vicinity of the single-chain (upper left region) to the
   double-chain transition (lower right region) for $J_1=3.41$ K at
   $T=0.5$ K. The corresponding parameters in the $J_1\mbox{-}J_{3a}$
   space are marked by black dots.  The intensity shifts from $(0\ 0\
   3)$ to $(\frac{3}{2}\ \frac{3}{2}\ \frac{3}{2})$ as one crosses
   from the single-chain region ({\bf c}) to the double-chain region
   ({\bf h}). }
\label{neutron_boundary}
\end{center}
\end{figure*}
As neutron scattering measurements provide direct information about
the spin-spin correlations, an analysis of available neutron
scattering data is a natural starting point for positioning
Dy$_2$Ti$_2$O$_7$ in the $J_1\mbox{-}J_{3a}$ phase diagram introduced
in the previous subsection.

Before we begin presenting our results, we briefly comment on pinch
point singularities in the neutron scattering pattern of spin ice.
The topic of pinch point singularities in the equal time (energy
integrated) neutron scattering data of spin ice had been the subject
of much theoretical\cite{Henley_AnnRev,Isakov_SS,Sen_dip} and
experimental\cite{Fennell_Science,Morris2009,Clancy} discussion.
Pinch points in dipolar spin ice arise from the combination of the
divergence-free condition of the coarse-grained magnetization field
due to the two-in/two-out ice rules as well as the long-range
dipole-dipole interaction\cite{Isakov_SS,Sen_dip}, even at high
temperature well above the formation of the spin ice manifold. As
pinch points result from these two generic phenomenologies, they are
not, apart from their ultimate disappearance\cite{Conlon_ZBS}, weighty
signatures of the competing subleading exchange interactions beyond
nearest neighbor that the present study aims to expose.  We therefore
henceforth omit pinch points in our analysis and discussion of the
neutron scattering data of Dy$_2$Ti$_2$O$_7$.

The main result of our analysis is shown in
Fig.~\ref{neutron_boundary} where we consider the evolution of the
neutron structure factor $S(\boldsymbol{q})$ upon crossing the
single/double-chain phase boundary and compare our numerical results
to existing experimental data.  We display in
Fig.~\ref{neutron_boundary}a the sub-kelvin measurement of
$S(\boldsymbol{q})$ reported in Ref.~[\onlinecite{Fennell04}] and
recorded at $0.3$ K.  We note two distinguishing features: pronounced
peaks of roughly equal intensity at $(0\ 0\ 3)$ and $(\frac{3}{2}\
\frac{3}{2}\ \frac{3}{2})$ and a ridge of rather flat scattering
intensity spanning the Brillouin zone boundary.  We refer to this
feature as {\it zone boundary scattering} (ZBS). The same fundamental
features can also be discerned in the measurement recorded at $1.3$ K
in Ref.~[\onlinecite{Fennell04}], but less clearly because of the
weaker correlations at this higher temperature.

We first analyze the ZBS, whose defining signature observed here is a
scattering intensity along the Brillouin zone boundary that is close
to constant.  A previous study\cite{Yavo08} showed that ZBS in a
dipolar spin ice, in particular Dy$_2$Ti$_2$O$_7$, is an indication of
competing Ising exchange interactions $J_{ij}$ beyond nearest-neighbor
$J_1$, originally the only exchange coupling taken into account along
with the long-range dipolar interactions in the DSM\cite{denHertog}.
A similar phenomenology has been discussed in the context of the
pyrochlore lattice with classical Heisenberg spins subject to beyond
nearest-neighbor competing superexchange
interactions\cite{Conlon_ZBS}.

As a quantitative measure of the ZBS feature in the Monte Carlo
calculations, we use the deviation from flatness,
\begin{equation}
\sigma_{\rm ZBS}^2=\frac{1}{N_q}\sum_{i=1}^{N_q}
[S_{\text{MC}}(q_i)-\langle S_{\text{MC}}\rangle]^2,
\label{eq:ZBS} 
\end{equation} 
along a zone boundary line segment (ZB) indicated in the experimental
neutron scattering in Fig.~\ref{neutron_boundary}a. The average of the
scattering intensity along the segment is denoted $\langle
S_{\text{MC}}\rangle$ and the simulation was performed for a system
size $L=8$ with periodic boundary conditions, yielding $N_q=9$
distinct ${\bm q}$-points on the zone boundary.

We display in Fig.~\ref{neutron_boundary}b a color map of $\sigma_{\rm
  ZBS}$ in the constrained $J_1\mbox{-}J_{3a}$ parameter space and
note that the deviation from flatness is \emph{smallest} along the
phase boundary, indicating that flat ZBS is confined to the
$J_1\mbox{-}J_{3a}$ boundary region between the two competing ground
states identified in Subsection \ref{sec:phasediagram}

\textbf{\textit{The first main result }} of our study is that
\emph{only} in the vicinity of the phase boundary (white line in
Fig.\ref{neutron_boundary}b) is there significant ZBS \emph{as well
  as} peaks of roughly equal intensity at $(0\ 0\ 3)$ and
$(\frac{3}{2}\ \frac{3}{2}\ \frac{3}{2})$.  These wave vectors,
equivalent to $(1\ 1\ 0 )$ and $(\frac{1}{2}\ \frac{1}{2}\
\frac{1}{2})$ in the first Brillouin zone, are the main intensity
peaks for the single and double chain states, respectively. This
result can be deduced from Figs.~\ref{neutron_boundary}c-h, where we
display $S(\boldsymbol{q})$, calculated from Monte Carlo simulations,
along a cut across the phase boundary (the solid white line) at
$J_1=3.41$ K.  Note that the peak at $(0\ 0\ 3)$ is pronounced in the
single-chain region above the phase boundary, while the peak at
$(\frac{3}{2}\ \frac{3}{2}\ \frac{3}{2})$ dominates in the
double-chain region below the boundary, as expected. Only in panels
Fig.~\ref{neutron_boundary}d and Fig.~\ref{neutron_boundary}e, close
to the boundary, are the two peaks of similar intensity.  Comparison
with the experimental $S(\boldsymbol{q})$ data thus provides
compelling evidence that the appropriate set of exchange parameters
for Dy$_2$Ti$_2$O$_7$ puts the material quite close to the phase
boundary between the single- and double-chain ground states.  Note
that this realization was much facilitated by the dimensional
reduction of the $J_{ij}$ parameter space using the above analysis of
the $[112]$ magnetization experiment in Subsection \ref{sec:112}.

Considering the extraordinarily slow relaxation observed in
experiments upon cooling\cite{Revell13, Poma13}, one could ask whether
the neutron scattering data at $0.3$ K are adequately equilibrated and
therefore able to form the basis of a systematic analysis. It is
therefore necessary to consider what effects a possible freezing of
the sample may have on the properties of $S(\boldsymbol{q})$ that we
consider here. In a Monte Carlo exploration of these effects, we find
that there is \emph{no} significant change in the ZBS and \emph{no}
fundamental reciprocal space redistribution in the peak intensities
due to freezing.  The details of the Monte Carlo results supporting
this result are discussed in Appendix \ref{app:neutronscatt}.  We
therefore conclude that relative $ (0\ 0\ 3)$ and $(\frac{3}{2}
\frac{3}{2} \frac{3}{2})$ peak intensities \emph{and} ZBS in the
experimental neutron scattering data place Dy$_2$Ti$_2$O$_7$ firmly
near the boundary between the two competing states,
\emph{irrespective} of the possibility that the spin dynamics may have
frozen at $T>0.3$ K. Furthermore, and quite importantly, our analysis
of the ratio of the peak intensity at $ (0\ 0\ 3)$ and $(\frac{3}{2}\
\frac{3}{2}\ \frac{3}{2})$ discussed in Appendix
\ref{app:neutronscatt} reaches the very same conclusion \emph{already}
at $T= 1.3$ K, where Dy$_2$Ti$_2$O$_7$ has barely entered the spin ice
regime and equilibration is not contentious.

\subsection{Specific Heat Analysis}
\label{sec:specheat}

\begin{figure*}[ht]
\begin{center}
 \includegraphics[width=18cm]{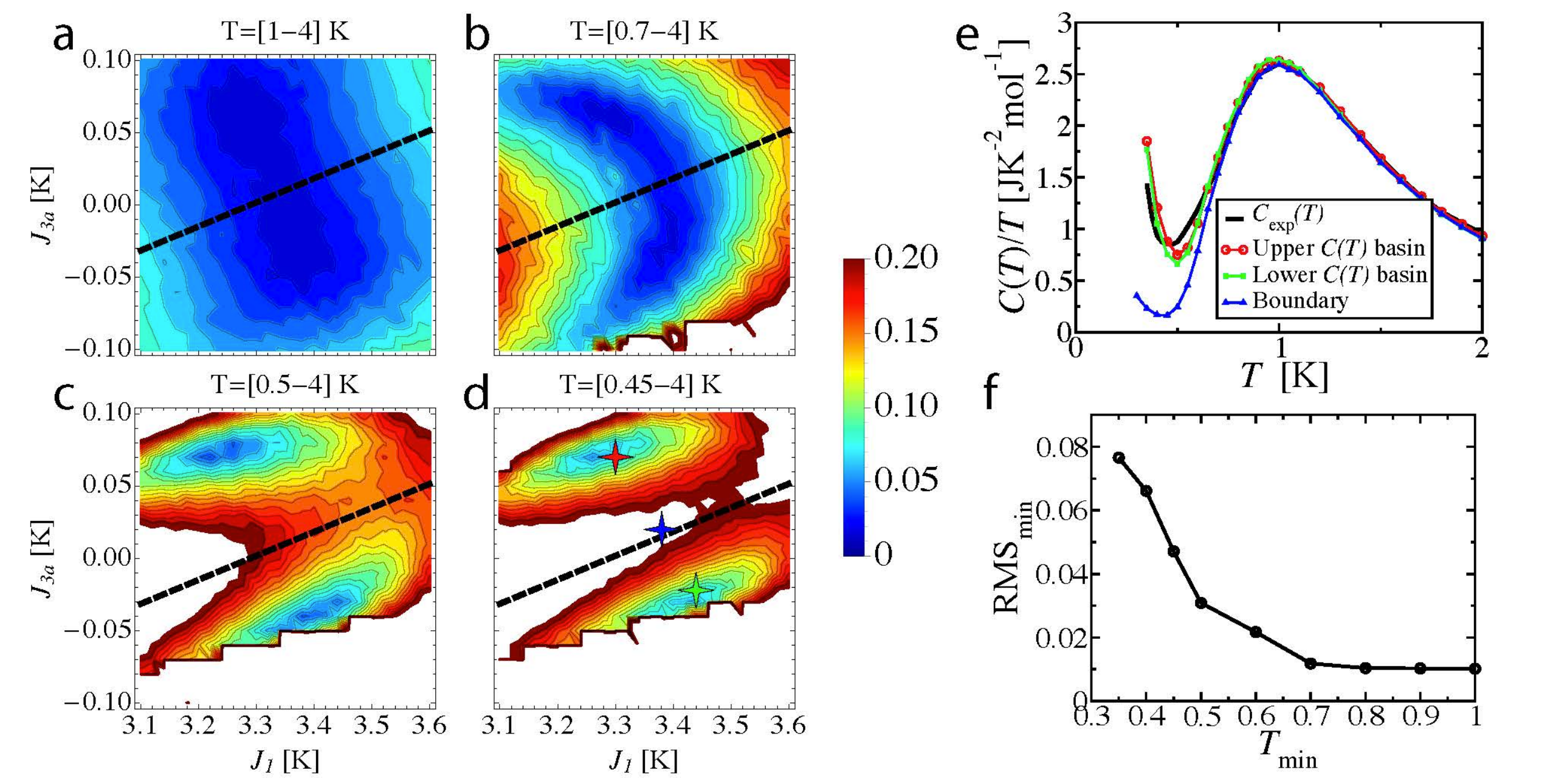}
 \caption{Goodness of fit of Monte Carlo data compared to experimental
   specific heat. {\bf a-d,} RMS error deviation, $\sigma_C$, of the
   dipolar spin-ice model in a constrained $J_1$-$J_{3a}$ parameter
   space compared to recently reported specific heat
   measurements\cite{Poma13}. The experimental specific heat data
   displayed in panel {\bf e } are the raw data minus the calculated
   nuclear portion, $C_{\text{exp}}(T) \equiv C_{\text{raw}}(T) -
   C_{\text{nuc}}(T)$ (see Appendix \ref{app:nuclearspecheat}).  The
   label above each panel gives the temperature range of the fit. A
   dark blue color indicates a good fit, while white regions are off
   the scale. {\bf e,} Specific heat divided by temperature for the
   experimental data (thick black line), a representative point in the
   upper basin (red open circles and red line), a representative point
   in the lower basin (filled green up triangles and green line) and a
   point on the boundary (filled blue squares and blue line). The
   corresponding $J_1$-$J_{3a}$ points are marked by correspondingly
   colored stars in panel {\bf d}. {\bf f,} The minimal RMS error in
   the entire $J_1$-$J_{3a}$ plane as a function of the lower limit of
   the temperature range of the fit, $T_{\text{min}}$. Note the rapid
   increase in the RMS error below $T \approx 0.7$ K, indicating the
   rapid onset in the failure of the model to describe the
   experimental specific heat data, $C_{\text{exp}}(T)$. }
\label{cv_rms}
\end{center}
\end{figure*}

The temperature dependence of the magnetic specific heat and, most
importantly, the magnetic entropy, is one of the key indicators of the
formation of the spin ice manifold.  It is therefore necessary to
consider how the recent specific heat measurements\cite{Poma13} on
Dy$_2$Ti$_2$O$_7$ fit in with the spin interaction model parameterized
above to describe the main features of the neutron scattering pattern.

In the previous subsection we showed that neutron scattering data for
Dy$_2$Ti$_2$O$_7$ are well-described by the dipolar spin ice model
with up to third-nearest-neighbor exchange. Our analysis of the
neutron scattering data indicates that the material sits near the
boundary between two ordered classical states and that our results are
in good agreement with a previous analysis that did not rely on
neutron scattering data\cite{Yavo08}.  On the boundary, we find that
the system exhibits a Pauling entropy plateau for temperatures
$200$~mK $\lesssim T \lesssim$ $600$~mK (see Fig.~\ref{entropy}).  In
possible contradiction with this conclusion stands the recent specific
heat measurement employing very long equilibration time
scales\cite{Poma13} which finds a rise of their raw
$C_{\text{raw}}(T)$ specific heat data below the unexpectedly high
temperature $T^*\sim 0.5$ K and a consequential lack of a Pauling
entropy plateau. We investigate whether the specific heat data can be
modeled with our constrained $J_{1}\mbox{-}J_{3a }$ parameter space.
However, before proceeding to do so, we revisit what is the expected
nuclear specific heat contribution, $C_{\text{nuc}}(T)$, to the
low-temperature ($T\lesssim 1$ K) specific heat data reported in
Ref.~[\onlinecite{Poma13}].  We then assess whether the results for
the electronic specific heat (alone) are consistent with our
conclusion based on our neutron scattering analysis.

\subsubsection{Nuclear specific heat in Dy$_2$Ti$_2$O$_7$}
\label{sec:nuclearspecheat}

Dy$_2$Ti$_2$O$_7$ and Ho$_2$Ti$_2$O$_7$ are the two prototypical
dipolar spin ice materials.  The latter, with its low neutron
absorption cross-section has been a favorite for neutron scattering
studies\cite{Harris,BramwellHo2Ti2O72001,Clancy}.  On the other hand,
the very large hyperfine contact interaction in Ho makes
Ho$_2$Ti$_2$O$_7$ less attractive for specific heat measurements, with
the nuclear contribution to the specific heat overshadowing the
electronic part for $T\lesssim 1.5$ K
\cite{arXivHTO,BramwellHo2Ti2O72001,Cornelius}.  Conversely,
Dy$_2$Ti$_2$O$_7$, because of the small hyperfine contact interaction
of the only two nuclear spin-carrying $^{161}$Dy and $^{163}$Dy
isotopes in Dy$_2$Ti$_2$O$_7$ with natural Dy isotope abundance, has
generally been viewed as a much better suited compound for
calorimetric measurements.

In their work, Pomaranski {\it et al.}\cite{Poma13} argued that the
nuclear contribution to their (presumed) equilibrated specific heat
measurements down to 0.35 K is entirely negligible. To reach that
conclusion, the authors of Ref.~[\onlinecite{Poma13}] referred to the
nuclear specific heat, $C_{{\text{nuc}}}(T)$, estimated for the
Dy$_3$Ga$_5$O$_{12}$ garnet in Ref.~[\onlinecite{Filippi}].  This
estimation, transferred to Dy$_2$Ti$_2$O$_7$ using the
high-temperature series expansion form for $C_{{\text{nuc}}}(T)$ of
Ref.~\onlinecite{Filippi} is quantitatively incorrect since the
low-temperature static Dy$^{3+}$ magnetic moment in
Dy$_3$Ga$_5$O$_{12}$ is about $\sim 4.5$ $\mu_{\text{B}}$, and thus
significantly smaller than the $\sim 9.8$ $\mu_{\text{B}}$ value in
Dy$_2$Ti$_2$O$_7$. In practice, this means that the net hyperfine
contact interaction, $A\langle J^z\rangle$, arising in the nuclear
hyperfine term $A {\bm I}\cdot {\bm J}$, where ${\bm I}$ is the
nuclear spin and ${\bm J}$ is the total electronic spin, was
implicitly taken to be roughly $9.9/4.5 \sim 2.2$ times too small in
Ref.~[\onlinecite{Poma13}].  Since Dy$^{3+}$ has very strong Ising
easy axis anisotropy in Dy$_2$Ti$_2$O$_7$, the nuclear partition
function fully factorizes out from the electronic one\cite{Mattis}.
As the nuclear specific heat scales as $\sim A^2 \langle
J^z\rangle^2/T^2$ for temperature $T \gg A \langle J^z\rangle$, we
argue that the nuclear specific heat estimated at $T\sim 0.5$ K in
Ref.~[\onlinecite{Poma13}] to be about 4.5 times too small.  Appendix
\ref{app:nuclearspecheat} discusses in more detail the subtraction of
the nuclear specific heat, $C_{\text{nuc}}(T)$, from the raw
experimental data, $C_{\text{raw}}(T)$, reported in
Ref.~[\onlinecite{Poma13}], with the magnitude of the correction
displayed in Fig.~\ref{nuclear}.  In what follows, we refer to the
experimental (electronic) magnetic specific heat, $C_{\text{exp}}(T)$,
to be compared with the Monte Carlo data, defined as
$C_{\text{exp}}(T) \equiv C_{\text{raw}}(T) - C_{\text{nuc}}(T)$.

\subsubsection{Magnetic specific heat}
\label{sec:magneticspecheat}

Similarly to the analysis of the neutron scattering data, a goodness
of fit parameter is required to assess the ability of the model to
describe the experimental specific heat data, $C_{\text{exp}}(T)$,
against the Monte Carlo simulation data $C_{\text{MC}}(T)$.  In this
study, we calculate the specific heat goodness of fit, $\sigma_C$,
according to
\begin{equation}
  \sigma_C^2=\sum_{i=1}^{N_T} 
  \frac{[C_{\text{MC}}(T_i)-C_{\text{exp}}(T_i)]^2}{T_i^2 N_T},
\end{equation}
and we use $N_T=42$ distinct temperatures $T_i$ between $T=0.45$ K and
$T=4$ K.  The experimental data, $C_{\text{exp}}(T_i)/T_i$, come from
three different sources: below $T=0.8$ K from Pomaranski {\it et
  al.}\cite{Poma13}, between $0.8$ K and $1.4$ K from Klemke {\it et
  al.}\cite{Klemke11}, and above $1.4$ K from Higashinaka {\it et
  al.}\cite{Higa02}. By applying a cubic spline fit to the
experimental data, suitable temperature points $T_i$ were selected.

We display in Fig.~\ref{cv_rms} the RMS deviation, $\sigma_C$, of
Monte Carlo data, $C_{\text{MC}}(T)$, from $C_{\text{exp}}(T)$ (See
Subsection \ref{sec:method}), which we refer to as RMS in
Figs.~\ref{cv_rms}(a-d) and Fig. \ref{cv_rms}f.  The temperature
interval for the comparison is extended from $[1\mbox{-}4]$ K in
Fig.~\ref{cv_rms}a to $[0.45\mbox{-}4]$ K in Fig.~\ref{cv_rms}d.  We
find a single wide basin of low $\sigma_C\sim 0.01$ for the
$[1\mbox{-}4]$ K high-temperature range (Fig.~\ref{cv_rms}a), but
observe that the quality of the fit deteriorates rapidly to
$\sigma_C\sim 0.08$ with two separate basins of lowest $\sigma_C$
developing when the temperature range is extended to include the
experimental upturn at $T\lesssim 0.5$ K (Fig.~\ref{cv_rms}d).

If the upturn in the experimental specific heat were caused by an
impending ordering transition in the material at a critical
temperature $T_c\sim [0.25-0.30]$ K, we would expect that points away
from the phase boundary would naturally yield a better fit to the
experimental data since the transition temperature is suppressed by
phase competition in the vicinity of the boundary, as was explicitly
demonstrated in Fig~\ref{energy_boundary}.  To further analyze this,
we display in Fig.~\ref{cv_rms}e the specific heat for a point in the
upper basin ($J_1$=$3.30$, $J_{3a}$=$0.07$), the lower basin
($J_1$=$3.44$, $J_{3a}$=$-0.02$) and on the boundary ($J_1$=$3.38$,
$J_{3a}$=$0.02$).  As expected, the specific heat at the two points
away from the boundary turn up at a higher temperature and therefore
yield a somewhat better match to the experimental data.  However, note
that as the temperature range of the fit is extended below
$T_{\text{min}}=0.7$ K there is really \emph{no} point in the
parameter space that matches the experimental data well. In
Fig.~\ref{cv_rms}f, we plot the minimal RMS error (RMS$_{\text{min}})$
in the entire $J_1\mbox{-}J_{3a}$ plane as a function of the lower
limit of the temperature range of the fit, $T_{\rm min}$.

The rapid increase of the RMS error below $0.7$ K indicates that the
observed rise of the specific heat is not caused by an impending
ordering transition in the material since the model of
Eq.~(\ref{eq:ham}) should naively be able to describe such a
transition while maintaining a not too strongly temperature dependent
RMS deviation between $C_{\text{MC}}(T)/T$ and $C_{\text{exp}}(T)/T$.
The reason for this increase in the RMS deviation is that the ordering
transition for parameter points far away from the boundary is strongly
first order, as was shown in Fig~\ref{energy_boundary}. We would
therefore expect only small pre-transitional fluctuations, and a very
rapid and sudden rise of the specific heat. The Monte Carlo results
for the two basins (red and green curves) fall deeper than the
experimental (black) curve, and rise more rapidly, indicating that the
experimental data are unable to keep up with the first-order
transition of the DSM with $J_{ij}$ parameters corresponding to the
upper or lower basin.  In the next section, we further elaborate on
this point by making a detailed comparison of the neutron scattering
and specific heat data, see in particular Fig.~\ref{entropy}.
 
\subsection{Comparison of Neutron data and Specific Heat}
\label{sec:specheatneutroncomp}

\begin{figure}[h!]
  \resizebox{\hsize}{!}{\includegraphics[clip=true]{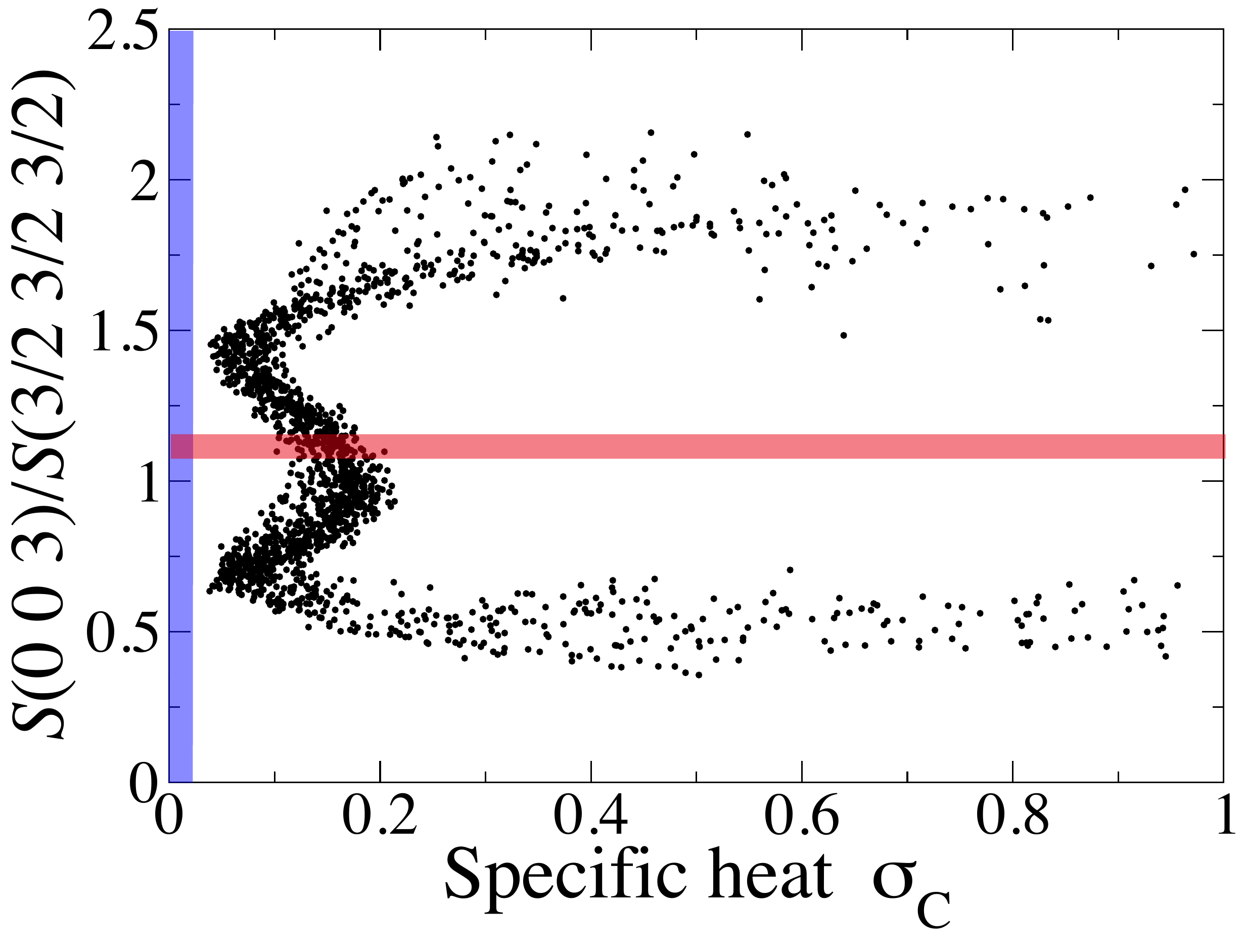}}
  \caption{Ratio of neutron intensity versus goodness of specific heat
    fit in Monte Carlo. The $J_1-J_{3a}$ space has been parametrized
    on a mesh. For each point, the structure factor ratio $r\equiv
    S(0\ 0\ 3)/S(\frac{3}{2}\ \frac{3}{2}\ \frac{3}{2})$ at $T=1.3$ K
    is displayed against the specific heat deviation from
    experiment\cite{Poma13}, $\sigma_C$ for the $[0.45\mbox{-}4]$ K
    temperature interval. The horizontal red band is centered at the
    experimental ratio of $r\cong1.13$, while the vertical blue band
    indicates the region of a good match with the experimental data
    ($\sigma_C<0.2$). Note that there are no points close to the
    intersection of the two bands. Such points would represent
    parameter sets that model \emph{both} the experimental neutron
    scattering and specific heat data well.}
\label{scatter}
\end{figure}

\begin{figure*}[ht]
\begin{center}
\includegraphics[width=18cm]{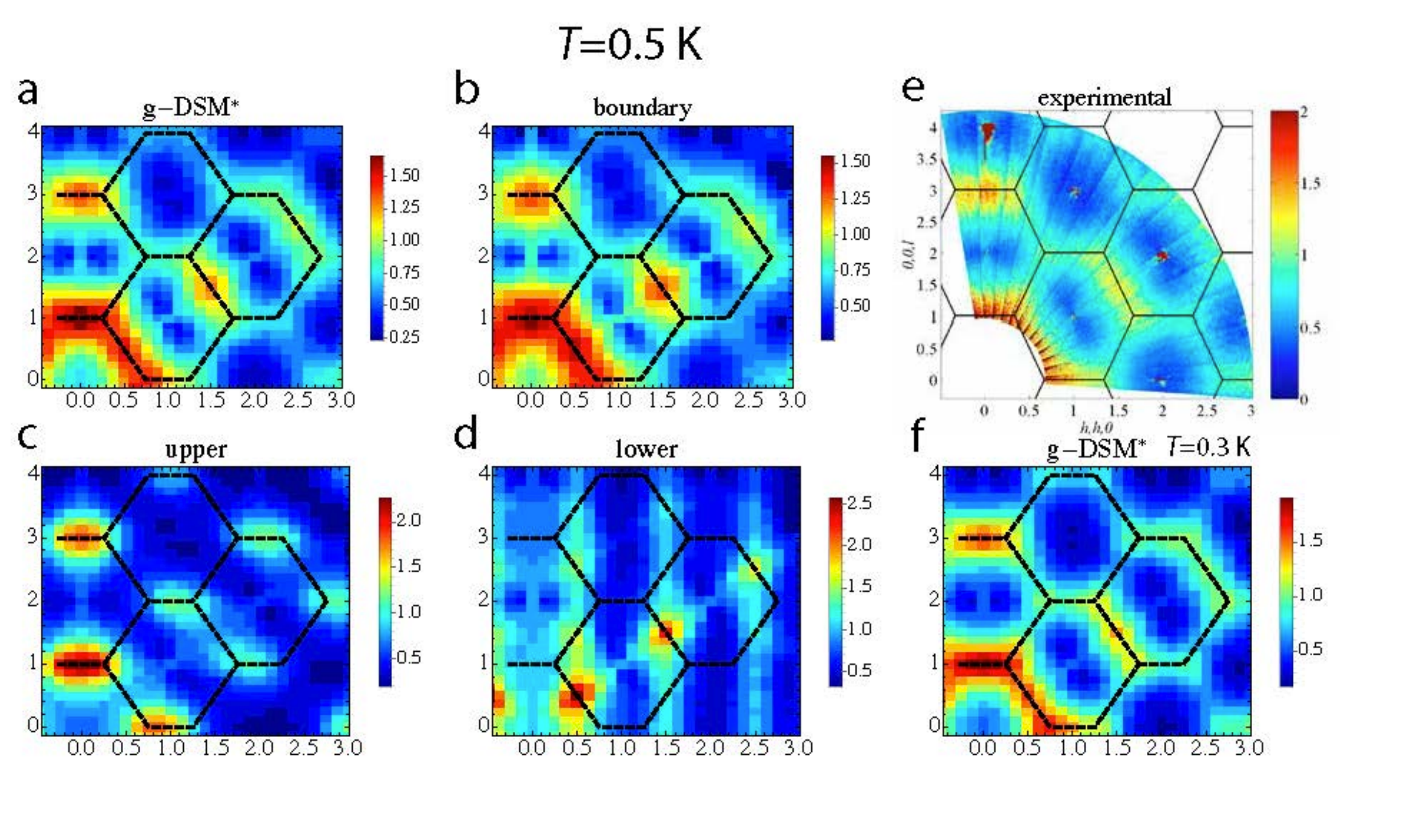}
\caption{Monte Carlo simulation results for the neutron structure
  factor $S(\boldsymbol{q})$ at $T=0.5$ K calculated for the four
  parameter sets in Table~\ref{t:models} in panels a-d. The maps can
  be compared to the experimentally measured structure factor at
  $T=0.3$~K in panel e (Ref.~[\onlinecite{Fennell04}]). Points of
  interest are the broad intensity maxima centered around
  $\boldsymbol{q}$=$(0\: 0\: 3)$ and
  $\boldsymbol{q}$=$(\frac{3}{2}\:\: \frac{3}{2}\:\:\frac{3}{2})$ and
  the ZBS pattern. Note that the experimental intensity maximum at
  $\boldsymbol{q}$=$(\frac{3}{2}\:\: \frac{3}{2}\:\:\frac{3}{2})$ is
  more closely matched by the structure factor for the g-DSM$^*$
  parameter set calculated at $T=0.3$ K, shown in panel f. }
\label{sq05}
\end{center}
\end{figure*}

\begin{figure*}[ht]
\begin{center}
 \includegraphics[width=18cm]{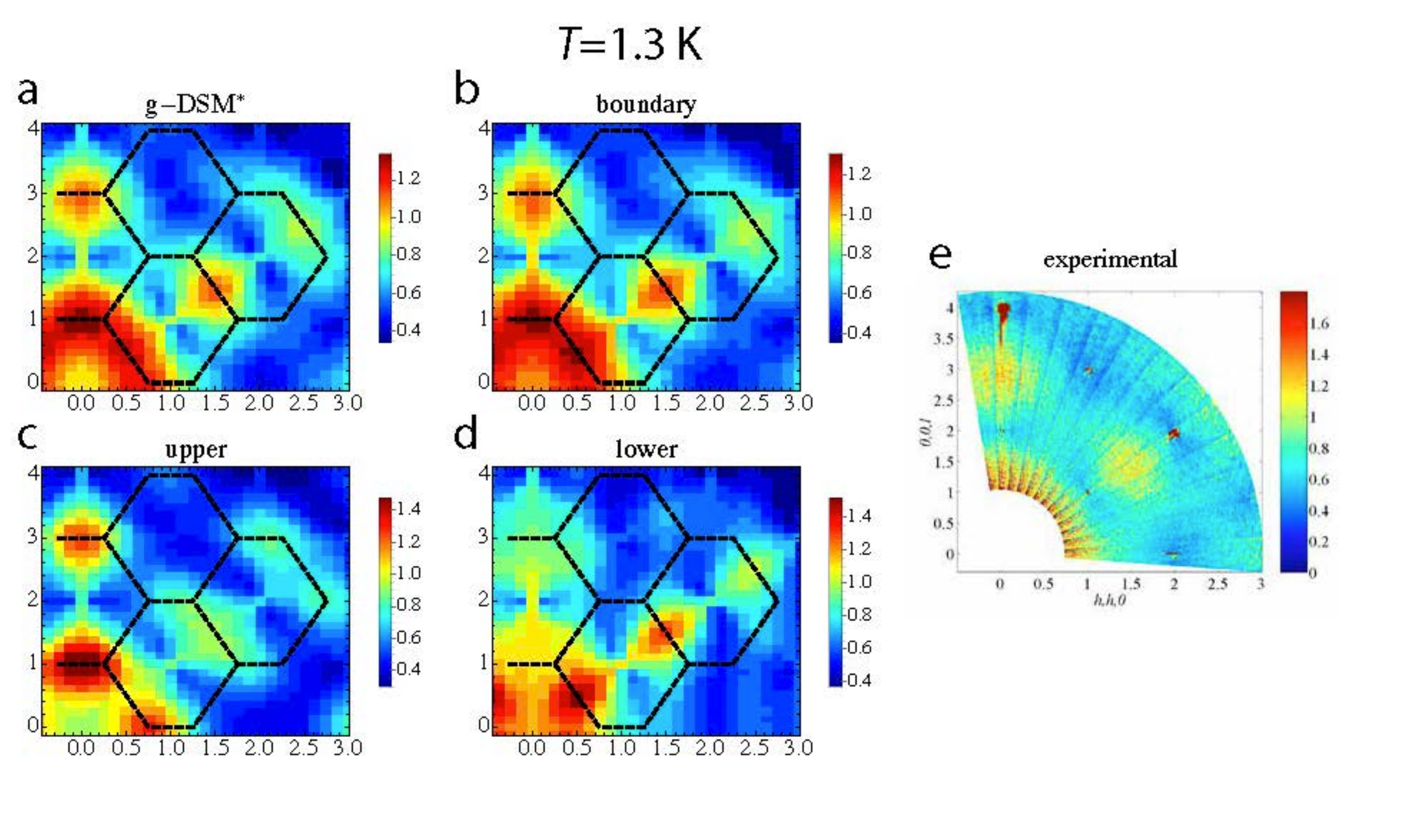}
 \caption{Monte Carlo simulation result for the neutron structure
   factor $S(\boldsymbol{q})$ at $T=1.3$ K calculated for the four
   parameter sets in Table~\ref{t:models}. The maps can be compared to
   the experimentally measured structure factor at $T=1.3$~K in panel
   e (Ref.~[\onlinecite{Fennell04}]). Points of interest are the broad
   intensity maxima centered around $\boldsymbol{q}$=$(0\: 0\: 3)$ and
   $\boldsymbol{q}$=$(\frac{3}{2}\:\: \frac{3}{2}\:\:\frac{3}{2})$ and
   the remnants of the ZBS pattern, more clearly seen in
   Fig~\ref{neutron_boundary}a. }
\label{sq13}
\end{center}
\end{figure*}

\begin{figure}[h!]
 \resizebox{\hsize}{!}{\includegraphics[clip=true]{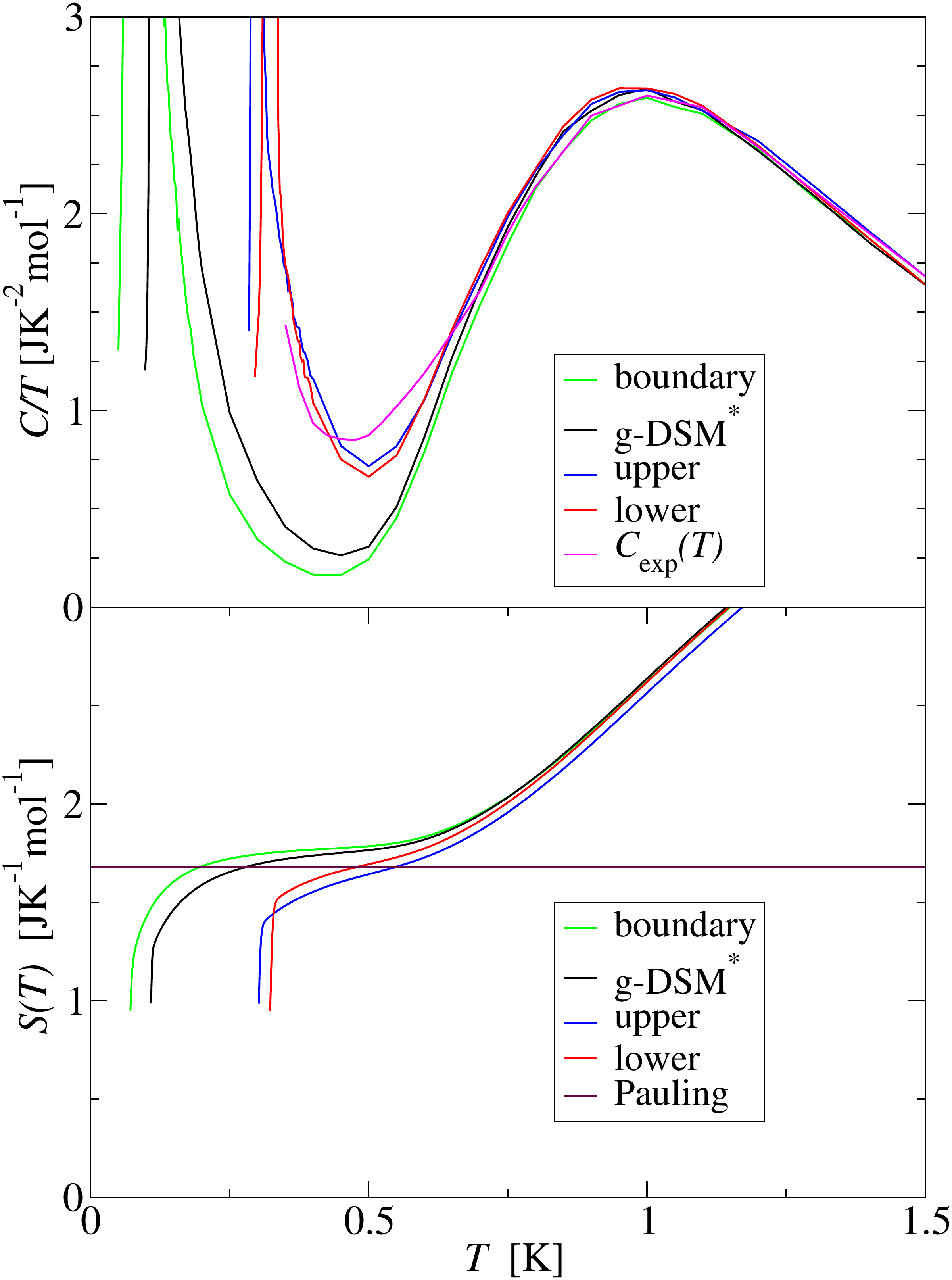}}
 \caption{The specific heat (upper panel) and corresponding entropy,
   $S(T)$, (lower panel) for the four parameter sets in
   Table~\ref{t:models}. The entropy value of the Pauling plateau at
   $S$=(R/2)$\ln$(3/2) is indicated by a brown horizontal line in the
   lower panel.}
 \label{entropy}
\end{figure}

In the previous two sections we found that the neutron scattering
structure factor $S(\boldsymbol{q})$ indicates an exchange parameter
set close to the phase boundary between the two relevant ordered
phases, while the recent specific heat data are more compatible with a
parameter set well within one of the two ordered phases.  To examine
which alternative is most likely, we begin by ruling out the existence
of a region in the parameter space that reconciles the
$S(\boldsymbol{q})$ and $C_{\text{exp}}(T)$ measurements.  We choose
to characterize $S(\boldsymbol{q})$ by the ratio $r\equiv S(0\ 0 \
3)/S(\frac{3}{2}\ \frac{3}{2}\ \frac{3}{2})$, which we found in
Subsection \ref{sec:neutronscatt} to be a revealing overall indicator
of the spin-spin correlations in Dy$_2$Ti$_2$O$_7$. For the specific
heat we consider the $C_{\text{exp}}(T)/T$ RMS deviation $\sigma_C$.
We divide the $J_1\mbox{-}J_{3a}$ space into a grid of points at which
we calculate the ratio $r$ and $\sigma_C$.  We choose to calculate the
neutron scattering ratio $r$ at $T=1.3$ K since we know that there are
no experimental equilibration issues at this temperature. The specific
heat deviation $\sigma_C$ we evaluate for the $[0.45\mbox{-}4]$ K
temperature interval in order to include the experimentally observed
upturn in the specific heat.  In Fig.~\ref{scatter}, we plot $r$
versus $\sigma_C$ for all the points in the $J_1\mbox{-}J_{3a}$ space.
In order to accommodate both experiments, the $r$ coordinate for a
point should be close to the experimental value $r\cong1.13$ at
$T=1.3$, while $\sigma_C$ should be small.  In Fig.~\ref{scatter}, we
have indicated the experimental value of $r$=1.13 with a red band, and
a good specific heat match ($\sigma_C<0.2$) with a blue band. We note
that there are points within the red band: these correspond to
parameter values in the boundary region. We also note that there are
\emph{no} points within the blue band, emphasizing the conclusion of
the previous Subsection that there is \emph{no} parameter set able to
match the recent specific heat data well, irrespective of the quality
of fit to the neutron scattering data.  The two branches of the
scattered points that reach a minimum $\sigma_C$ of about 0.04
correspond to points in the upper and lower basin, discussed in the
previous section.  Hence, the present model is unable to
\emph{simultaneously} describe the specific heat ($C_{\text{exp}}(T)$)
and neutron scattering ($S(\boldsymbol{q})$) experiments.  We have
performed extensive Monte Carlo simulations to verify that this
conclusion still holds if we add a nonzero fourth-nearest-neighbor
interaction, $J_4$, and relax the $[ 1 1 2]$ experiment-based
constraints well beyond the optimal Eqs.  \ref{eq:j3} and \ref{eq:j2},
as detailed in Appendix \ref{app:112}. For Fig.~\ref{scatter} we have
considered parameters in the range (all in units of kelvin):
$2.8<J_1<4.0$, $|J_{3a}|<0.15$ and $|J_4|<0.15$, with corresponding
$J_2$ and $J_{3b}$ values given by Eq.~\ref{eq:j2} and Eq.~\ref{eq:j3}
respectively. Note that the conclusion of this analysis also holds if
we calculate the neutron ratio $r$ at $T=0.3$ K and compare our Monte
Carlo result to the experimental value at this lower temperature.

\begin{table}[h]
\begin{tabular}{|c|c|c|c|c|}
  \hline
  \hline
  model& $J_1$ & $J_2$ & $J_{3a}$ & $J_{3b}$ \\
  \hline
  upper &  3.30 & -0.0949 & 0.07 & -0.0167  \\
  middle &  3.38 & -0.1349 & 0.02 & 0.0167\\
  lower &  3.44 & -0.1649 & -0.02 & 0.0433 \\
  g-DSM$^* $ & 3.41 & -0.14 & 0.025 &0.025 \\
  \hline
  \hline
\end{tabular}
\caption{Named parameter sets. The first three sets are indicated by
  stars in Fig.\ref{cv_rms}d. In all cases the dipolar constant
  $D=1.3224$ K  and all parameter values are given in kelvin.}   
\label{t:models}
\end{table}
To further assess the likelihood of a parameter set in the upper or
lower basin versus a parameter set on the boundary, we explicitly
calculate $S(\boldsymbol{q})$ for the three parameter points indicated
by stars in Fig.~\ref{cv_rms}d.  In addition, we show results for the
parameter set determined in Ref.~[\onlinecite{Yavo08}], referring to
this particular parameter set as the ``g-DSM$^*$''.  A summary of the
naming convention and parameter values we examine are given in
Table~\ref{t:models}.  In Figs.~\ref{sq05}a-d we display the structure
factor for these four parameter sets. The results can be compared to
the experimental structure factor, nominally measured at $T=0.3$
K. The experimental data agree remarkably well with the structure
factor calculated for the point on the boundary and the g-DSM$^*$
parameter set, while there are major differences between the
experimental data and the results for the parameter values in the
upper and lower basin. Also note that the Monte Carlo result for the
g-DSM$^*$ parameter set at $T=0.3$~K, shown in panel f appears even
closer to the experimental result than panel a. This is surprising,
since as argued in Section~\ref{sec:neutronscatt}, one could expect
the sample to freeze around $T=0.5$~K.  We therefore take
Fig.~\ref{sq05} as further strong evidence that the appropriate
parameter set for Dy$_2$Ti$_2$O$_7$ is located close to the phase
boundary.  We arrive at the very same conclusion by considering
Fig.~\ref{sq13}, which makes the same comparison as Figs.~\ref{sq05},
but at the elevated temperature $T=1.3$ K, where the sample is well
equilibrated.

We conclude this subsection by examining the low-temperature behavior
of the specific heat and entropy of the four parameter points examined
in the previous paragraph. Using the Monte Carlo method, we calculate
$C(T)/T$ and integrate this function to obtain the entropy,
$S(T)$. The result is shown in Fig.~\ref{entropy}. Consider first the
upper panel showing the specific heat. The specific heat for the
points in the upper and lower basin rise very abruptly due to the
strong first-order transition and lack of pretransitional
fluctuations, as discussed in the previous subsection. The transition
temperature is in the $0.30$-$0.32$ K interval, immediately below the
last experimentally measured temperature of 0.34 K in
Ref.~\onlinecite{Poma13}.  This suggests that if specific heat
measurements could be carried out to slightly lower temperature than
the one considered in Ref.~[\onlinecite{Poma13}], one could
experimentally resolve whether the upturn is indeed caused by an
ordering transition as described by the g-DSM.  In the lower panel of
Fig.~\ref{entropy}, we find that the Pauling plateau is not developed
over any significant temperature interval for the parameter points in
the upper and lower basin, while it is clearly visible in the 0.2-0.6
K range for the points close to the boundary (including the
g-DSM$^*$).

While we find that there is no parameter set that is compatible with
both neutron scattering data and the recent specific heat measurements
down to $0.34$ K, we do find that neutron scattering data down to
nominally 0.3 K and specific heat data \emph{above} $0.7$ K are
\emph{both} consistent with a parameter set placing Dy$_2$Ti$_2$O$_7$
near the boundary between two competing long-range ordered ice-rule
obeying states.  While the low-temperature rise in the specific heat
could indicate a parameter set in the upper or lower basin, we find
this unlikely since our analysis shows that these parameter points
yield spin-spin correlations that are incompatible with neutron
scattering data already at 1.3 K.

While it was not obvious before initiating the present study, we have
discovered that the g-DSM$^*$ parameter set of
Ref.~[\onlinecite{Yavo08} almost satisfies our $J_1\mbox{-}J_2$ and
$J_{3a}\mbox{-}J_{3b}$ constraints, and is also located very close to
the phase boundary. It therefore appears that the $J_{3a}=J_{3b}$
constraint assumed in Ref.~[\onlinecite{Yavo08} is almost realized in
Dy$_2$Ti$_2$O$_7$, although this was not a priori evident.  One should
note that the main goal of Ref.~[\onlinecite{Yavo08}] was to
investigate to what extent a model of finite-size clusters of magnetic
moments can describe the neutron scattering in the spin ice regime of
Dy$_2$Ti$_2$O$_7$.  It was not aimed at determining an optimal
exchange parameter set for this compound. That notwithstanding, the
present analysis confirms that the g-DSM$^*$ parameter set of
Ref.~[\onlinecite{Yavo08}] appears fairly appropriate for
Dy$_2$Ti$_2$O$_7$.  For example, one notes that the g-DSM$^*$
parameter set has also been found to accurately model a $[ 1 1 1]$
field experiment\cite{Yavo08}, as well as the specific heat for a
number of diamagnetically diluted samples of
Dy$_{2-x}$Y$_x$Ti$_2$O$_7$ \cite{Lin14} for temperatures above $T\sim
0.5$ K.

To summarize: the fact that the parametrization of the g-DSM obtained
on the basis of the main generic neutron scattering features does not
match the recent carefully equilibrated specific heat
measurements\cite{Poma13} in the \emph{lower} temperature range
$0.35\mbox{-}0.7$ K leads to \textbf{\textit{the second main result }}
of our study: the upturn in the specific heat is a strong indication
that, at temperatures below $~0.7$ K, some new physics becomes
relevant which cannot be readily exposed by the model of
Eq.~(\ref{eq:ham}). Exploring two such possible causes is the topic of
the next subsection.

\subsection{Quantum Effects and Random Disorder}
\label{sec:quantumdisorder}

From our analysis of the neutron scattering data, we conclude that the
recently observed upturn in the specific heat is not caused by an
impending ordering transition within the g-DSM. We therefore consider
the next most likely causes of the specific heat upturn below $T\sim
0.5$ K: quantum effects and random disorder.

\subsubsection{Quantum effects \& non-Ising exchange}

An intriguing possibility is that quantum effects could be responsible
for the increasing specific heat below 0.5 K as found at low
temperatures in simulations of a spin-$1/2$ XXZ model on the
pyrochlore lattice\cite{Kato14}.  In this context,
Ref.~[\onlinecite{McClarty}] considers the effects of a hexagonal
``ring exchange'' tunneling term on the g-DSM, and finds that if the
tunneling amplitude $g$ is greater than the classical ordering
temperature $T_c$ for $g=0$, a quantum spin liquid state may be
stabilized. Given our conclusion that Dy$_2$Ti$_2$O$_7$ is located at
the phase boundary between two classical ordered states, where the
critical temperature is suppressed to about $T_c\sim 70$~mK (see
Fig.~\ref{energy_boundary}), it is perhaps conceivable that quantum
(non-Ising) terms become relevant at higher temperatures than
previously thought.  We note that a simple spin-1/2 XXZ model results
in a tunneling strength $g=12\frac{J_{\pm}^3}{(J_{zz})^2}$, where
$J_{zz}$ and $J_{\pm}$ are the longitudinal (Ising) and transverse
(XY) exchange couplings\cite{Savary_QSI}.  The Monte Carlo work of
Ref.~[\onlinecite{Kato14}] finds that the specific heat in such a
model starts to increase below a crossover temperature $T^*\sim g$. To
assess whether $g$ is large enough in Dy$_2$Ti$_2$O$_7$ to cause the
upturn in the specific heat at a temperature $T^* \sim 0.5$ K, we
analyze the expected scale of the quantum corrections to the Ising
part of the Hamiltonian (1).

One possible route to transverse couplings comes from quantum
corrections to the effective Hamiltonian through virtual crystal field
excitations\cite{Molavian07}.  Due to the large gap $\Delta \sim$300 K
to the first excited crystal field level and the $J_{\rm eff} \equiv
[5D-J_1]/3 \sim 1$ K scale of the interactions acting within the full
crystal field manifold, quantum corrections can be expected to appear
perturbatively at a temperature of order $J_{\pm} \sim J_{\rm
  eff}^2/\Delta \sim 3$~mK and are therefore not detectable at a
temperature of 0.5~K.  Off-diagonal terms can also appear through the
interaction of high rank multipoles\cite{santini2009multipolar,Rau15}
involving the ${\bm J}=15/2$ angular momentum components within the
${}^6 H_{15/2}$ ground state electronic manifold of Dy$^{3+}$.  Due to
the dominant $|\pm 15/2\rangle$ composition of the ground state
crystal field doublet\cite{Bramwell_Science,Gingras11}, the largest
contributions to the transverse $S^\pm$ effective spin-$1/2$ operators
would originate from rank-15 multipolar interactions between the
Dy$^{3+}$ ions\cite{Rau15}.  The most significant source of such
multipolar interactions are super-exchange processes mediated through
the neighboring oxygen ions\cite{Rau15}.  However, these
super-exchange processes predominantly generate interactions between
multipoles of rank seven or less
\cite{onoda2011quantum,santini2009multipolar,Rau15}.  Such multipole
interactions will thus only connect the subleading spectral components
of the crystal field ground state doublet, $C_{m_J}\vert
J=15/2,m_J\rangle$, $m_J\ne \pm 15/2$. From the experimentally
observed large magnetic moment $\mu\sim 10$ $\mu_{\rm B}$ of Dy$^{3+}$
in Dy$_2$Ti$_2$O$_7$, we can infer that these small $C_{m_J}$
components are at most $C_{m_J} \sim 10\%$ of the leading $C_{m_J=\pm
  15/2}\approx 1$ component\cite{Rau15}.

With $J_{zz} \sim 4 J_{\rm eff}$ and $J_\perp \sim 4 (C_{m_J})^2
J_{\rm eff}/(C_{15/2})^2$, where the prefactor 4 arises from moving
from the Ising $S_i^{z}=\pm 1$ convention in the present work to the
$S=1/2$ convention of Ref.~[\onlinecite{Savary_QSI}], we therefore
expect $g$ to be at the scale $g\sim 0.05$~mK.  Thus, despite the much
suppressed critical temperature $T_c\sim 70$~mK that we exposed above
(see Fig.~\ref{energy_boundary}), we argue that quantum effects are
unlikely to be responsible for the rise in the specific heat detected
at a temperature $T^* \sim 0.5$ K given that $T_c/g \sim 10^4$, hence
far up in the classical dipolar spin ice regime, unlike in the
proposal of Ref.~[\onlinecite{McClarty}].  While we expect that
further calculations would lead to a more accurate estimation of
$J_\perp$, and thus of $g$, it would seem unlikely to lead to a
rescaling of $g$ by four orders of magnitude.

After having argued that the rise of $C_{\text{exp}(}T)$ below
$T^*~\sim~0.5$ K cannot be explained by a classical dipolar spin ice
model that consistently describes the neutron scattering data, we have
now argued that the same rise cannot originate from the development of
a quantum coherent regime below $T^*$.

\begin{figure*}[ht]
\begin{center}
 \includegraphics[width=18cm]{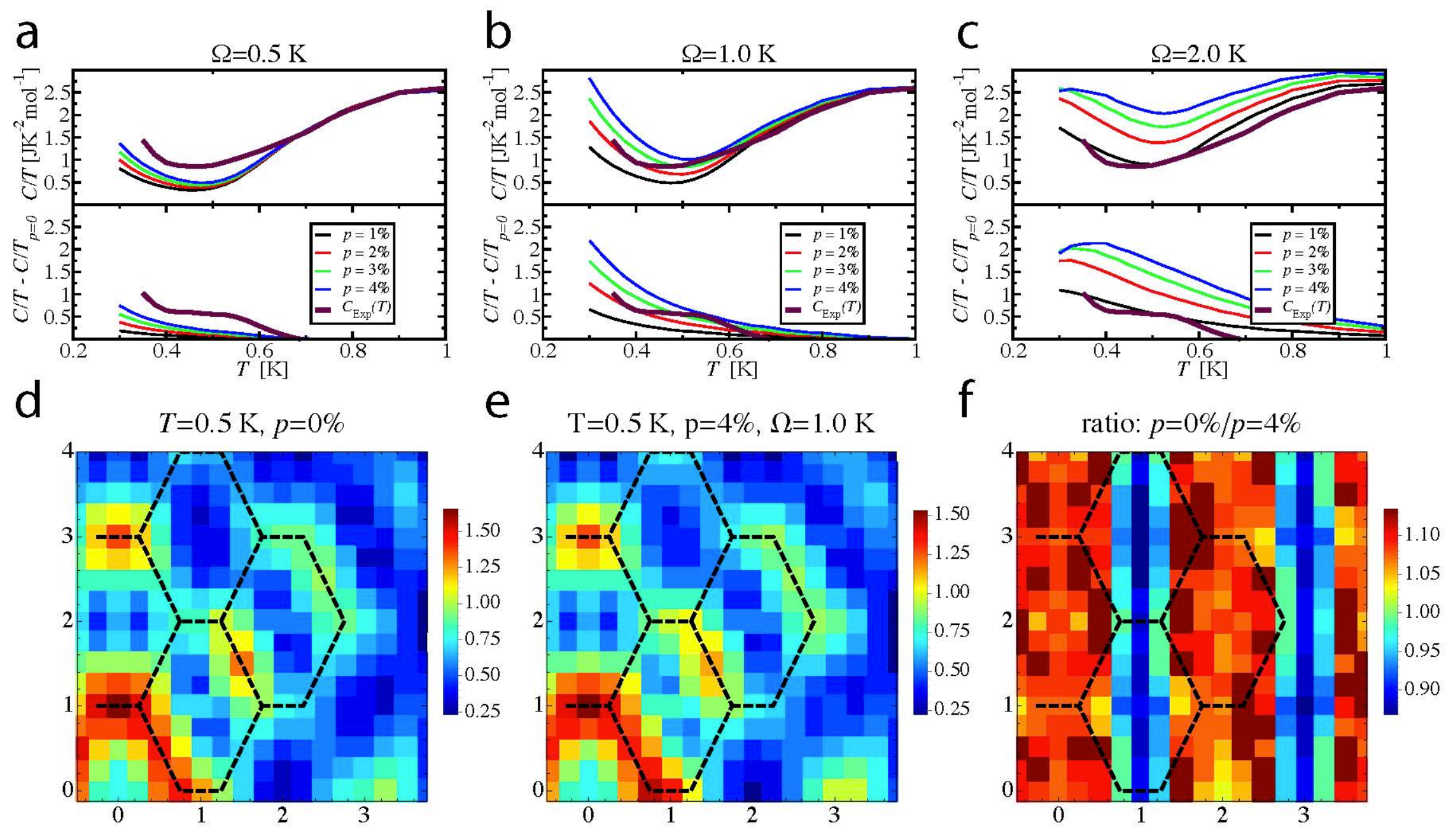}
 \caption{ Effects of impurities on specific heat and neutron
   scattering.  Specific heat, $C(T)/T$ for the $\Omega$-model with
   {\bf a} $\Omega$= 0.5 K, {\bf b} $\Omega$= 1.0 K and {\bf c}
   $\Omega$= 2.0 K with the percentage of stuffed spins $p$=1,2,3, and
   4\%. The upper panel shows the total specific heat, while in the
   lower panel the specific heat of the clean system ($p=0$) has been
   subtracted from the disorder average of the upper panel revealing
   the impurity contribution. The neutron structure factor
   $S(\boldsymbol{q})$ for the disorder-free ($p=0$) g-DSM
   model\cite{Yavo08} at $T=0.5$ K is shown in {\bf d,} and with $p=
   4\%$ stuffed spins included in {\bf e}.  The last panel, {\bf f},
   shows the ratio of the structure factor for the clean ({\bf d}) and
   stuffed ({\bf e}) systems.}
\label{stuff}
\end{center}
\end{figure*}

\subsubsection{Random disorder} 

Another potential origin for the rise of the electronic specific heat
at $T\lesssim 0.5$ K is a Schottky-type response induced by random
disorder. It was recently demonstrated that disorder, in the form of
local magnetic impurities, may explain the long-time relaxation in
Dy$_2$Ti$_2$O$_7$ (Ref.~[\onlinecite{Revell13}]). Possible physical
realizations of disorder include stuffed spins\cite{Ross12,Cava15} or
local Dy$^{3+}$ moments with easy-plane anisotropy due to oxygen
vacancies affecting the crystal field\cite{Sala14} or the concomitant
magnetic Ti$^{3+}$ magnetic impurities.  We now illustrate that such a
generic phenomenon could explain the rise of $C(T)$ for $T\lesssim
0.5$ K.

The quantitative microscopic description of the effects of disorder in
highly frustrated magnetic materials is a complex and rich
problem\cite{Villain79,Sen}.  In the present case, the complexity of
such a task is further compounded by the fact that all pertinent forms
of random disorder (e.g. oxygen vacancies, stuffing, etc) have not yet
been fully identified and their effects quantified in
Dy$_2$Ti$_2$O$_7$.  It appears that the problem of dilution of the
magnetic Dy$^{3+}$ and Ho$^{3+}$ ions by non-magnetic Y$^{3+}$ in
Dy$_2$Ti$_2$O$_7$ and Ho$_2$Ti$_2$O$_7$ is the simplest case of random
disorder, with a description in terms of a mere site-diluted version
of a dipolar spin ice model accounting well for the specific heat data
for $T\gtrsim 0.5$ K in Dy$_{1-x}$Y$_x$Ti$_2$O$_7$ and
Ho$_{1-x}$Y$_x$Ti$_2$O$_7$\cite{Lin14} (see, however, the recent work
in Ref.~[\onlinecite{Scharffe}]).  On the other hand, we expect the
microscopic description of oxygen vacancies or rare-earth ions
stuffing the pyrochlore $B$ site normally occupied by non-magnetic
transition-metal ions, and its ultimate quantitative description via a
controlled numerical calculation, to be significantly more
complicated.  Indeed, one would expect that deformation of the
superexchange pathways and modifications of the local crystal field
would occur.  This would result in a randomization of the $J_{ij}$
exchange couplings and of the dipolar coupling ($D \rightarrow
D_{ij}$) as well as possibly induce non-Ising (transverse) exchange
terms coupling the other components of the effective spin-$1/2$
describing the crystal field doublet of Dy$^{3+}$.  In the present
work, we consider a minimal model of dilute random disorder in
Dy$_2$Ti$_2$O$_7$.  Our goal is not to develop a quantitative
description of the role of disorder on the low-temperature properties
of Dy$_2$Ti$_2$O$_7$. Rather, we wish to illustrate that the
\emph{generic} effects of dilute random disorder with a realistic
energy scale $\Omega$ could perhaps naturally explain the rise of
$C_{\text{exp}}(T)$ for $T\lesssim T^*$.

With this agenda spelled out, we now proceed and consider the effects
of stuffed spins in the model defined by Eq~\ref{eq:ham} for
Dy$_2$Ti$_2$O$_7$.  The stuffed spins are coupled to the Dy$^{3+}$
pyrochlore backbone through Eq.~(\ref{eq:Omega}), the
$\Omega$-model. . We perform loop Monte Carlo simulations for varying
stuffing percentages $p$ and coupling constants $\Omega$ and display
the results in Fig.~\ref{stuff}.  Disorder averages have been
performed over 32 independent disorder configurations, since the
variation between different configurations is quite small. Panels {\bf
  a-c} contain the specific heat data for $\Omega=0.5, 1.0$ and $2.0$
K respectively. Results for stuffing percentage $p= 1, 2, 3$ and $4\%$
along with the experimental result are shown in the upper subpanels,
while in the lower subpanels we have subtracted the specific heat for
the clean g-DSM model, exposing the impurity contribution to the
specific heat. We see that, at least for $\Omega=0.5$ K and
$\Omega=1.0$ K the impurity contribution is roughly linear in $p$,
indicating predominantly independent impurities.  The impurity
contribution also shifts to higher temperature as $\Omega$ is
increased, as we would expect from a simple two-state Schottky
model. In this case the scaling is more complicated since each
impurity spin interacts with the electronic spin on the six nearest
neighbors, and we would expect the stuffed spin to effectively
renormalize the local couplings.  Again, our goal is not to perform an
exhaustive study of this specific model.  Rather, we wish to
illustrate that such a simple effective local impurity model can cause
an upturn in rough qualitative agreement with the measured specific
heat with a stuffing ratio at the one percent level and a realistic
$\Omega=1$ K energy scale.  Interestingly, we also note that this
level of stuffing does not produce a very strong response in the
neutron structure factor. In Figs.~\ref{stuff}d and e, we display the
calculated $S(\boldsymbol{q})$ for the g-DSM model with $p=0\%$ and
$p=4\%$ stuffing, respectively.  The disorder average has here been
performed over 96 disorder configurations.  The effects are rather
hard to discern in an energy-integrated scattering profile (i.e. equal
time correlations computed through the present classical Monte Carlo
simulations) with the ratio of the two scattering intensity profiles,
plotted in Figs.~\ref{stuff}f, showing a change of about $10\%$.  With
the recent realization that stuffing of the heavy rare-earth (RE) ions
(RE=Yb, Er, Ho) is at play in Yb$_2$Ti$_2$O$_7$, Er$_2$Ti$_2$O$_7$ and
Ho$_2$Ti$_2$O$_7$ (i.e. RE$^{3+}$ replacing
Ti$^{3+}$)\cite{Ross12,Cava15}, it would appear plausible that some
level of stuffing may also occur in Dy$_2$Ti$_2$O$_7$ given that the
ionic radius of Dy$^{3+}$ is only $\sim 1\%$ larger than
Ho$^{3+}$~\cite{Shannon_radii}.  Further experimental investigations
are required to assess whether or not this is the case.  While $1\%$
stuffing is possibly a bit too high for Dy$_2$Ti$_2$O$_7$ samples, our
results do show that \emph{effective} local random magnetic disorder
in the g-DSM associated with an energy scale $\Omega\sim 1$ K can in
principle capture the essential features observed in calorimetric
measurements without creating a significant contradiction with
available neutron scattering data.

\section{Conclusion}
\label{sec:conclusion}

In summary, the present study gives a natural interpretation of the
main features of the structure factor $S(\boldsymbol{q})$ observed in
neutron scattering measurements on the Dy$_2$Ti$_2$O$_7$ spin ice
material in terms of competing phases.  Remarkably, we find that
neutron scattering experiments on a single crystal of
Dy$_2$Ti$_2$O$_7$ place the material at the most interesting point in
the phase diagram $-$ precisely near the boundary between competing
single- and double-chain long-range-ordered phases. This region
displays unusual properties including masked ground state order and
extremely slow magnetic relaxation associated with the coarsening of
stacking defects.  We expect the equilibration of these defects to be
a further mechanism, beyond the mere dynamical arrest associated with
the ice rule formation, impinging on the approach to equilibrium of a
real material.

The observation of such accidental competing ground states suggests
that Dy$_2$Ti$_2$O$_7$ could be extremely sensitive to random disorder
or quantum fluctuations.  Furthermore, our analysis of the recently
observed upturn in the specific heat\cite{Poma13} shows that this is
caused by terms not present in the classical generalized dipolar spin
ice model.  Quantum corrections appear to be much too small in
magnitude to become noticeable at temperatures as high as $T^*\sim
0.5$ K where recent thermally equilibrated specific heat
measurements\cite{Poma13} find an upturn, even after correcting for
the nuclear specific heat.  We therefore believe that random disorder,
either in the form of low-level of stuffing\cite{Cava15} or oxygen
vacancies\cite{Sala14}, or both, is the most likely source of the
upturn.  There is also an interesting scenario in which random
disorder could, by lowering the local symmetry of the crystal field,
destroy the local Ising nature of the moments and induce quantum
fluctuations in the nearby Dy$^{3+}$ moments.  A study that would
assess the level of Dy$^{3+}$ stuffing onto the B site otherwise
occupied by Ti$^{4+}$ in Dy$_2$Ti$_2$O$_7$, as was done in
Ref.~[\onlinecite{Cava15}] for Ho$_2$Ti$_2$O$_7$, Er$_2$Ti$_2$O$_7$
and Yb$_2$Ti$_2$O$_7$, would therefore be highly desirable.  More
well-equilibrated experimental studies of a variety of samples will be
required to determine the exact nature of such impurities and their
specific role on the low-temperature thermodynamic properties of
Dy$_2$Ti$_2$O$_7$.  As a corollary, our work suggests that a (fairly)
well-defined Pauling plateau may be observed in well-equilibrated
measurements on samples of high stoichiometric purity.  Perhaps the
most important conclusion of our work is the following one:
notwithstanding the fact that the two disorder-free models that are
partially able to describe the specific heat data between 0.35 K and
0.7 K are inconsistent with the main neutron scattering features, both
models predict a phase transition to long-range order near 0.30 K.
Consequently, it would seem imperative to push the low-temperature
limit of well-equilibrated calorimetric measurements down to, say, 250
mK.  The present work predicts that the disagreement with the best
classical dipolar spin ice model for Dy$_2$Ti$_2$O$_7$ would then
become vividly manifest, with such an experiment providing important
clues as to the ultimate low-temperature fate of the spin ice state in
this archetypal spin ice material.

\begin{acknowledgments}
  It is a pleasure to thank our experimental colleagues D. Pomaranski
  and J. Kycia (specific heat), Sakakibara and Hiroi ($[ 1 1 2]$
  experiment), and T. Fennell (neutron scattering) for sharing their
  data and for stimulating discussions.  We also acknowledge
  N. Shannon and P. McClarty for useful discussions.  This work was
  supported in part by the NSERC of Canada and the Canada Research
  Chair program (M.G., Tier 1).  P.H. is grateful for the computer
  support of PDC-HPC (Ferlin) at KTH and the financial support by the
  Stenb{\"a}ck Foundation and the Swedish Research Council.
  J.A. acknowledges support from the Natural Sciences and Engineering
  Research Council of Canada. F. F. acknowledges support from a
  Lindemann Trust Fellowship of the English Speaking Union. The
  numerical work at the University of Waterloo was made possible by
  the facilities of the Shared Hierarchical Academic Research
  Computing Network (SHARCNET:www.sharcnet.ca) and Calcul
  Canada. M.G. acknowledges support from the Canada Council for the
  Arts and the Perimeter Institute (PI) for Theoretical Physics.
  Research at PI is supported by the Government of Canada through
  Industry Canada and by the Province of Ontario through the Ministry
  of Economic Development \& Innovation.
\end{acknowledgments}

\appendix

\section{Paramagnetic Constraints from [112] Magnetization Measurements}
\label{app:112}

The constraints that we invoke between the $J_{ij}$ couplings to
analyze the specific heat and neutron data in the main part of the
paper result from considerations of an experiment in which a strong
magnetic field is applied near the cubic $[112]$
direction\cite{Ruff,Higa_112,Sato06}. The main effect of the field is
to maximally polarize the moments along their local $[111]$ Ising
directions on three of the four FCC sublattices that comprise the
pyrochlore lattice and which form \emph{kagome} layers. The spins on
the remaining sublattice form \emph{triangular}
layers\cite{Gardner10}, as illustrated in Fig.~\ref{112_3D}.  Spins on
these triangular layers interact with each other only through the
dipolar and third-nearest-neighbor interactions $J_{3a}$ and $J_{3b}$
since they reside on the same FCC sublattice\cite{transverse}.  While
the spins on the ``triangular sites'' are perpendicular to, and thus
decoupled from, the $[112]$ field component, they are still subject to
an internal field, $h_{\text{int}}$, from the polarized spins on the
kagome layers. This internal field enforces the ice rules so that the
spins in the triangular layers point ``in'' along the $[\bar 1\bar 1
1]$ direction (see Fig.~\ref{112_3D}).  Experimentally, this internal
``ice rule" enforcing field can be cancelled by rotating the applied
external field away from the $[112]$ direction
\cite{Ruff,Sato06,horizvert_foot}, so that it acquires a component,
$h^{[\bar{1}\bar{1}1]}$, in the $[\bar{1}\bar{1}1]$ direction parallel
to the Ising axis of the ``triangular spins'' (see Fig.~\ref{112_3D}).
By tuning this $[\bar{1}\bar{1} 1]$ field component, one eventually
reaches a critical field value, $h_{\text{c}} ^{[\bar{1}\bar{1}1]}$,
when the $[\bar {1} \bar {1} 1]$ field component cancels out the ice
rule enforcing field $h_{\rm int}$. The experimentally measured value
of the cancellation field is $h_{\text{c}} ^{[\bar{1}\bar{1}1]} \equiv
-h_{\text{int}}=-0.28\pm 0.02$ T (the negative sign arises because the
applied field has to oppose the ice-rule enforcing field
$h_{\text{int}}$ which is along $[\bar{1}\bar{1}
1]$)\cite{Higa_112,Sato06}.  Theoretically, the cancellation field can
be expressed in terms of our model Hamiltonian, Eq.~(\ref{eq:ham}) of
the main text, as
\begin{equation} 
  h_{\text{c}}^{[\bar{1}\bar{1}1]} ({\text{T}})  \mu_{\rm Dy} k_{\text{B}}
  =\frac{2}{3} J_1+\frac{4}{3}J_2 
  -2.972 D, 
  \tag{A1}
\end{equation} 
keeping exchange couplings $J_{ij}$ up to third-nearest-neighbor
exchange interactions. Since $J_{3a}$ and $J_{3b}$ connect spins on
the same sublattice they do not contribute to the internal field
$h_{\text{int}}$.  The dipolar contribution, $-2.972D$, was computed
using the Ewald summation method\cite{Melko04}.  Using the above
equation we obtain Eq.~(\ref{eq:j3}) of the main text,
\begin{equation} 
  J_2=-\frac{1}{2} J_1 + \frac{3}{4}h_1,
  \tag{A2}
\end{equation} 
where $h_1= \mu_{\rm Dy} k_{\text{B}} h_{\text{c}}
^{[\bar{1}\bar{1}1]} ({\text{T}}) + 2.972 D$. Inserting
$h_{\text{c}}^{[\bar{1}\bar{1}1]} = -0.28\pm 0.02$ T
(Ref.~[\onlinecite{Sato06}]), $\mu_{\rm Dy}=9.87\mu_{\text{B}}$ and
$D=1.322$ (Ref.~[\onlinecite{Yavo08}]) we find that $h_1=2.07\pm 0.13$
K. In Fig.~\ref{scatter}, the full range of $h_1$ within the
experimental uncertainty was allowed, while for the other figures the
optimal value of $h_1=2.07$ K was used.

A second constraint can be obtained for the third-nearest-neighbor
interactions $J_{3a}$ and $J_{3b}$ by considering the susceptibility
of the spins on the triangular layers. Again, experimentally, the
field component in the [$\bar{1}\bar{1}$1] direction can either be
varied by a rotation of the applied field or by varying the vertical
${[\bar{1}\bar{1}1]}$ field.  The susceptibility of the spins on the
triangular planes (which form an FCC lattice) can therefore be
measured as a function of that tunable [$\bar{1}\bar{1}$1] field
component. The susceptibility depends only on $J_{3a}$, $J_{3b}$ and
$D$ in our model Hamiltonian. Through this
manipulation\cite{Ruff,Higa_112,Sato06} of the applied magnetic field,
it is thus possible to directly probe the effect of the
third-nearest-neighbor exchange parameters on the thermodynamic
properties of Dy$_2$Ti$_2$O$_7$ in a strong magnetic field near the
$[112]$ direction.  We also note that, at the decoupling field
$h_{\text{c}}^{[\bar{1}\bar{1}1]}$, the above experiment results in an
Ising face-centered cubic magnet, which is normally not possible since
a unique global easy axis direction cannot be defined for a system
with cubic global symmetry.

In the experiment of Ref.~[\onlinecite{Sato06}], the susceptibility as
a function of the [$\bar{1}\bar{1}$1] field component was measured at
four different temperatures, $T= 0.29$, $0.41$, $0.70$ and $1.08$
K. With the field component in the $[\bar{1}\bar{1}1]$ direction set
precisely to the cancellation field value,
$h_{\text{c}}^{[\bar{1}\bar{1}1]}$, the magnetic moments on the
triangular layer undergo a transition to long-range ferromagnetic
order at $T_{\text{c}} \cong 0.26$ K
(Refs.~[\onlinecite{Ruff,Higa_112}]). As a result, the
${[\bar{1}\bar{1}1]}$ susceptibility associated with this transition
is divergent at $T_{\text{c}}$ when the ${[\bar{1}\bar{1}1]}$ field
$h^{[\bar{1}\bar{1}1]}$ is at its canceling value
$h_{\text{c}}^{[\bar{1}\bar{1}1]}$. When fitting the experimental
susceptibility to our simulation data, the system needs to be
sufficiently far away from the critical temperature to avoid strong
finite-size effects in the simulations.  However, the higher the
temperature, the less ``structure'' the susceptibility has due to the
decreasing correlation length, with the fit consequently becoming less
constrained. Based on these two concerns, we select the data of
Ref.~[\onlinecite{Sato06}] measured at $0.70$ K.

\begin{figure}[h!]
\resizebox{\hsize}{!}{\includegraphics[clip=true]{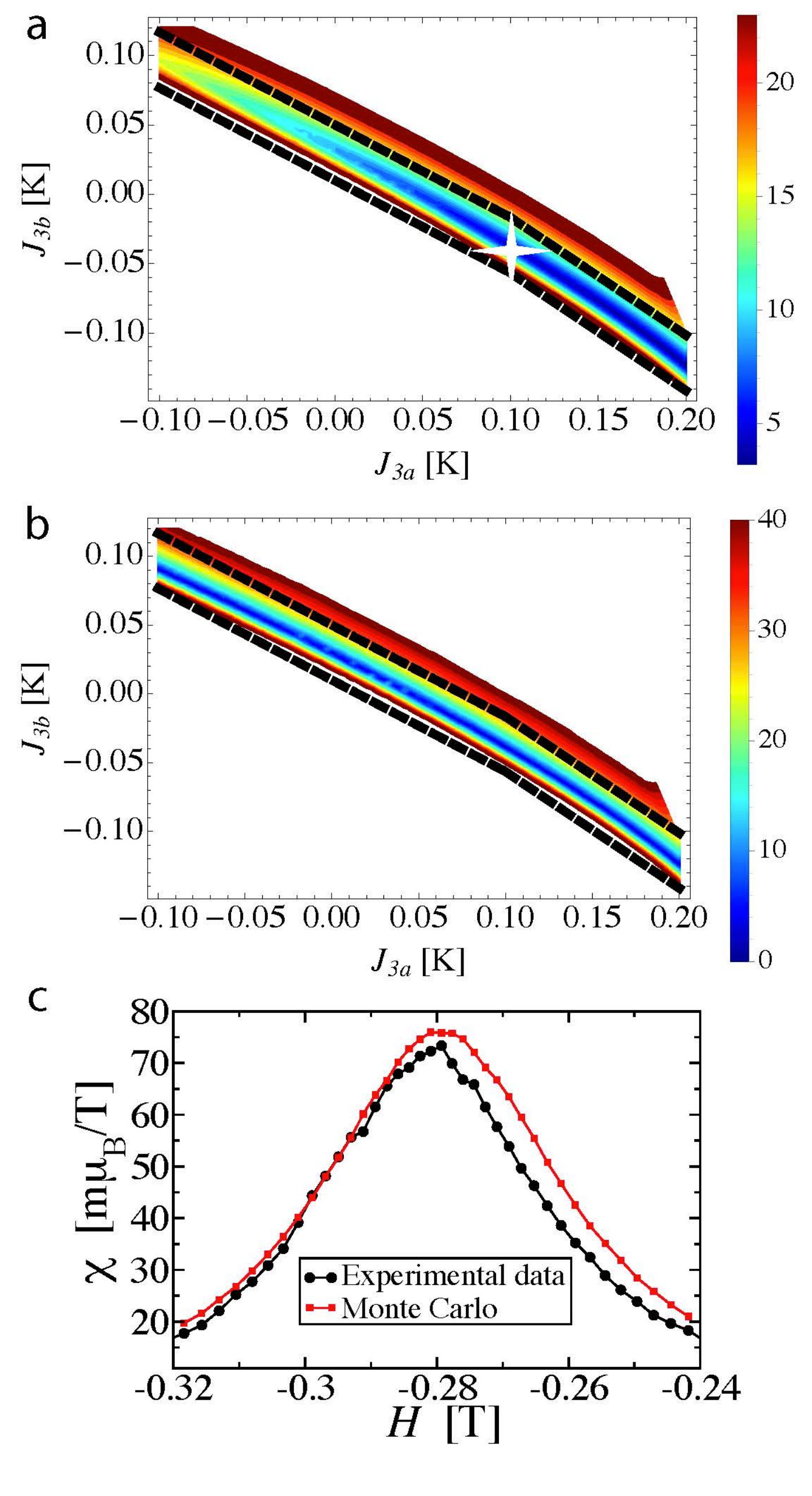}}
\caption{Determination of the $J_{3a}-J_{3b}$ constraining
  equation. Deviations of the susceptibility for the FCC sublattice
  (defined by the spins on triangular planes), calculated in the Monte
  Carlo simulations, from the experimental data\cite{Sato06} as a
  function of $J_{3a}$ and $J_{3b}$. In {\bf a}, the RMS deviation
  $\sigma_{\chi}$, calculated according to Eq.~(\ref{eq:sigma_chi}),
  is shown, while in {\bf b} the absolute value of the height
  difference, determined using Eq.~(\ref{eq:Delta}), is displayed. The
  dashed black lines show the parameter range used in
  Fig.\ref{scatter} of the main text, while the optimal relationship
  lies precisely between the two lines. In {\bf c}, the susceptibility
  measured in the experiment and calculated in the Monte Carlo
  simulation at $J_{3a}=0.10$ K and $J_{3b}=-0.04$ K (indicated by a
  white star in panel {\bf a}) is shown on an absolute scale as a
  representative example.  }
\label{112}
\end{figure}

We calculate the ${[\bar{1}\bar{1}1]}$ susceptibility in Monte Carlo
simulations using a system size $L=4$, and evaluate the goodness of
fit according to
\begin{equation}
  \sigma_{\chi}^2 =
  \frac{1}{N_h} \sum_{i=1}^{N_h} [\chi_{\text{MC}}
  (h_i^{[\bar{1}\bar{1}1]}) - \chi_{\text{exp}}
  (h_i^{[\bar{1}\bar{1}1]}) ]^2, 
  \label{eq:sigma_chi}
\tag{A3}
\end{equation}
where $\chi_{\text{MC}} (h_i^{[\bar{1}\bar{1}1]})$ and
$\chi_{\text{exp}} (h_i^{[\bar{1}\bar{1}1]})$ are the Monte Carlo and
experimental susceptibilities, respectively, determined at the applied
[$\bar{1}\bar{1}$1] field component $h_i^{[\bar{1}\bar{1}1]}$. The RMS
deviation, $\sigma_{\chi} $, is calculated for a total of $N_h=38$
different values of $h_i^{[\bar{1}\bar{1}1]}$ ranging from $-0.3184$ T
to $-0.2418$ T.  We display $\sigma_{\chi} $ in Fig.~\ref{112}a as a
function of $J_{3a}$ and $J_{3b}$. In order to illustrate the nature
and quality of the fit at a representative point, the experimental
susceptibility, $\chi_{\text {Exp}}$, is displayed along with
$\chi_{\text{MC}}$ calculated at $J_1=0.10$ K and $J_{3b} = - 0.04$ K
in Fig.~\ref{112}c.  We also calculate and plot in Fig.~\ref{112}b the
difference between the maximum of the Monte Carlo and experimental
susceptibilities precisely at the cancellation field
\begin{equation} 
\Delta = | \chi_{\text MC} - \chi_{\text{exp}}
(h_i^{[\bar{1}\bar{1}1]} = h_{\text{c}}^{[\bar{1}\bar{1}1]} ) |.
\label{eq:Delta}
\tag{A4}
\end{equation} 
As we see in Fig.~\ref{112}a and Fig.~\ref{112}b, the optimal fit
falls on a slightly bent curve. The appearance of such a line can be
understood within a mean-field interpretation. Since the
susceptibility at a given temperature depends on the sum of the
interactions, a decrease in the value of $J_{3a}$ can be compensated
for by an increase in the value of $J_{3b}$. This would keep the
susceptibility constant and lead to a line with negative slope.

The optimal relations between $J_{3a}$ and $J_{3b}$ that result from
this analysis are (all values in kelvin):
\begin{equation}
  J_{3b}=\left\{ \begin{array} {r@{\quad:\quad}l}
      -0.667 J_{3a} +0.03& J_{3a}<0.1\\ -0.842 J_{3a} +0.0474 &
      J_{3a}>0.1, \end{array} \right. 
  \tag{A5}
\end{equation} 
which is Eq.~(\ref{eq:j3}). As a precaution, we also explored the
parameter range beyond the above optimal equations and allowed
$J_{3b}$ to lie within $\pm 0.02$ K of the optimal relation (quite a
large range) as shown by the dashed lines in Fig.~\ref{112}a and
Fig.~\ref{112}b. This extended parameter space was used to generate
Fig.~\ref{scatter} of the main text.

The constraint Eqs. (A2) and (A5) derived above apply to the case of a
sufficiently large $[112]$ magnetic field component such that, for the
temperature $T = 0.7$ K considered to fit the data, one can safely
ignore the thermal fluctuations of the spins on the kagome layers. For
the temperatures relevant to this $[112]$ experiment, this is readily
achieved for a $[112]$ field component larger than about 2
Tesla\cite{Ruff,Sato06,Higa_112}.  One may then ask whether the
constrained $J_{ij}$ couplings extracted through such in-field
experiment would be significantly renormalized compared to the bare
$J_{ij}$ needed to describe the zero-field specific heat and neutron
scattering data? To address this issue we first note that with an
excited crystal field doublet at an energy of order 300 K above the
ground state doublet, one can safely neglect the field-induced
admixing between the ground doublet and excited doublet for a $[112]$
field of order of 2 Tesla.  This field strength corresponds to a
Zeeman splitting of about 20 K for a magnetic moment of 10 $\mu_{\rm
  B}$ (ignoring geometrical factors arising from projection of the
local $\langle 111 \rangle $ Ising direction of the spins on the
triangular layer with the $[112]$ direction).  Secondly, as the
$g$-tensor for the crystal-field Dy$^{3+}$ ion in Dy$_2$Ti$_2$O$_7$ is
strictly Ising-like,\cite{Gingras11} the projection of the microscopic
interionic Hamiltonian into the crystal field doublet would not be
modified by field-induced perturbative corrections to the
crystal-field ground state doublet wavefunctions.  Finally, one notes
that in the case of the Er$^{3+}$ and Yb$^{3+}$ Kramers ions in
Er$_2$Ti$_2$O$_7$ and Yb$_2$Ti$_2$O$_7$, the anisotropic spin-spin
coupling parameters describing the interactions between all the
components of the effective ${\bm S}=1/2$ spin determined in large
magnetic field \cite{Ross_ETO,Ross_YbTO} can be successfully used to
describe, without any adjustable parameters or field-induced
renormalization of the couplings, the zero-field properties of these
compounds \cite{Applegate,Hayre,Oitmaa_ETO}.  On the basis of those
three arguments, we would thus expect that the constrained $J_{ij}$
parameters for Dy$_2$Ti$_2$O$_7$ determined in a $[112]$ field of 2
Tesla do not suffer from important field-induced renormalization
compared to the zero-field values that we ultimately seek.

\section{Phase Diagram Calculations}
\label{app:phasediagram}

\begin{figure}[h!]
\resizebox{\hsize}{!}{\includegraphics[clip=true]{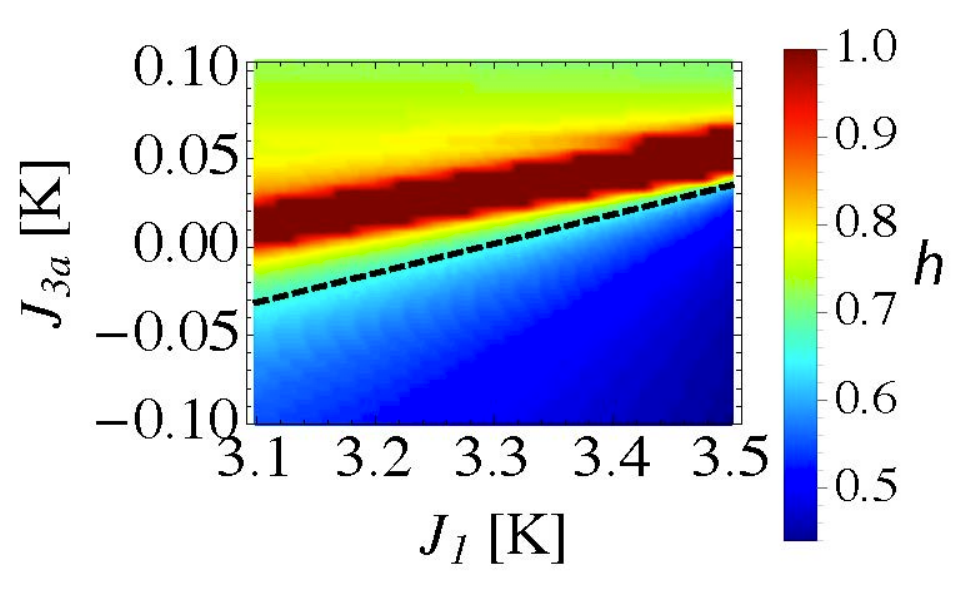}}
\caption{ Mean-field ordering wave vector. Map of the mean-field
  ordering wave vector, $\boldsymbol{q}$, in the constrained
  $J_1\mbox{-}J_{3a}$ parameter space, where the system selects a wave
  vector along the $(h h 0)$ direction. The broad red band across the
  middle indicates the single-chain phase with $\boldsymbol{q}=(1\ 1\
  0)$. In the lower part of the figure the order is close to the
  double-chain ordering wave vector $\boldsymbol{q}=(\frac{1}{2}\
  \frac{1}{2}\ 0)$. The phase boundary according to
  Eq.~\ref{eq:boundary} is shown as a dashed black line.  }
\label{mf_order}
\end{figure}

In order to determine the possible ordered state(s) that may arise in
the system, we first apply mean-field theory\cite{Enjalran_mft} and
then carry out Monte Carlo simulations. We employ mean-field theory to
perform a general survey of the entire parameter range and follow up
with a direct inspection of the low-temperature ordered states found
in the Monte Carlo calculation.

Considering the entire Brillouin zone, we find the mean-field ordering
wave vector to be in the $(hk0)$ plane for the entire parameter space
relevant for for Dy$_2$Ti$_2$O$_7$, namely $3.1\ \text{K}<J_1<3.5\
\text{K}$ and $-0.1\ \text{K}<J_{3a}<0.1\ \text{K}$.  We also find
that $h=k$ and show in Fig.~\ref{mf_order} the mean-field ordering
wave vector at the mean-field critical temperature, $T_c^{\rm{MF}}$
along the $(hh0)$ direction as a function of $J_1$ and $J_{3a}$. We
note a horizontal band of $(1\ 1\ 0)$ order below which there is a
gradual shift through incommensurate $h$-values to, ultimately, a
region with ordering at $(\frac{1}{2}\ \frac{1}{2}\ 0)$. At this
point, it is important to realize that the equal-moment constraint
does not apply to the mean-field theory, but limits the admissible
ordering wave vectors of the real material and our model. Furthermore,
we note that this constraint applies also to states with mean-field
ordering wave vectors that are commensurate with the Monte Carlo
simulation cells, such as, for example, a wave vector $(\frac{3}{4} \
\frac{3}{4} \ 0)$ (Ref.~[\onlinecite{Viertio93}]). Inspecting the
low-temperature states in the Monte Carlo simulation we find that the
gradual shift from ordering wave vector $(1\ 1\ 0)$ to $(\frac{1}{2}\
\frac{1}{2}\ 0)$, observed at $T_c^{\rm{MF}}$ in mean-field theory, is
replaced by a direct transition between these two ordered states, as
illustrated in Fig.~\ref{energy_boundary}. We now discuss the nature
of the two phases with ground state ordering wave vectors $(1\ 1\ 0)$
and $(\frac{1}{2}\ \frac{1}{2}\ 0)$ found in the Monte Carlo
simulations. These correspond to the upper and lower basins of
Fig.~\ref{cv_rms}c and Fig.~\ref{cv_rms}d in the main text,
respectively. We first focus on the ordered structures at $T=0$ before
considering finite temperature behavior.

\begin{figure}[h!]
\resizebox{\hsize}{!}{\includegraphics[clip=true]{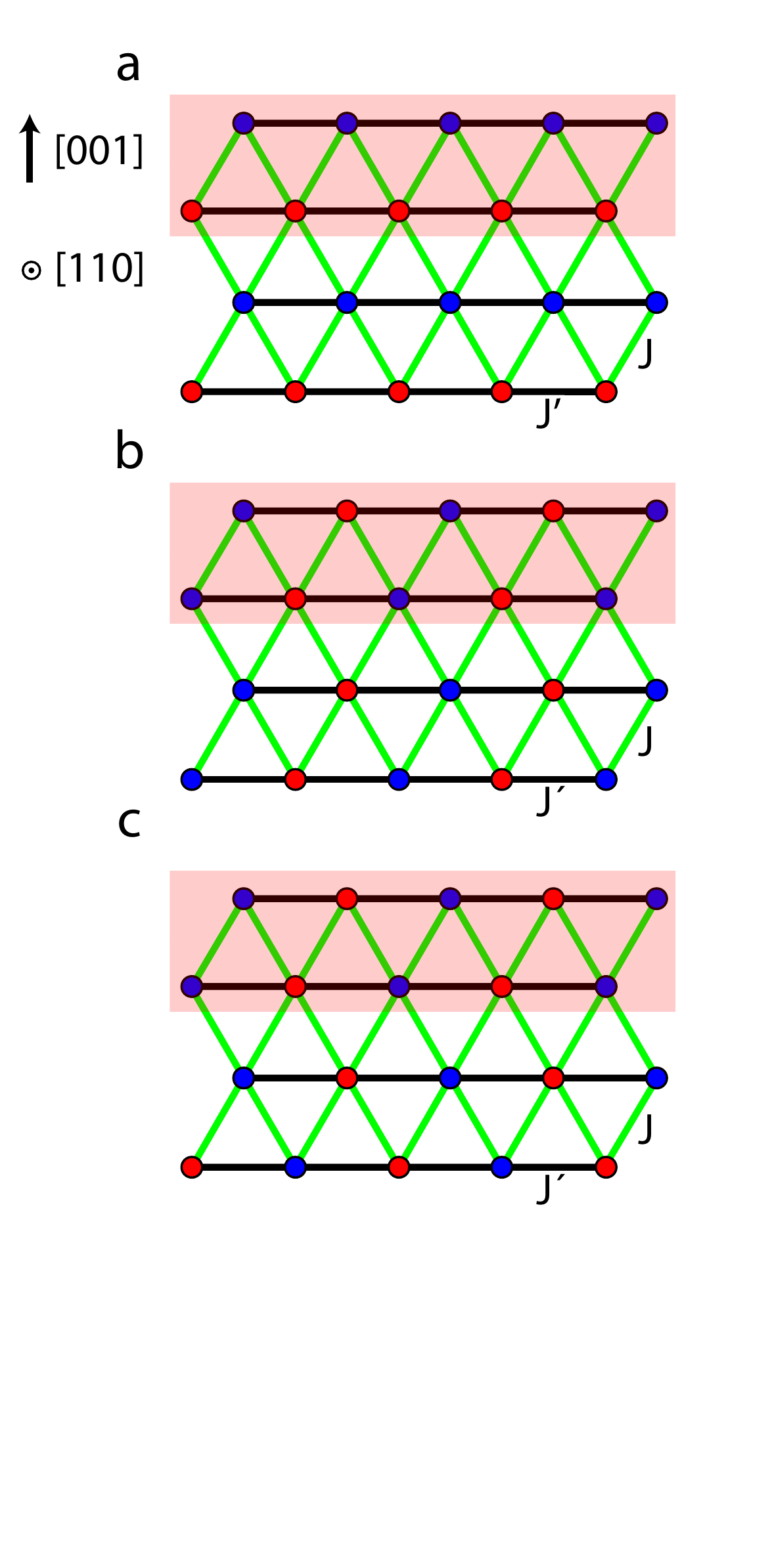}}
\caption{ Spin configuration in the ordered state.  View along spin
  chains in the $[110]$ direction. A red circle denotes an entire,
  ferromagnetically ordered, chain pointing out of the page, while a
  blue circle denotes a chain pointing into the page. A black bond
  segment indicate an exchange coupling $J'=J_{3b}$ while a green bond
  segment denotes coupling $J=J_{3a}+J_{3b}+J_2/3$. The
  semitransparent red bar indicates a layer of spin chains, commonly
  viewed along the $[001]$ direction, such as in Fig.~1f and Fig.~1g
  of the main text. The single-chain state is shown in {\bf a}, while
  a double-chain long-range-ordered state is depicted in {\bf b}. An
  example of a random stacking of the double-chain state is shown in
  {\bf c}.  }
\label{chains}
\end{figure}

As previously found for the simple dipolar spin ice model with
$J_2=J_{3a}=J_{3b}=0$ (Ref.~[\onlinecite{Melko01}]), the state at $(1\
1\ 0)$, or equivalently $(0\ 0\ 1)$, corresponds to the
``single-chain" state.  In this state, parallel chains of spins order
antiferromagnetically when viewed along a cubic $\langle100\rangle$
axis (see Fig.~\ref{chains}a). Inspection of the Monte Carlo spin
configuration of the novel state with propagation vector
$(\frac{1}{2}\ \frac{1}{2}\ 0)$ reveals a pattern where \emph{pairs}
of adjacent chains have spins aligned parallel, but pointing in the
opposite direction of the adjacent pairs on either side (see
Fig.~\ref{chains}b).  This period doubling in the $x-y$ plane is the
direct space origin of why the ordering wave vector is reduced from
$(1\ 1\ 0)$ to $(\frac{1}{2}\ \frac{1}{2}\ 0)$. We thus refer to this
as the ``double-chain'' state.

\begin{figure}[h!]
\resizebox{\hsize}{!}{\includegraphics[clip=true]{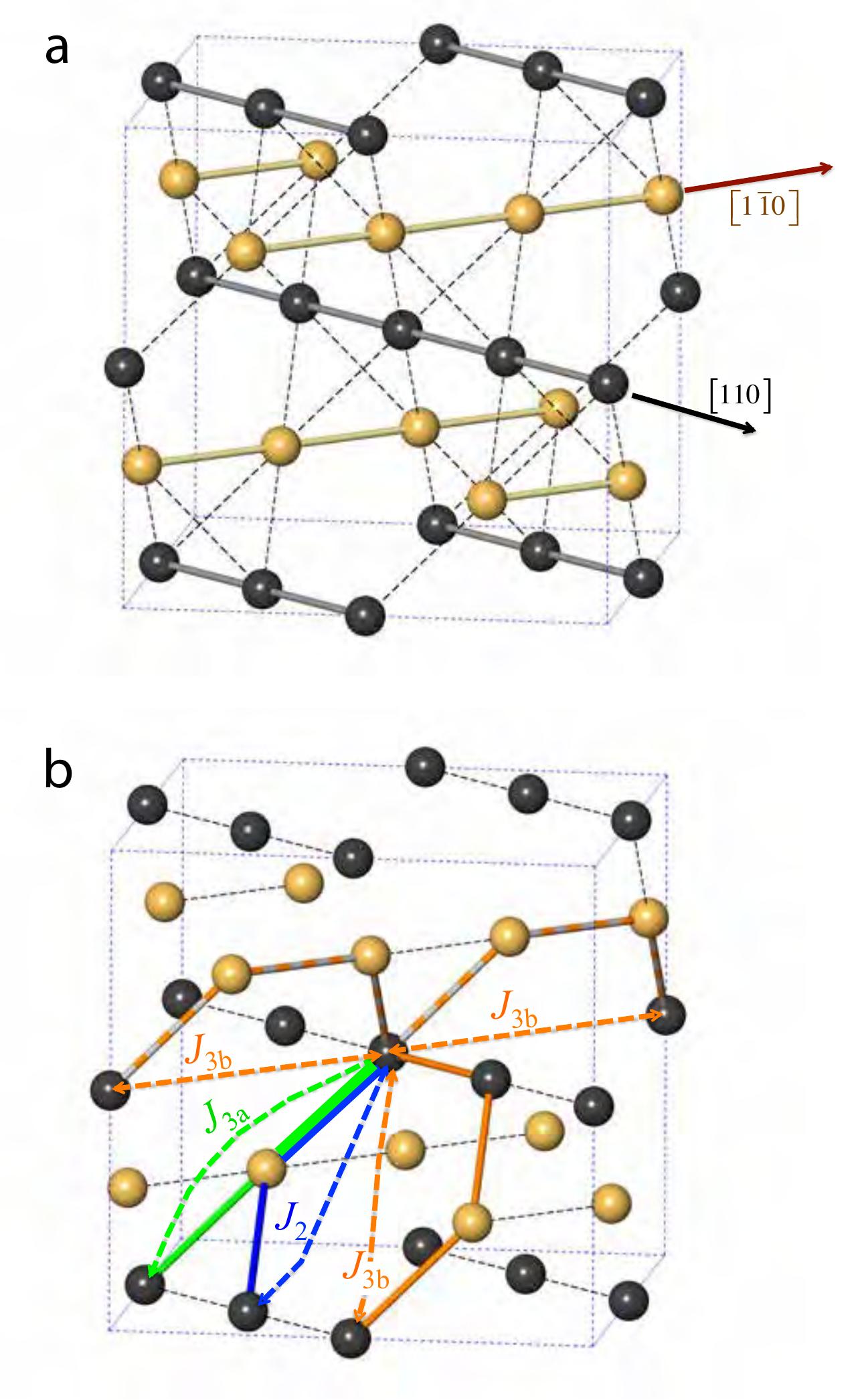}}
\caption{Spin chains. {\bf a.} The pyrochlore lattice is represented
  as a collection of spin chains along the $[110]$ (black atoms and
  bonds) and $[1\bar{1}0]$ (yellow atoms and bonds) directions. Dashed
  black bonds couple different chains together. {\bf b.}
  Further-neighbor interactions $J_{2}$ (blue dashed) $J_{3a}$ (green
  dashed) and $J_{3b}$ (orange dashed) couple different $[110]$ chains
  together. Two neighboring chains in the same $(001)$ plane are
  coupled by $J^\prime\equiv J_{3b}$ (per spin) while a pair of
  neighboring chains on two adjacent $(001)$ planes are coupled by
  $J\equiv J_{3a}+J_{3b}+J_{2}/3$ (per spin).  }
\label{chains3D}
\end{figure}

To gain a better understanding of how the competition between these
two states arises from third-nearest-neighbor interactions\cite{j3},
we invoke yet another description of the magnetic sites of the
pyrochlore lattice.  After having viewed it as a collection of
alternating kagome and triangular layers in the previous section, we
now view it as an array of two sets of one-dimensional spin chains
along the $[110]$ and $[1\bar{1}0]$ directions (see
Fig.~\ref{chains3D}a). In the long-ranged ordered states observed in
the Monte Carlo simulation the spins on each individual $[110]$ and
$[1\bar{1}0]$ chains are ferromagnetically correlated, and we
therefore treat these chains as elementary units. In a plane
perpendicular to the respective direction of the chains, the two sets
of chains form two triangular lattices, which are decoupled from each
other if only exchange couplings $J_1$, $J_2$, $J_{3a}$ and $J_{3b}$
are considered. Therefore, we first consider a triangular lattice
formed by spin chains along the $[110]$ direction and ignore (for the
moment) the long-range dipolar interactions. Each chain couples to two
of its nearest neighbors with the same $z$ coordinate with exchange
coupling $J^\prime=J_{3b}$, while the chain couples to the other four
nearest neighbors with exchange coupling $J=J_{3a}+J_{3b}+J_{2}/3$, as
illustrated in Fig.~\ref{chains3D}b and Fig.~\ref{chains}. Both
couplings are determined using the convention used in
Ref.~[\onlinecite{Yavo08}] with $\hat{z}_i\cdot\hat{z}_j$ explicitly
included. As stated above, two third-neighbor coupled sites belong to
the same sublattice with $\hat{z}_i\cdot\hat{z}_j=1$, while two
second-neighbor sites reside on two different sublattices so that
$\hat{z}_i\cdot\hat{z}_j=-1/3$. The negative sign does not appear in
the definition of $J$ thanks to an additional negative sign coming
from the alternating directions of spins, represented in their local
components along $\hat{z}_i$, along a ``ferromagnetically'' ordered
chain in the global $[110]$ direction.  In summary, the $(001)$ planes
of $[110]$ spin chains have intra-plane nearest-neighbor coupling
$J^\prime$ and inter-plane coupling $J$.

We thus end up mapping the competing states observed in the Monte
Carlo simulations of the three-dimensional pyrochlore lattice to a
two-dimensional triangular lattice, for which the ground state phase
diagram can be obtained by straightforward energy arguments. For
$J^\prime<J$, the planes of spin chains form ferromagnetic sheets
whose directions alternate between planes. This state corresponds to
the single-chain phase (Fig.~\ref{chains}a).  For $J^\prime>J$, the
$[110]$ spin chains within the same $(001)$ plane form an
antiferromagnetic state, corresponding to the double-chain state
(Fig.~\ref{chains}b). The inter-plane coupling is frustrated and these
planes are decoupled with different stackings of the antiferromagnetic
planes being degenerate (Fig.~\ref{chains}c).  Moreover, due to the
antiferromagnetic ordering of the spin chains within a given $(001)$
plane, the long-range dipolar interaction is well-shielded for an
arbitrary stacking of the planes (\emph{i.e.}  with a propagation
vector along $[001]$).  In Fig.~\ref{chains}, the structure of the
single and double-chain states is shown from the spin-chain
perspective.

The boundary between the two phases is determined by the condition
$J=J^\prime$ if the long-range dipolar interaction is not
included. The dipolar interaction $D$ shifts the boundary marginally
by a small constant.  By determining this constant numerically using
the Ewald summation method, we obtain Eq.~\ref{eq:boundary} for the
boundary:
\begin{equation}
  J_{3a}+\frac{J_{2}}{3} + 0.02D=0. 
\label{eq:boundary2}
\tag{A6}
\end{equation} 
This equation was used to draw the phase boundary in
Fig.~\ref{neutron_boundary}b and Fig.~\ref{cv_rms}a-d as well as for
Figs.~\ref{mf_order} and \ref{neutron_ratio}. Note the small $0.02$
prefactor in front of the dipolar contribution term $D$, which \emph{a
  posteriori} justifies the discussion above in terms of approximately
independent sets of $[110]$ and $[1\bar{1}0]$ chains.

\begin{figure}[h!]
\resizebox{\hsize}{!}{\includegraphics[clip=true]{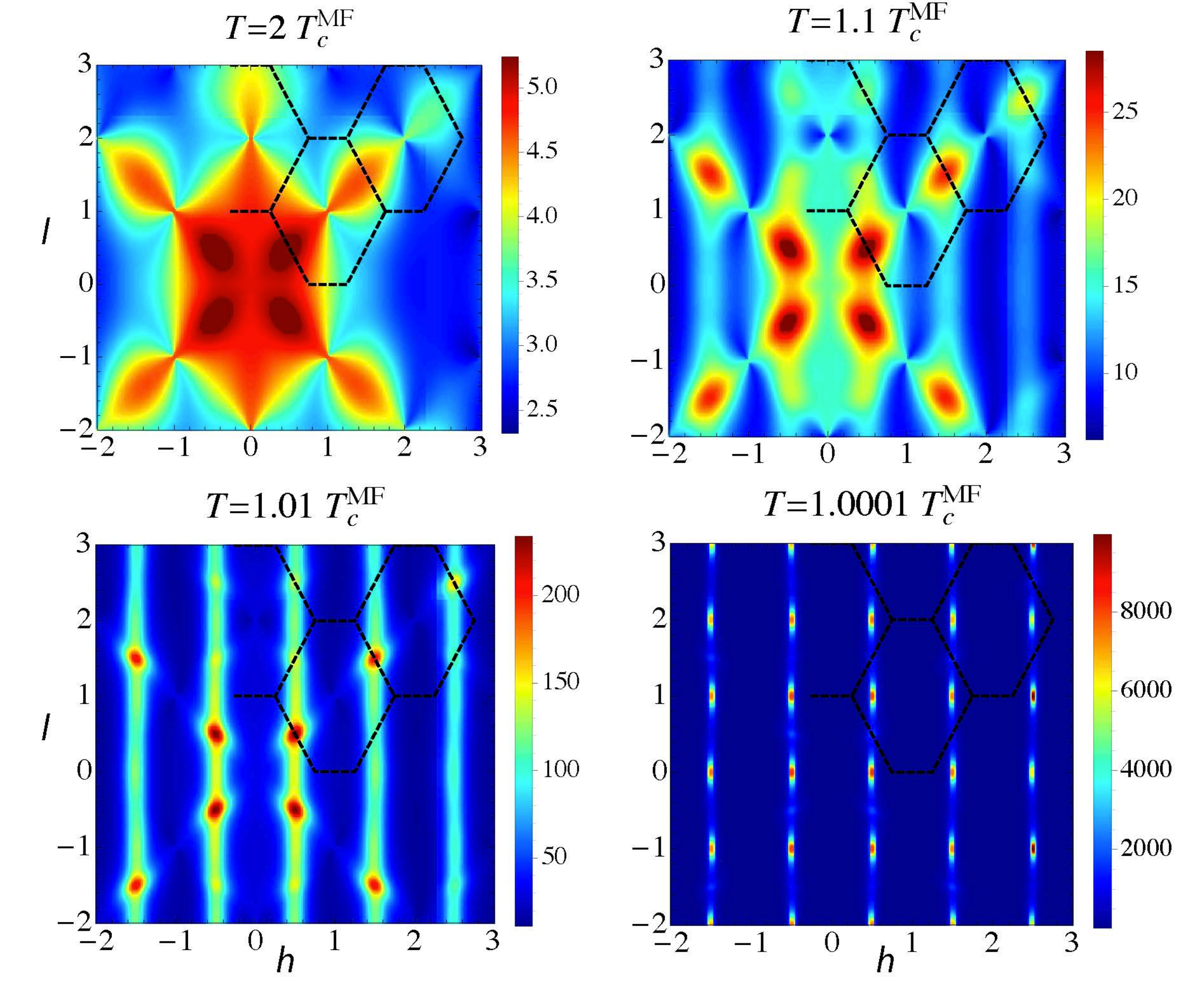}}
\caption{ Mean-field neutron scattering structure factor
  $\bm{S(}\boldsymbol{q}\bm{)}$ in the double chain region. The
  various panels show the structure factor $S(\boldsymbol{q})$ in the
  $(hhl)$ plane calculated in mean-field theory as the mean-field
  critical temperature, $T_{\rm{c}}^{\rm{MF}}=2.9614$ K is approached
  from the paramagnetic phase for the set of constrained exchange
  couplings with $J_1=3.41$ K and $J_{3a}=-0.04$ K, a point deep in
  the double-chain region in the phase diagram of Fig.~\ref{mf_order}.
  The weight shifts from $(\frac{1}{2}\ \frac{1}{2}\ \frac{1}{2})$ to
  the true ordering wave vector $(\frac{1}{2}\ \frac{1}{2}\ 0)$ at
  $T=1.0001\, T_{\rm{c}}^{\rm{MF}}$, extremely close to the transition
  temperature.  }
\label{mf_sq}
\end{figure}

Finally, we consider some experimentally observable finite-temperature
consequences of the stacking degeneracy of the double-chain
state. From Figs.~\ref{neutron_boundary}f,g,h it is clear that the
signature of the double-chain phase in the Monte Carlo calculation of
the neutron scattering $S(\boldsymbol{q})$ occurs at $(\frac{1}{2}\
\frac{1}{2}\ \frac{1}{2})$, and not at the mean-field ordering wave
vector $(\frac{1}{2}\ \frac{1}{2}\ 0)$. In order to rationalize the
temperature evolution of $S(\boldsymbol{q})$ we again first turn to a
mean-field analysis.  In Fig.~\ref{mf_sq}, we show $S(\boldsymbol{q})$
calculated in mean-field theory at $J_1=3.41$ K and $ J_{3a}=-0.04$ K,
a point deep in the double-chain region. The temperature is expressed
in terms of the critical mean-field temperature, $T_{\rm{c}}^{\rm{MF}}
= 2.9614$ K, for these ($J_1$, $ J_{3a}$) values. Note how the main
response stays at $(\frac{1}{2}\ \frac{1}{2}\ \frac{1}{2})$ until
extremely close to the transition temperature, $T= 1.0001
T_{\rm{c}}^{\rm{MF}}$, where the weight finally shifts to the ordering
wave vector $(\frac{1}{2}\ \frac{1}{2}\ 0)$.  The true order parameter
is therefore effectively hidden by the near degeneracy of different
stackings in the $z$-direction down to very close to the transition
temperature. In the Monte Carlo simulation, the ordering proceeds in a
different manner. At high temperature, the Monte Carlo results for
$S(\boldsymbol{q})$ agree perfectly with mean-field theory as it
should\cite{Enjalran_mft}. Below $T \approx 10\,
T_{\rm{c}}^{\rm{MF}}$, differences start to emerge. At the transition
temperature, $T_{\rm{c}}^{\rm{MC}}\sim0.13$ K, the Monte Carlo system
freezes into a state of long-range order in the $x\mbox{-}y$ $(110)$
plane, but with random stacking in the ($[001]$) $z$-direction (see
Fig.~\ref{chains}c).  Only by using parallel tempering Monte Carlo
methods with a very fine temperature mesh (see Subsection
\ref{sec:method}) could the true perfectly stacked double-chain ground
state (Fig.~\ref{chains}b) be resolved in the simulation performed
with 1024 ($L=4$) spins.

\section{Neutron Scattering Analysis}
\label{app:neutronscatt}

We begin by analyzing how a possible freezing of the sample would
affect the neutron scattering structure factor $S(\boldsymbol{q})$.
From numerous experiments, is is clear that thermal equilibration is
adequately fast above $0.7$ K, while the samples rapidly fall out of
equilibrium below that temperature\cite{Poma13, Higa02,
  Klemke11}. Considering the typical time duration of neutron
scattering data accumulation at a given temperature ($\sim 10^0 -
10^1$ hours) we assume that the experimental neutron scattering data
were well equilibrated down to $0.7$ K, while below this temperature
we cannot say with certainty whether the sample was fully
equilibrated, partially equilibrated or frozen.

\begin{figure}[h!]
 \resizebox{\hsize}{!}{\includegraphics[clip=true]{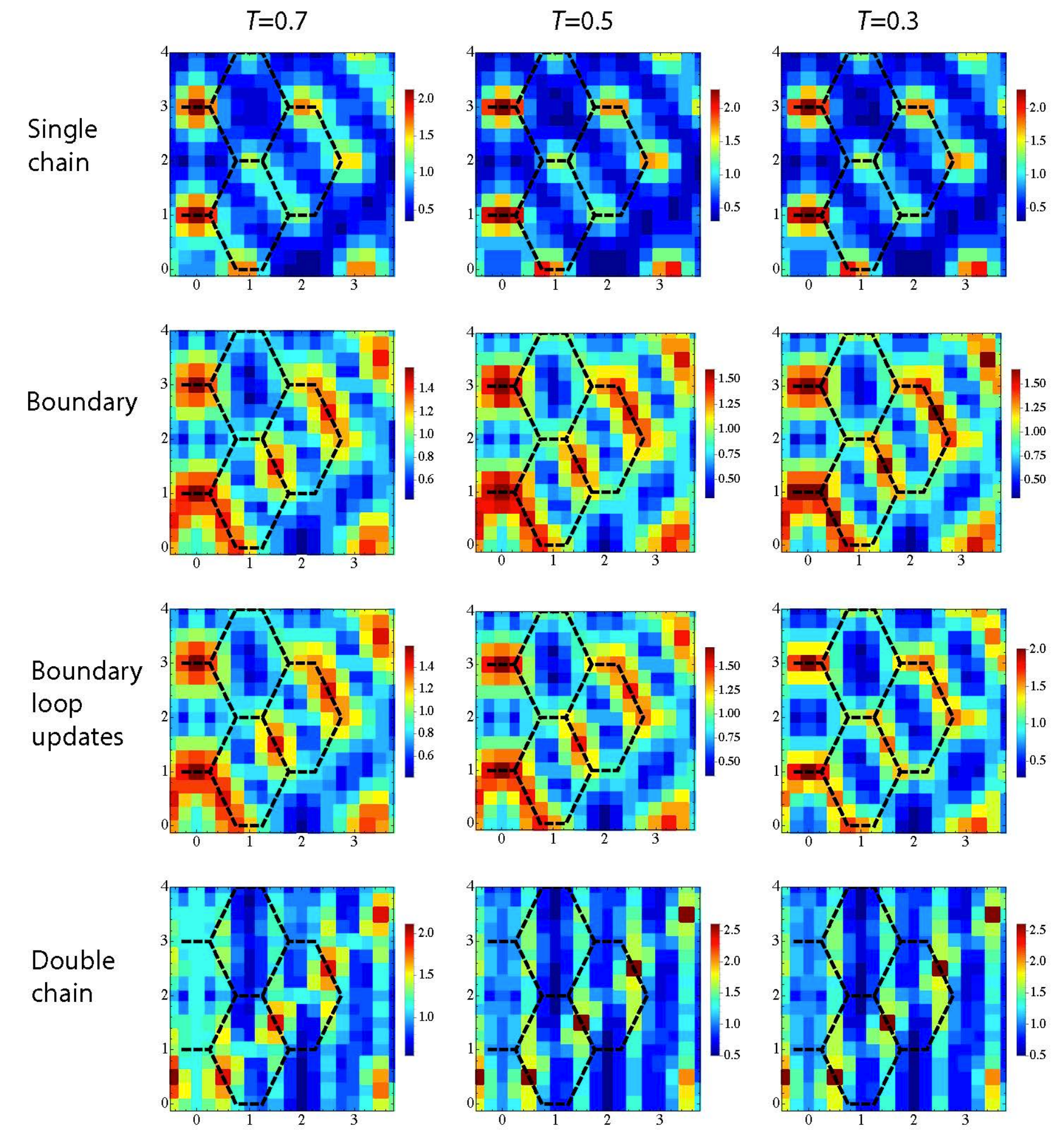}}
 \caption{ Simulation of structure factor as the system freezes.
   Simulation of neutron structure factor using 10$^5$ MC steps
   consisting of only attempted single spin flips for a system of 1024
   spins ($L=4$). The three columns stand for $T$=0.7, 0.5 and 0.3
   K. The top row represents a point in the single chain region
   ($J_1$=3.3 K, $J_{3a}$=0.07 K), the second row a point close to the
   boundary ($J_1=3.41$ K, $J_2=-0.14$ K, $J_{3a}=J_{3b}=0.025$ K,
   Ref.~[\onlinecite{Yavo08}]), and the bottom row a point in the
   double chain region ($J_1$=3.44 K, $J_{3a}$=-0.02 K).  Only few
   updates are accepted below $T=0.6$ K causing little further change
   in the structure factor below this temperature. This indicates that
   the structure factor in a frozen material is an imprint of the
   pattern at the temperature at which the sample froze.  The third
   row represents the same parameter point as the second row, but now
   with loop updates employed to achieve complete equilibration at all
   temperatures. Also in this case the scattering patterns at $T$=0.5
   K and $T$=0.3 K are quite similar. The displayed structure factor
   in the first, second and last rows is an average over 400 different
   simulations that have frozen in different configurations, modeling
   the spatial variations in a macroscopic sample used in the
   experiments and which neutron scattering measurements average out.
 }
\label{frozen}
\end{figure}

The effects of freezing can be explored in a Monte Carlo simulation.
Below $0.7$ K, the number of single spin flips required to equilibrate
the sample increases exponentially and loop updates are generally
needed to speed up the thermalization\cite{Melko01,Melko04}. We thus
study the effects of freezing by measuring $S(\boldsymbol{q})$ at
successively lower temperatures by using \emph{only} single spin flip
updates and keeping the number of updates at each temperature constant
to mimic a fixed experimental equilibration time scale. To this end,
we use $10^5$ MC steps at each temperature. With this choice, the
simulated system effectively freezes slightly below a temperature of
$0.6$ K. In Fig.~\ref{frozen} we display $S(\boldsymbol{q})$ where the
columns represent $T=0.7 , 0.5$ and $0.3$ K, moving left to right. The
first, second and fourth rows represent a parameter point in the
single-chain region, on the boundary and in the double chain region,
respectively.  One clearly sees that, as the system freezes, as
assessed by the vanishing Monte Carlo spin flip rate (not shown), the
neutron scattering pattern also freezes. Most importantly, there is
\emph{no change} in the ZBS and \emph{no} fundamental reciprocal space
redistribution in the peak intensities. We therefore conclude that the
pattern recorded at $T=0.3$ K looks like the pattern at the last
temperature the sample was properly thermally equilibrated,
independently of the dynamical state of the experimental data
nominally measured at $0.3$ K

\begin{figure}[h!]
 \resizebox{\hsize}{!}{\includegraphics[clip=true]{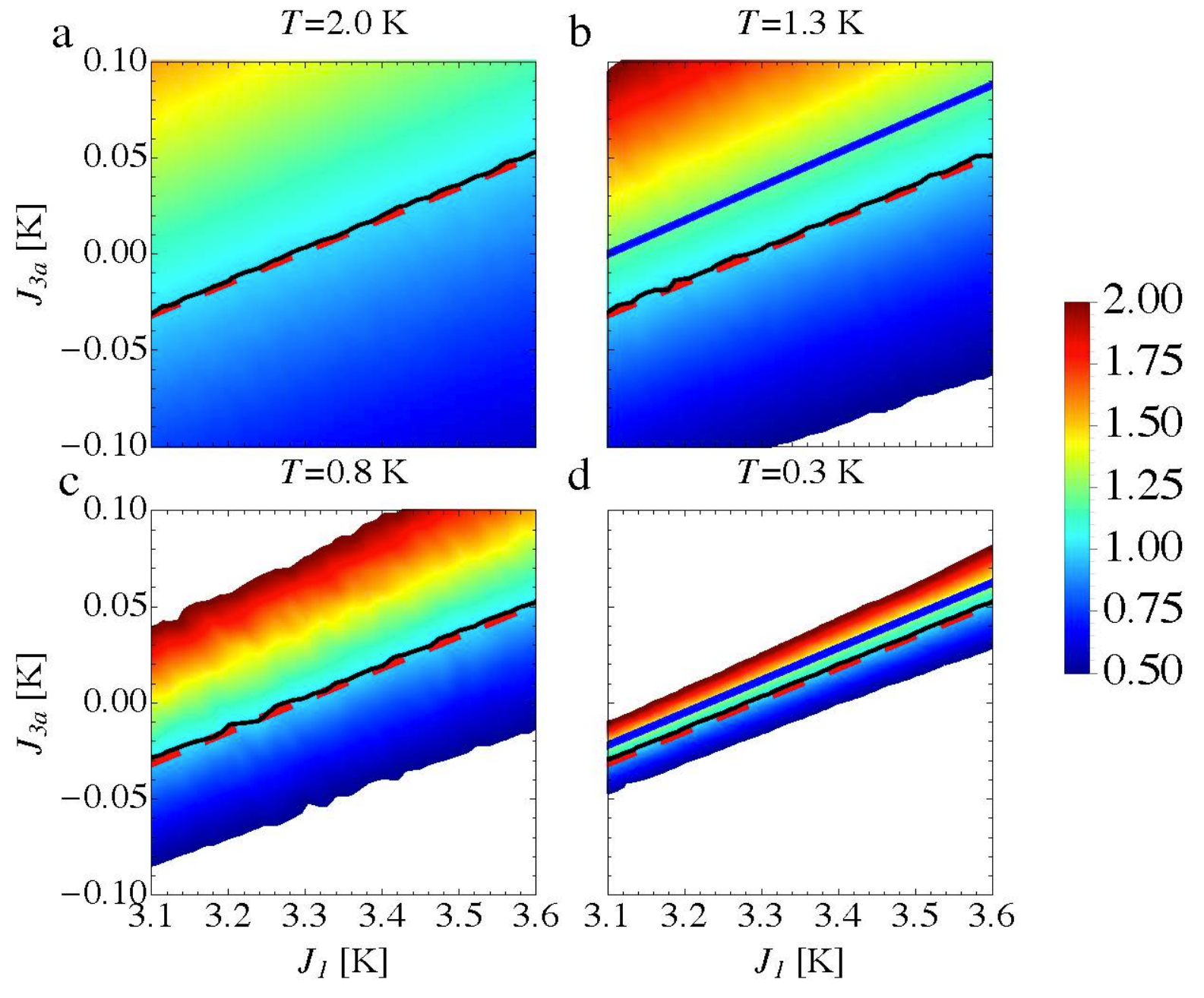}}
 \caption{ Ratio of structure factor for single-chain and double-chain
   order.  The ratio $r\equiv S(0\ 0\ 3)/S(\frac{3}{2}\ \frac{3}{2}\
   \frac{3}{2})$ in the constrained $J_1\mbox{-}J_{3a}$ plane
   calculated in the Monte Carlo simulations, without the Dy$^{3+}$
   form factor. The phase boundary is marked by a dashed red line, and
   the $r=1$ contour is indicated by a thin black line, which overlaps
   the phase boundary in all panels. For all temperatures, the ratio
   is therefore unity at the phase boundary, while it is greater than
   one in the single-chain region above the boundary and less than one
   in the double-chain region below the boundary. The blue lines in
   {\bf b} and {\bf d} indicate the ratio for the experimental neutron
   scattering data at $T=1.3$ K and $T=0.3$ K.  }
\label{neutron_ratio}
\end{figure}

Next, we perform a more detailed analysis of the intensity maxima in
the structure factor.  As noted in the main text, the single-chain
phase neutron scattering response, $S(\boldsymbol{q})$, occurs at the
ordering wave vector $(1\ 1\ 0)$, or equivalently at $(0\ 0\ 3)$ in
the experimental neutron scattering data where the second Brillouin
zone is probed. On the other hand, the double-chain phase response
occurs at the ordering wave vector $(\frac{1}{2}\ \frac{1}{2}\
\frac{1}{2})$, or equivalently $(\frac{3}{2}\ \frac{3}{2}\
\frac{3}{2})$ in the experiments, rather than at $(\frac{1}{2}\
\frac{1}{2}\ 0)$ because of the aforementioned random stacking of
ordered spin planes. In order to quantify this observation, we show in
Fig.~\ref{neutron_ratio} the ratio $r$ $\equiv$ $S(0\ 0\
3)/S(\frac{3}{2}\ \frac{3}{2}\ \frac{3}{2})$ at temperatures $T=2.0$,
$1.3$, $0.8$ and $0.3$ K (without the Dy$^{3+}$ magnetic form factor)
computed using Monte Carlo simulations.  Interestingly, $r$ is
precisely unity at the boundary determined by Eq.~(\ref{eq:boundary})
and is shown as a dashed red line in the four panels of
Fig.~\ref{neutron_ratio}.  In the single-chain region, above the phase
boundary, $r>1$, while $r<1$ in the double-chain region
\emph{independently} of the temperature. The experimental ratio, after
dividing out the Dy$^{3+}$ form factor, is $r\cong1.18$ at $T=1.3$ K
and $r\cong 1.29$ at $T=0.3$ K and is marked by a dark blue line in
Fig.~\ref{neutron_ratio}b and Fig.~\ref{neutron_ratio}d,
respectively. The present neutron scattering experiments therefore
position Dy$_2$Ti$_2$O$_7$ very close to, but slightly above the
ground state phase boundary line, that is in the single-chain region
of the phase diagram. Note that the experimental ratio at $T=0.3$~K
locates Dy$_2$Ti$_2$O$_7$ closer to the phase boundary than does the
ratio at $T=1.3$ K. However, small differences in the measured value
of $r$ have a large effect on the position of the corresponding
contour line in Fig.~\ref{neutron_ratio}d. To determine the location
of Dy$_2$Ti$_2$O$_7$ in the phase diagram with high precision one
would need a sequence of high-resolution and well-equilibrated
experimental neutron scattering measurements of $S(\boldsymbol{q})$ at
several temperatures between $0.3$ K and $1.3$ K. Also, note the
remarkable implication of the analysis above: neutron scattering
experiments performed at a high temperature allows one to
``anticipate'' the ground state of the material (excluding quantum,
and other disruptive low-$T$ phenomena) simply by analyzing the peak
ratio $r$.

\section{Nuclear Specific Heat}
\label{app:nuclearspecheat}

\begin{figure}[h!]
\resizebox{\hsize}{!}{\includegraphics[clip=true]{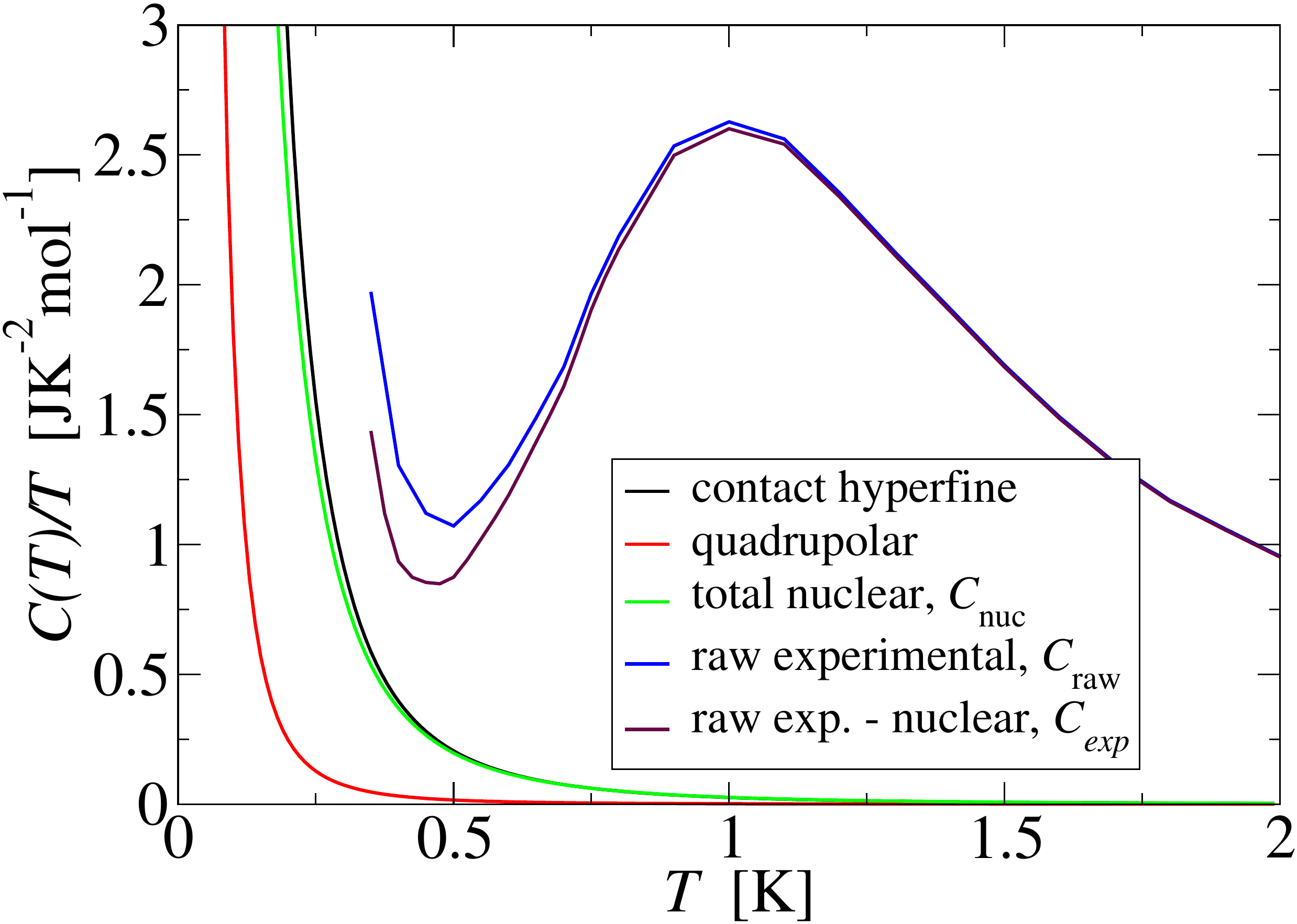}}
\caption{Nuclear contact contribution (black curve), quadrupolar (red
  curve ) and total, $C_{\text {nuc}}/T$ (green curve) nuclear
  contribution to the specific heat. The blue curve shows the raw
  experimental $C_{\text {raw}}/T$ data while the purple curves shows
  the residual electronic data, $C_{\text{exp}}(T)/T$ data.  }
\label{nuclear}
\end{figure}

Whereas it is common knowledge that the nuclear contribution dominates
over the electronic part of the specific heat of Holmium-based
compounds at $T\lesssim 1.5$
K\cite{arXivHTO,BramwellHo2Ti2O72001,Cornelius}, the nuclear specific
heat of Dy$_2$Ti$_2$O$_7$ has generally been ignored.  Here we show
that, for a quantitative analysis, this is not justified below
$T\approx0.5$ K.  Since the main nuclear contribution to the specific
heat comes from the interaction of the deep lying 4$f$ electrons and
the nucleus this interaction is considered not to be directly affected
by the ionic environment.\cite{Anderson69} While the crystal field can
alter the symmetry of the crystal field doublet, one expects that if
the magnitude and symmetry symmetry of the moment, $\langle
J^z\rangle$, are similar in two different compounds, the effective
hyperfine field $A\langle J^z\rangle$ interaction should be is
expected to be roughly the same.  In Dy$_2$Ti$_2$O$_7$ the
Ising\cite{Rau15} magnetic moment of each ion is nearly saturated,
with $\langle J^z\rangle \approx 7.4$, which is close the $J=15/2$
value for Dy$^{3+}$.  We therefore use the results of a calorimetric
investigation of the nuclear interactions in metallic Dy, which also
has a fully developed electronic moment.\cite{Anderson69} Following
this investigation we write the nuclear hyperfine (hf) Hamiltonian as
  \begin{equation}
H_{\text{hf}}= \tilde AI^z + P\left[(I^z)^2-\frac{1}{3}I(I+1)\right],
\end{equation}
with $\tilde A \equiv A \langle J^z\rangle$ where $A$ and $P$ are the
contact hyperfine term and electric quadrupole constants,
respectively.  Assuming the natural abundance of Dy, the relevant
magnetic isotopes are 18.88\% $^{161}$Dy and 24.97\% $^{163}$Dy, both
with nuclear spin $I=\frac{5}{2}$. We use the parameters $\tilde
A_{161}=0.0399$ K, $P_{161}=0.0093$ K, $\tilde A_{163}=0.0559$ K and
$P_{163}=0.0098$ K from Ref.~[\onlinecite{Anderson69}].  These values
are also roughly ($\pm 10$\%) from those determined in various
Dy-based insulating
salts\cite{Mattis,Dy_salt_1,Dy_salt_2,Dy_salt_3,Dy_salt_4}.  The
hyperfine partition function is given by
$Z_{\text{hf}}=\sum_{I^z=-5/2}^{I^z=+5/2}e^{-\beta H_{\text{hf}}}$,
and the calculation of the nuclear contribution to the specific heat
is straightforward. The result is shown in Fig.~\ref{nuclear}, where
we display the total nuclear specific heat as well as the separate
contact hyperfine and electric quadrupolar terms. Notice that electric
quadrupolar term, which becomes somewhat noticeable below $T\approx
0.2$~K, {\it reduces} the total nuclear contribution. It is clear that
there is a significant nuclear contribution to the total
$C_{\text{raw}}(T)/T$ specific heat below $T\approx0.5$ K, and the
nuclear specific heat causes part of the shoulder that starts to
develop already at $T\approx0.7$ K. Given this non-negligible
hyperfine contribution, it would be very interesting to see the result
of a well-equilibrated measured specific heat measurement on a
$^{162}$Dy enriched sample used for neutron scattering
studies\cite{Fennell04,Morris2009,Klemke11}, since there should be no
nuclear specific heat for such a sample.

\bibliography{comp_prb}

\begin{thebibliography}{85}%
\makeatletter
\providecommand \@ifxundefined [1]{%
 \@ifx{#1\undefined}
}%
\providecommand \@ifnum [1]{%
 \ifnum #1\expandafter \@firstoftwo
 \else \expandafter \@secondoftwo
 \fi
}%
\providecommand \@ifx [1]{%
 \ifx #1\expandafter \@firstoftwo
 \else \expandafter \@secondoftwo
 \fi
}%
\providecommand \natexlab [1]{#1}%
\providecommand \enquote  [1]{``#1''}%
\providecommand \bibnamefont  [1]{#1}%
\providecommand \bibfnamefont [1]{#1}%
\providecommand \citenamefont [1]{#1}%
\providecommand \href@noop [0]{\@secondoftwo}%
\providecommand \href [0]{\begingroup \@sanitize@url \@href}%
\providecommand \@href[1]{\@@startlink{#1}\@@href}%
\providecommand \@@href[1]{\endgroup#1\@@endlink}%
\providecommand \@sanitize@url [0]{\catcode `\\12\catcode `\$12\catcode
  `\&12\catcode `\#12\catcode `\^12\catcode `\_12\catcode `\%12\relax}%
\providecommand \@@startlink[1]{}%
\providecommand \@@endlink[0]{}%
\providecommand \url  [0]{\begingroup\@sanitize@url \@url }%
\providecommand \@url [1]{\endgroup\@href {#1}{\urlprefix }}%
\providecommand \urlprefix  [0]{URL }%
\providecommand \Eprint [0]{\href }%
\providecommand \doibase [0]{http://dx.doi.org/}%
\providecommand \selectlanguage [0]{\@gobble}%
\providecommand \bibinfo  [0]{\@secondoftwo}%
\providecommand \bibfield  [0]{\@secondoftwo}%
\providecommand \translation [1]{[#1]}%
\providecommand \BibitemOpen [0]{}%
\providecommand \bibitemStop [0]{}%
\providecommand \bibitemNoStop [0]{.\EOS\space}%
\providecommand \EOS [0]{\spacefactor3000\relax}%
\providecommand \BibitemShut  [1]{\csname bibitem#1\endcsname}%
\let\auto@bib@innerbib\@empty
\bibitem [{\citenamefont {Lacroix}\ \emph {et~al.}(2011)\citenamefont
  {Lacroix}, \citenamefont {Mendels},\ and\ \citenamefont
  {Mila}}]{Springer_frust_book}%
  \BibitemOpen
  \bibinfo {editor} {\bibfnamefont {C.}~\bibnamefont {Lacroix}}, \bibinfo
  {editor} {\bibfnamefont {P.}~\bibnamefont {Mendels}}, \ and\ \bibinfo
  {editor} {\bibfnamefont {F.}~\bibnamefont {Mila}},\ eds.,\ \href@noop {}
  {\emph {\bibinfo {title} {Highly Frustrated Magnetism}}},\ \bibinfo {series}
  {Springer Series in Solid-State Sciences}, Vol.\ \bibinfo {volume} {164}\
  (\bibinfo  {publisher} {Springer},\ \bibinfo {year} {2011})\BibitemShut
  {NoStop}%
\bibitem [{\citenamefont {Gardner}\ \emph {et~al.}(2010)\citenamefont
  {Gardner}, \citenamefont {Gingras},\ and\ \citenamefont
  {Greedan}}]{Gardner10}%
  \BibitemOpen
  \bibfield  {author} {\bibinfo {author} {\bibfnamefont {J.~S.}\ \bibnamefont
  {Gardner}}, \bibinfo {author} {\bibfnamefont {M.~J.~P.}\ \bibnamefont
  {Gingras}}, \ and\ \bibinfo {author} {\bibfnamefont {J.~E.}\ \bibnamefont
  {Greedan}},\ }\bibfield  {title} {\enquote {\bibinfo {title} {Magnetic
  pyrochlore oxides},}\ }\href@noop {} {\bibfield  {journal} {\bibinfo
  {journal} {Rev. Mod. Phys.}\ }\textbf {\bibinfo {volume} {82}},\ \bibinfo
  {pages} {53--107} (\bibinfo {year} {2010})}\BibitemShut {NoStop}%
\bibitem [{\citenamefont {Balents}(2010)}]{Balents_Nature}%
  \BibitemOpen
  \bibfield  {author} {\bibinfo {author} {\bibfnamefont {L.}~\bibnamefont
  {Balents}},\ }\bibfield  {title} {\enquote {\bibinfo {title} {Spin liquids in
  frustrated magnets},}\ }\href@noop {} {\bibfield  {journal} {\bibinfo
  {journal} {Nature}\ }\textbf {\bibinfo {volume} {464}},\ \bibinfo {pages}
  {199} (\bibinfo {year} {2010})}\BibitemShut {NoStop}%
\bibitem [{\citenamefont {Anderson}({1987})}]{Anders87}%
  \BibitemOpen
  \bibfield  {author} {\bibinfo {author} {\bibfnamefont {P.~W.}\ \bibnamefont
  {Anderson}},\ }\bibfield  {title} {\enquote {\bibinfo {title} {{The
  resonating valence bond state in La$_2$CuO$_4$ and superconductivity}},}\
  }\href@noop {} {\bibfield  {journal} {\bibinfo  {journal} {{Science}}\
  }\textbf {\bibinfo {volume} {{235}}},\ \bibinfo {pages} {{1196}} (\bibinfo
  {year} {{1987}})}\BibitemShut {NoStop}%
\bibitem [{\citenamefont {Gingras}\ and\ \citenamefont
  {McClarty}(2014)}]{Gingras14}%
  \BibitemOpen
  \bibfield  {author} {\bibinfo {author} {\bibfnamefont {M.~J.~P.}\
  \bibnamefont {Gingras}}\ and\ \bibinfo {author} {\bibfnamefont {P.~A.}\
  \bibnamefont {McClarty}},\ }\bibfield  {title} {\enquote {\bibinfo {title}
  {{Quantum spin ice: a search for gapless quantum spin liquids in pyrochlore
  magnets}},}\ }\href@noop {} {\bibfield  {journal} {\bibinfo  {journal} {Rep.
  Prog. Phys.}\ }\textbf {\bibinfo {volume} {77}},\ \bibinfo {pages} {056501}
  (\bibinfo {year} {2014})}\BibitemShut {NoStop}%
\bibitem [{\citenamefont {Chandra}\ and\ \citenamefont
  {Doucot}(1988)}]{Chandra}%
  \BibitemOpen
  \bibfield  {author} {\bibinfo {author} {\bibfnamefont {P.}~\bibnamefont
  {Chandra}}\ and\ \bibinfo {author} {\bibfnamefont {B.}~\bibnamefont
  {Doucot}},\ }\bibfield  {title} {\enquote {\bibinfo {title} {Possible
  spin-liquid state at large $s$ for the frustrated square heisenberg
  lattice},}\ }\href@noop {} {\bibfield  {journal} {\bibinfo  {journal} {Phys.
  Rev. B}\ }\textbf {\bibinfo {volume} {38}},\ \bibinfo {pages} {9335--9338}
  (\bibinfo {year} {1988})}\BibitemShut {NoStop}%
\bibitem [{\citenamefont {Melzi}\ \emph {et~al.}(2000)\citenamefont {Melzi},
  \citenamefont {Carretta}, \citenamefont {Lascialfari}, \citenamefont
  {Mambrini}, \citenamefont {Troyer}, \citenamefont {Millet},\ and\
  \citenamefont {Mila}}]{Melzi}%
  \BibitemOpen
  \bibfield  {author} {\bibinfo {author} {\bibfnamefont {R.}~\bibnamefont
  {Melzi}}, \bibinfo {author} {\bibfnamefont {P.}~\bibnamefont {Carretta}},
  \bibinfo {author} {\bibfnamefont {A.}~\bibnamefont {Lascialfari}}, \bibinfo
  {author} {\bibfnamefont {M.}~\bibnamefont {Mambrini}}, \bibinfo {author}
  {\bibfnamefont {M.}~\bibnamefont {Troyer}}, \bibinfo {author} {\bibfnamefont
  {P.}~\bibnamefont {Millet}}, \ and\ \bibinfo {author} {\bibfnamefont
  {F.}~\bibnamefont {Mila}},\ }\bibfield  {title} {\enquote {\bibinfo {title}
  {{Li$_2$VO(Si,Ge)O$_4$, a prototype of a two-dimensional frustrated quantum
  Heisenberg antiferromagnet}},}\ }\href@noop {} {\bibfield  {journal}
  {\bibinfo  {journal} {Phys. Rev. Lett.}\ }\textbf {\bibinfo {volume} {85}},\
  \bibinfo {pages} {1318} (\bibinfo {year} {2000})}\BibitemShut {NoStop}%
\bibitem [{\citenamefont {Bombardi}\ \emph {et~al.}(2004)\citenamefont
  {Bombardi}, \citenamefont {Rodriguez-Carvajal}, \citenamefont {Di~Matteo},
  \citenamefont {de~Bergevin}, \citenamefont {Paolasini}, \citenamefont
  {Carretta}, \citenamefont {Millet},\ and\ \citenamefont
  {Caciuffo}}]{Bombardi}%
  \BibitemOpen
  \bibfield  {author} {\bibinfo {author} {\bibfnamefont {A.}~\bibnamefont
  {Bombardi}}, \bibinfo {author} {\bibfnamefont {J.}~\bibnamefont
  {Rodriguez-Carvajal}}, \bibinfo {author} {\bibfnamefont {S.}~\bibnamefont
  {Di~Matteo}}, \bibinfo {author} {\bibfnamefont {F.}~\bibnamefont
  {de~Bergevin}}, \bibinfo {author} {\bibfnamefont {L.}~\bibnamefont
  {Paolasini}}, \bibinfo {author} {\bibfnamefont {P.}~\bibnamefont {Carretta}},
  \bibinfo {author} {\bibfnamefont {P.}~\bibnamefont {Millet}}, \ and\ \bibinfo
  {author} {\bibfnamefont {R.}~\bibnamefont {Caciuffo}},\ }\bibfield  {title}
  {\enquote {\bibinfo {title} {{Direct determination of the magnetic ground
  state in the square lattice $S=1/2$ antiferromagnet Li$_2$VOSIO$_4$}},}\
  }\href@noop {} {\bibfield  {journal} {\bibinfo  {journal} {Phys. Rev. Lett.}\
  }\textbf {\bibinfo {volume} {93}},\ \bibinfo {pages} {027202} (\bibinfo
  {year} {2004})}\BibitemShut {NoStop}%
\bibitem [{\citenamefont {F\aa{}k}\ \emph {et~al.}(2012)\citenamefont
  {F\aa{}k}, \citenamefont {Kermarrec}, \citenamefont {Messio}, \citenamefont
  {Bernu}, \citenamefont {Lhuillier}, \citenamefont {Bert}, \citenamefont
  {Mendels}, \citenamefont {Koteswararao}, \citenamefont {Bouquet},
  \citenamefont {Ollivier}, \citenamefont {Hillier}, \citenamefont {Amato},
  \citenamefont {Colman},\ and\ \citenamefont {Wills}}]{Kapel}%
  \BibitemOpen
  \bibfield  {author} {\bibinfo {author} {\bibfnamefont {B.}~\bibnamefont
  {F\aa{}k}}, \bibinfo {author} {\bibfnamefont {E.}~\bibnamefont {Kermarrec}},
  \bibinfo {author} {\bibfnamefont {L.}~\bibnamefont {Messio}}, \bibinfo
  {author} {\bibfnamefont {B.}~\bibnamefont {Bernu}}, \bibinfo {author}
  {\bibfnamefont {C.}~\bibnamefont {Lhuillier}}, \bibinfo {author}
  {\bibfnamefont {F.}~\bibnamefont {Bert}}, \bibinfo {author} {\bibfnamefont
  {P.}~\bibnamefont {Mendels}}, \bibinfo {author} {\bibfnamefont
  {B.}~\bibnamefont {Koteswararao}}, \bibinfo {author} {\bibfnamefont
  {F.}~\bibnamefont {Bouquet}}, \bibinfo {author} {\bibfnamefont
  {J.}~\bibnamefont {Ollivier}}, \bibinfo {author} {\bibfnamefont {A.~D.}\
  \bibnamefont {Hillier}}, \bibinfo {author} {\bibfnamefont {A.}~\bibnamefont
  {Amato}}, \bibinfo {author} {\bibfnamefont {R.~H.}\ \bibnamefont {Colman}}, \
  and\ \bibinfo {author} {\bibfnamefont {A.~S.}\ \bibnamefont {Wills}},\
  }\bibfield  {title} {\enquote {\bibinfo {title} {Kapellasite: A kagome
  quantum spin liquid with competing interactions},}\ }\href {\doibase
  10.1103/PhysRevLett.109.037208} {\bibfield  {journal} {\bibinfo  {journal}
  {Phys. Rev. Lett.}\ }\textbf {\bibinfo {volume} {109}},\ \bibinfo {pages}
  {037208} (\bibinfo {year} {2012})}\BibitemShut {NoStop}%
\bibitem [{\citenamefont {Reimers}\ \emph {et~al.}(1991)\citenamefont
  {Reimers}, \citenamefont {Berlinsky},\ and\ \citenamefont {Shi}}]{Reimers91}%
  \BibitemOpen
  \bibfield  {author} {\bibinfo {author} {\bibfnamefont {J.~N.}\ \bibnamefont
  {Reimers}}, \bibinfo {author} {\bibfnamefont {A.~J.}\ \bibnamefont
  {Berlinsky}}, \ and\ \bibinfo {author} {\bibfnamefont {A.-C.}\ \bibnamefont
  {Shi}},\ }\bibfield  {title} {\enquote {\bibinfo {title} {Mean-field approach
  to magnetic ordering in highly frustrated pyrochlores},}\ }\href@noop {}
  {\bibfield  {journal} {\bibinfo  {journal} {Phys. Rev. B}\ }\textbf {\bibinfo
  {volume} {43}},\ \bibinfo {pages} {865--878} (\bibinfo {year}
  {1991})}\BibitemShut {NoStop}%
\bibitem [{\citenamefont {Moessner}\ and\ \citenamefont
  {Chalker}(1998)}]{Moessner98}%
  \BibitemOpen
  \bibfield  {author} {\bibinfo {author} {\bibfnamefont {R.}~\bibnamefont
  {Moessner}}\ and\ \bibinfo {author} {\bibfnamefont {J.~T.}\ \bibnamefont
  {Chalker}},\ }\bibfield  {title} {\enquote {\bibinfo {title} {Low-temperature
  properties of classical geometrically frustrated antiferromagnets},}\
  }\href@noop {} {\bibfield  {journal} {\bibinfo  {journal} {Phys. Rev. B}\
  }\textbf {\bibinfo {volume} {58}},\ \bibinfo {pages} {12049--12062} (\bibinfo
  {year} {1998})}\BibitemShut {NoStop}%
\bibitem [{\citenamefont {Gingras}\ and\ \citenamefont {den
  Hertog}({2001})}]{Gingras_CJP}%
  \BibitemOpen
  \bibfield  {author} {\bibinfo {author} {\bibfnamefont {M.~J.~P.}\
  \bibnamefont {Gingras}}\ and\ \bibinfo {author} {\bibfnamefont {B.~C.}\
  \bibnamefont {den Hertog}},\ }\bibfield  {title} {\enquote {\bibinfo {title}
  {{Origin of spin-ice behavior in Ising pyrochlore magnets with long-range
  dipole interactions: an insight from mean-field theory}},}\ }\href@noop {}
  {\bibfield  {journal} {\bibinfo  {journal} {{Canadian Journal of Physics}}\
  }\textbf {\bibinfo {volume} {{79}}},\ \bibinfo {pages} {{1339}} (\bibinfo
  {year} {{2001}})}\BibitemShut {NoStop}%
\bibitem [{\citenamefont {Villain}(1979)}]{Villain79}%
  \BibitemOpen
  \bibfield  {author} {\bibinfo {author} {\bibfnamefont {J.}~\bibnamefont
  {Villain}},\ }\bibfield  {title} {\enquote {\bibinfo {title} {Insulating spin
  glasses},}\ }\href@noop {} {\bibfield  {journal} {\bibinfo  {journal} {Z.
  Phys B}\ }\textbf {\bibinfo {volume} {33}},\ \bibinfo {pages} {31--42}
  (\bibinfo {year} {1979})}\BibitemShut {NoStop}%
\bibitem [{\citenamefont {Sen}\ \emph {et~al.}(2011)\citenamefont {Sen},
  \citenamefont {Damle},\ and\ \citenamefont {Moessner}}]{Sen}%
  \BibitemOpen
  \bibfield  {author} {\bibinfo {author} {\bibfnamefont {A.}~\bibnamefont
  {Sen}}, \bibinfo {author} {\bibfnamefont {K.}~\bibnamefont {Damle}}, \ and\
  \bibinfo {author} {\bibfnamefont {R.}~\bibnamefont {Moessner}},\ }\bibfield
  {title} {\enquote {\bibinfo {title} {{Fractional spin textures in the
  frustrated magnet SrCr$_{9p}$Ga$_{12-9p}$O$_{19}$}},}\ }\href@noop {}
  {\bibfield  {journal} {\bibinfo  {journal} {Phys. Rev. Lett.}\ }\textbf
  {\bibinfo {volume} {106}},\ \bibinfo {pages} {127203} (\bibinfo {year}
  {2011})}\BibitemShut {NoStop}%
\bibitem [{\citenamefont {Gingras}(2011)}]{Gingras11}%
  \BibitemOpen
  \bibfield  {author} {\bibinfo {author} {\bibfnamefont {M.~J.~P.}\
  \bibnamefont {Gingras}},\ }\bibfield  {title} {\enquote {\bibinfo {title}
  {Spin ice},}\ }in\ \href@noop {} {\emph {\bibinfo {booktitle} {Highly
  Frustrated Magnetism}}},\ \bibinfo {series} {Springer Series in Solid-State
  Sciences}, Vol.\ \bibinfo {volume} {164},\ \bibinfo {editor} {edited by\
  \bibinfo {editor} {\bibfnamefont {C}~\bibnamefont {Lacroix}}, \bibinfo
  {editor} {\bibfnamefont {P}~\bibnamefont {Mendels}}, \ and\ \bibinfo {editor}
  {\bibfnamefont {F}~\bibnamefont {Mila}}}\ (\bibinfo  {publisher} {Springer},\
  \bibinfo {year} {2011})\BibitemShut {NoStop}%
\bibitem [{\citenamefont {Bramwell}\ and\ \citenamefont
  {Gingras}(2001)}]{Bramwell_Science}%
  \BibitemOpen
  \bibfield  {author} {\bibinfo {author} {\bibfnamefont {S.~T.}\ \bibnamefont
  {Bramwell}}\ and\ \bibinfo {author} {\bibfnamefont {M.~J.~P.}\ \bibnamefont
  {Gingras}},\ }\bibfield  {title} {\enquote {\bibinfo {title} {Spin ice state
  in frustrated magnetic pyrochlore materials},}\ }\href@noop {} {\bibfield
  {journal} {\bibinfo  {journal} {Science}\ }\textbf {\bibinfo {volume}
  {294}},\ \bibinfo {pages} {1495} (\bibinfo {year} {2001})}\BibitemShut
  {NoStop}%
\bibitem [{\citenamefont {Harris}\ \emph {et~al.}(1997)\citenamefont {Harris},
  \citenamefont {Bramwell}, \citenamefont {McMorrow}, \citenamefont {Zeiske},\
  and\ \citenamefont {Godfrey}}]{Harris}%
  \BibitemOpen
  \bibfield  {author} {\bibinfo {author} {\bibfnamefont {M.~J.}\ \bibnamefont
  {Harris}}, \bibinfo {author} {\bibfnamefont {S.~T.}\ \bibnamefont
  {Bramwell}}, \bibinfo {author} {\bibfnamefont {D.~F.}\ \bibnamefont
  {McMorrow}}, \bibinfo {author} {\bibfnamefont {T.}~\bibnamefont {Zeiske}}, \
  and\ \bibinfo {author} {\bibfnamefont {K.~W.}\ \bibnamefont {Godfrey}},\
  }\bibfield  {title} {\enquote {\bibinfo {title} {{Geometrical Frustration in
  the Ferromagnetic Pyrochlore Ho$_2$Ti$_2$O$_7$}},}\ }\href@noop {} {\bibfield
   {journal} {\bibinfo  {journal} {Phys. Rev. Lett.}\ }\textbf {\bibinfo
  {volume} {79}},\ \bibinfo {pages} {2554} (\bibinfo {year}
  {1997})}\BibitemShut {NoStop}%
\bibitem [{\citenamefont {Ramirez}\ \emph {et~al.}(1999)\citenamefont
  {Ramirez}, \citenamefont {Hayashi}, \citenamefont {Cava}, \citenamefont
  {Siddharthan},\ and\ \citenamefont {Shastry}}]{Ramirez}%
  \BibitemOpen
  \bibfield  {author} {\bibinfo {author} {\bibfnamefont {A.~P.}\ \bibnamefont
  {Ramirez}}, \bibinfo {author} {\bibfnamefont {A.}~\bibnamefont {Hayashi}},
  \bibinfo {author} {\bibfnamefont {R.~J.}\ \bibnamefont {Cava}}, \bibinfo
  {author} {\bibfnamefont {R.}~\bibnamefont {Siddharthan}}, \ and\ \bibinfo
  {author} {\bibfnamefont {B.~S.}\ \bibnamefont {Shastry}},\ }\bibfield
  {title} {\enquote {\bibinfo {title} {Zero-point entropy in `spin ice'},}\
  }\href@noop {} {\bibfield  {journal} {\bibinfo  {journal} {Nature}\ }\textbf
  {\bibinfo {volume} {399}},\ \bibinfo {pages} {333} (\bibinfo {year}
  {1999})}\BibitemShut {NoStop}%
\bibitem [{\citenamefont {Siddharthan}\ \emph {et~al.}(1999)\citenamefont
  {Siddharthan}, \citenamefont {Shastry}, \citenamefont {Ramirez},
  \citenamefont {Hayashi}, \citenamefont {Cava},\ and\ \citenamefont
  {Rosenkranz}}]{Siddharthan}%
  \BibitemOpen
  \bibfield  {author} {\bibinfo {author} {\bibfnamefont {R.}~\bibnamefont
  {Siddharthan}}, \bibinfo {author} {\bibfnamefont {B.~S.}\ \bibnamefont
  {Shastry}}, \bibinfo {author} {\bibfnamefont {A.~P.}\ \bibnamefont
  {Ramirez}}, \bibinfo {author} {\bibfnamefont {A.}~\bibnamefont {Hayashi}},
  \bibinfo {author} {\bibfnamefont {R.~J.}\ \bibnamefont {Cava}}, \ and\
  \bibinfo {author} {\bibfnamefont {S.}~\bibnamefont {Rosenkranz}},\ }\bibfield
   {title} {\enquote {\bibinfo {title} {Ising pyrochlore magnets:
  Low-temperature properties, ``ice rules,'' and beyond},}\ }\href@noop {}
  {\bibfield  {journal} {\bibinfo  {journal} {Phys. Rev. Lett.}\ }\textbf
  {\bibinfo {volume} {83}},\ \bibinfo {pages} {1854} (\bibinfo {year}
  {1999})}\BibitemShut {NoStop}%
\bibitem [{\citenamefont {den Hertog}\ and\ \citenamefont
  {Gingras}(2000)}]{denHertog}%
  \BibitemOpen
  \bibfield  {author} {\bibinfo {author} {\bibfnamefont {B.~C.}\ \bibnamefont
  {den Hertog}}\ and\ \bibinfo {author} {\bibfnamefont {M.~J.~P.}\ \bibnamefont
  {Gingras}},\ }\bibfield  {title} {\enquote {\bibinfo {title} {Dipolar
  interactions and origin of spin ice in ising pyrochlore magnets},}\
  }\href@noop {} {\bibfield  {journal} {\bibinfo  {journal} {Phys. Rev. Lett.}\
  }\textbf {\bibinfo {volume} {84}},\ \bibinfo {pages} {3430} (\bibinfo {year}
  {2000})}\BibitemShut {NoStop}%
\bibitem [{\citenamefont {Isakov}\ \emph {et~al.}(2005)\citenamefont {Isakov},
  \citenamefont {Moessner},\ and\ \citenamefont {Sondhi}}]{Isakov_SS}%
  \BibitemOpen
  \bibfield  {author} {\bibinfo {author} {\bibfnamefont {S.~V.}\ \bibnamefont
  {Isakov}}, \bibinfo {author} {\bibfnamefont {R.}~\bibnamefont {Moessner}}, \
  and\ \bibinfo {author} {\bibfnamefont {S.~L.}\ \bibnamefont {Sondhi}},\
  }\bibfield  {title} {\enquote {\bibinfo {title} {Why spin ice obeys the ice
  rules},}\ }\href@noop {} {\bibfield  {journal} {\bibinfo  {journal} {Phys.
  Rev. Lett.}\ }\textbf {\bibinfo {volume} {95}},\ \bibinfo {pages} {217201}
  (\bibinfo {year} {2005})}\BibitemShut {NoStop}%
\bibitem [{\citenamefont {Melko}\ \emph {et~al.}(2001)\citenamefont {Melko},
  \citenamefont {den Hertog},\ and\ \citenamefont {Gingras}}]{Melko01}%
  \BibitemOpen
  \bibfield  {author} {\bibinfo {author} {\bibfnamefont {R.~G.}\ \bibnamefont
  {Melko}}, \bibinfo {author} {\bibfnamefont {B.~C.}\ \bibnamefont {den
  Hertog}}, \ and\ \bibinfo {author} {\bibfnamefont {M.~J.~P.}\ \bibnamefont
  {Gingras}},\ }\bibfield  {title} {\enquote {\bibinfo {title} {{Long-range
  order at low temperatures in dipolar spin ice}},}\ }\href@noop {} {\bibfield
  {journal} {\bibinfo  {journal} {Phys. Rev. Lett.}\ }\textbf {\bibinfo
  {volume} {87}},\ \bibinfo {pages} {067203} (\bibinfo {year}
  {2001})}\BibitemShut {NoStop}%
\bibitem [{\citenamefont {Melko}\ and\ \citenamefont
  {Gingras}(2004)}]{Melko04}%
  \BibitemOpen
  \bibfield  {author} {\bibinfo {author} {\bibfnamefont {R.~G.}\ \bibnamefont
  {Melko}}\ and\ \bibinfo {author} {\bibfnamefont {M.~J.~P.}\ \bibnamefont
  {Gingras}},\ }\bibfield  {title} {\enquote {\bibinfo {title} {{Monte Carlo
  studies of the dipolar spin ice model}},}\ }\href@noop {} {\bibfield
  {journal} {\bibinfo  {journal} {J. Phys.: Condens. Matter}\ }\textbf
  {\bibinfo {volume} {16}},\ \bibinfo {pages} {R1277} (\bibinfo {year}
  {2004})}\BibitemShut {NoStop}%
\bibitem [{\citenamefont {Harris}\ \emph {et~al.}(1998)\citenamefont {Harris},
  \citenamefont {Bramwell}, \citenamefont {Zeiske}, \citenamefont {McMorrow},\
  and\ \citenamefont {King}}]{Harris98}%
  \BibitemOpen
  \bibfield  {author} {\bibinfo {author} {\bibfnamefont {M.~J.}\ \bibnamefont
  {Harris}}, \bibinfo {author} {\bibfnamefont {S.~T.}\ \bibnamefont
  {Bramwell}}, \bibinfo {author} {\bibfnamefont {T.}~\bibnamefont {Zeiske}},
  \bibinfo {author} {\bibfnamefont {D.~F.}\ \bibnamefont {McMorrow}}, \ and\
  \bibinfo {author} {\bibfnamefont {P.~J.~C.}\ \bibnamefont {King}},\
  }\bibfield  {title} {\enquote {\bibinfo {title} {Magnetic structures of
  highly frustrated pyrochlores},}\ }\href@noop {} {\bibfield  {journal}
  {\bibinfo  {journal} {J. Magn. Magn. Mater.}\ }\textbf {\bibinfo {volume}
  {177/181}},\ \bibinfo {pages} {757--762} (\bibinfo {year}
  {1998})}\BibitemShut {NoStop}%
\bibitem [{\citenamefont {Fukazawa}\ \emph {et~al.}(2002)\citenamefont
  {Fukazawa}, \citenamefont {Melko}, \citenamefont {Higashinaka}, \citenamefont
  {Maeno},\ and\ \citenamefont {Gingras}}]{Fuka02}%
  \BibitemOpen
  \bibfield  {author} {\bibinfo {author} {\bibfnamefont {H.}~\bibnamefont
  {Fukazawa}}, \bibinfo {author} {\bibfnamefont {R.~G.}\ \bibnamefont {Melko}},
  \bibinfo {author} {\bibfnamefont {R.}~\bibnamefont {Higashinaka}}, \bibinfo
  {author} {\bibfnamefont {Y.}~\bibnamefont {Maeno}}, \ and\ \bibinfo {author}
  {\bibfnamefont {M.~J.~P.}\ \bibnamefont {Gingras}},\ }\bibfield  {title}
  {\enquote {\bibinfo {title} {{Magnetic anisotropy of the spin-ice compound
  Dy$_2$Ti$_2$O$_7$}},}\ }\href@noop {} {\bibfield  {journal} {\bibinfo
  {journal} {Phys. Rev. B}\ }\textbf {\bibinfo {volume} {65}},\ \bibinfo
  {pages} {054410} (\bibinfo {year} {2002})}\BibitemShut {NoStop}%
\bibitem [{\citenamefont {Castelnovo}\ \emph {et~al.}({2008})\citenamefont
  {Castelnovo}, \citenamefont {Moessner},\ and\ \citenamefont
  {Sondhi}}]{Castelnovo_Nature}%
  \BibitemOpen
  \bibfield  {author} {\bibinfo {author} {\bibfnamefont {C.}~\bibnamefont
  {Castelnovo}}, \bibinfo {author} {\bibfnamefont {R.}~\bibnamefont
  {Moessner}}, \ and\ \bibinfo {author} {\bibfnamefont {S.~L.}\ \bibnamefont
  {Sondhi}},\ }\bibfield  {title} {\enquote {\bibinfo {title} {{Magnetic
  monopoles in spin ice}},}\ }\href {\doibase {10.1038/nature06433}} {\bibfield
   {journal} {\bibinfo  {journal} {{Nature}}\ }\textbf {\bibinfo {volume}
  {{451}}},\ \bibinfo {pages} {{42--45}} (\bibinfo {year}
  {{2008}})}\BibitemShut {NoStop}%
\bibitem [{\citenamefont {Castelnovo}\ \emph {et~al.}(2010)\citenamefont
  {Castelnovo}, \citenamefont {Moessner},\ and\ \citenamefont
  {Sondhi}}]{Castel10}%
  \BibitemOpen
  \bibfield  {author} {\bibinfo {author} {\bibfnamefont {C.}~\bibnamefont
  {Castelnovo}}, \bibinfo {author} {\bibfnamefont {R.}~\bibnamefont
  {Moessner}}, \ and\ \bibinfo {author} {\bibfnamefont {S.~L.}\ \bibnamefont
  {Sondhi}},\ }\bibfield  {title} {\enquote {\bibinfo {title} {Thermal quenches
  in spin ice},}\ }\href@noop {} {\bibfield  {journal} {\bibinfo  {journal}
  {Phys. Rev. Lett.}\ }\textbf {\bibinfo {volume} {104}},\ \bibinfo {pages}
  {107201} (\bibinfo {year} {2010})}\BibitemShut {NoStop}%
\bibitem [{\citenamefont {Paulsen}\ \emph {et~al.}(2014)\citenamefont
  {Paulsen}, \citenamefont {Jackson}, \citenamefont {Lhotel}, \citenamefont
  {Canals}, \citenamefont {Prabhakaran}, \citenamefont {Matsuhira},
  \citenamefont {Giblin},\ and\ \citenamefont {Bramwell}}]{Paulsen14}%
  \BibitemOpen
  \bibfield  {author} {\bibinfo {author} {\bibfnamefont {C.}~\bibnamefont
  {Paulsen}}, \bibinfo {author} {\bibfnamefont {M.~J.}\ \bibnamefont
  {Jackson}}, \bibinfo {author} {\bibfnamefont {E.}~\bibnamefont {Lhotel}},
  \bibinfo {author} {\bibfnamefont {B.}~\bibnamefont {Canals}}, \bibinfo
  {author} {\bibfnamefont {D.}~\bibnamefont {Prabhakaran}}, \bibinfo {author}
  {\bibfnamefont {K.}~\bibnamefont {Matsuhira}}, \bibinfo {author}
  {\bibfnamefont {S.~R.}\ \bibnamefont {Giblin}}, \ and\ \bibinfo {author}
  {\bibfnamefont {S.~T.}\ \bibnamefont {Bramwell}},\ }\bibfield  {title}
  {\enquote {\bibinfo {title} {Far-from-equilibrium monopole dynamics in spin
  ice},}\ }\href@noop {} {\bibfield  {journal} {\bibinfo  {journal} {Nature
  Physics}\ }\textbf {\bibinfo {volume} {10}},\ \bibinfo {pages} {135}
  (\bibinfo {year} {2014})}\BibitemShut {NoStop}%
\bibitem [{\citenamefont {Revell}\ \emph {et~al.}({2013})\citenamefont
  {Revell}, \citenamefont {Yaraskavitch}, \citenamefont {Mason}, \citenamefont
  {Ross}, \citenamefont {Noad}, \citenamefont {Dabkowska}, \citenamefont
  {Gaulin}, \citenamefont {Henelius},\ and\ \citenamefont {Kycia}}]{Revell13}%
  \BibitemOpen
  \bibfield  {author} {\bibinfo {author} {\bibfnamefont {H.~M.}\ \bibnamefont
  {Revell}}, \bibinfo {author} {\bibfnamefont {L.~R.}\ \bibnamefont
  {Yaraskavitch}}, \bibinfo {author} {\bibfnamefont {J.~D.}\ \bibnamefont
  {Mason}}, \bibinfo {author} {\bibfnamefont {K.~A.}\ \bibnamefont {Ross}},
  \bibinfo {author} {\bibfnamefont {H.~M.~L.}\ \bibnamefont {Noad}}, \bibinfo
  {author} {\bibfnamefont {H.~A.}\ \bibnamefont {Dabkowska}}, \bibinfo {author}
  {\bibfnamefont {B.~D.}\ \bibnamefont {Gaulin}}, \bibinfo {author}
  {\bibfnamefont {P.}~\bibnamefont {Henelius}}, \ and\ \bibinfo {author}
  {\bibfnamefont {J.~B.}\ \bibnamefont {Kycia}},\ }\bibfield  {title} {\enquote
  {\bibinfo {title} {{Evidence of impurity and boundary effects on magnetic
  monopole dynamics in spin ice}},}\ }\href@noop {} {\bibfield  {journal}
  {\bibinfo  {journal} {{Nature Physics}}\ }\textbf {\bibinfo {volume} {{9}}},\
  \bibinfo {pages} {{34}} (\bibinfo {year} {{2013}})}\BibitemShut {NoStop}%
\bibitem [{\citenamefont {Pomaranski}\ \emph {et~al.}({2013})\citenamefont
  {Pomaranski}, \citenamefont {Yaraskavitch}, \citenamefont {Meng},
  \citenamefont {Ross}, \citenamefont {Noad}, \citenamefont {Dabkowska},
  \citenamefont {Gaulin},\ and\ \citenamefont {Kycia}}]{Poma13}%
  \BibitemOpen
  \bibfield  {author} {\bibinfo {author} {\bibfnamefont {D.}~\bibnamefont
  {Pomaranski}}, \bibinfo {author} {\bibfnamefont {L.~R.}\ \bibnamefont
  {Yaraskavitch}}, \bibinfo {author} {\bibfnamefont {S.}~\bibnamefont {Meng}},
  \bibinfo {author} {\bibfnamefont {K.~A.}\ \bibnamefont {Ross}}, \bibinfo
  {author} {\bibfnamefont {H.~M.~L.}\ \bibnamefont {Noad}}, \bibinfo {author}
  {\bibfnamefont {H.~A.}\ \bibnamefont {Dabkowska}}, \bibinfo {author}
  {\bibfnamefont {B.~D.}\ \bibnamefont {Gaulin}}, \ and\ \bibinfo {author}
  {\bibfnamefont {J.~B.}\ \bibnamefont {Kycia}},\ }\bibfield  {title} {\enquote
  {\bibinfo {title} {{Absence of Pauling's residual entropy in thermally
  equilibrated Dy$_2$Ti$_2$O$_7$}},}\ }\href@noop {} {\bibfield  {journal}
  {\bibinfo  {journal} {{Nature Physics}}\ }\textbf {\bibinfo {volume} {{9}}},\
  \bibinfo {pages} {{353}} (\bibinfo {year} {{2013}})}\BibitemShut {NoStop}%
\bibitem [{\citenamefont {Sala}\ \emph {et~al.}({2014})\citenamefont {Sala},
  \citenamefont {Gutmann}, \citenamefont {Prabhakaran}, \citenamefont
  {Pomaranski}, \citenamefont {Mitchelitis}, \citenamefont {Kycia},
  \citenamefont {Porter}, \citenamefont {Castelnovo},\ and\ \citenamefont
  {Goff}}]{Sala14}%
  \BibitemOpen
  \bibfield  {author} {\bibinfo {author} {\bibfnamefont {G.}~\bibnamefont
  {Sala}}, \bibinfo {author} {\bibfnamefont {M.~J.}\ \bibnamefont {Gutmann}},
  \bibinfo {author} {\bibfnamefont {D.}~\bibnamefont {Prabhakaran}}, \bibinfo
  {author} {\bibfnamefont {D.}~\bibnamefont {Pomaranski}}, \bibinfo {author}
  {\bibfnamefont {C.}~\bibnamefont {Mitchelitis}}, \bibinfo {author}
  {\bibfnamefont {J.~B.}\ \bibnamefont {Kycia}}, \bibinfo {author}
  {\bibfnamefont {D.~G.}\ \bibnamefont {Porter}}, \bibinfo {author}
  {\bibfnamefont {C.}~\bibnamefont {Castelnovo}}, \ and\ \bibinfo {author}
  {\bibfnamefont {J.~P.}\ \bibnamefont {Goff}},\ }\bibfield  {title} {\enquote
  {\bibinfo {title} {{Vacancy defects and monopole dynamics in oxygen-deficient
  pyrochlores}},}\ }\href@noop {} {\bibfield  {journal} {\bibinfo  {journal}
  {{Nature Materials}}\ }\textbf {\bibinfo {volume} {{13}}},\ \bibinfo {pages}
  {{488}} (\bibinfo {year} {{2014}})}\BibitemShut {NoStop}%
\bibitem [{\citenamefont {Zhou}\ \emph {et~al.}(2012)\citenamefont {Zhou},
  \citenamefont {Cheng}, \citenamefont {Hallas}, \citenamefont {Wiebe},
  \citenamefont {Li}, \citenamefont {Balicas}, \citenamefont {Zhou},
  \citenamefont {Goodenough}, \citenamefont {Gardner},\ and\ \citenamefont
  {Choi}}]{Zhou_chem_press}%
  \BibitemOpen
  \bibfield  {author} {\bibinfo {author} {\bibfnamefont {H.~D.}\ \bibnamefont
  {Zhou}}, \bibinfo {author} {\bibfnamefont {J.~G.}\ \bibnamefont {Cheng}},
  \bibinfo {author} {\bibfnamefont {A.~M.}\ \bibnamefont {Hallas}}, \bibinfo
  {author} {\bibfnamefont {C.~R.}\ \bibnamefont {Wiebe}}, \bibinfo {author}
  {\bibfnamefont {G.}~\bibnamefont {Li}}, \bibinfo {author} {\bibfnamefont
  {L.}~\bibnamefont {Balicas}}, \bibinfo {author} {\bibfnamefont {J.~S.}\
  \bibnamefont {Zhou}}, \bibinfo {author} {\bibfnamefont {J.~B.}\ \bibnamefont
  {Goodenough}}, \bibinfo {author} {\bibfnamefont {J.~S.}\ \bibnamefont
  {Gardner}}, \ and\ \bibinfo {author} {\bibfnamefont {E.~S.}\ \bibnamefont
  {Choi}},\ }\bibfield  {title} {\enquote {\bibinfo {title} {{Chemical pressure
  effects on pyrochlore spin ice}},}\ }\href@noop {} {\bibfield  {journal}
  {\bibinfo  {journal} {Phys. Rev. Lett.}\ }\textbf {\bibinfo {volume} {108}},\
  \bibinfo {pages} {207206} (\bibinfo {year} {2012})}\BibitemShut {NoStop}%
\bibitem [{\citenamefont {Klemke}\ \emph {et~al.}({2011})\citenamefont
  {Klemke}, \citenamefont {Meissner}, \citenamefont {Strehlow}, \citenamefont
  {Kiefer}, \citenamefont {Grigera},\ and\ \citenamefont {Tennant}}]{Klemke11}%
  \BibitemOpen
  \bibfield  {author} {\bibinfo {author} {\bibfnamefont {B.}~\bibnamefont
  {Klemke}}, \bibinfo {author} {\bibfnamefont {M.}~\bibnamefont {Meissner}},
  \bibinfo {author} {\bibfnamefont {P.}~\bibnamefont {Strehlow}}, \bibinfo
  {author} {\bibfnamefont {K.}~\bibnamefont {Kiefer}}, \bibinfo {author}
  {\bibfnamefont {S.~A.}\ \bibnamefont {Grigera}}, \ and\ \bibinfo {author}
  {\bibfnamefont {D.~A.}\ \bibnamefont {Tennant}},\ }\bibfield  {title}
  {\enquote {\bibinfo {title} {{Thermal relaxation and heat transport in the
  spin ice material Dy$_2$Ti$_2$O$_7$}},}\ }\href@noop {} {\bibfield  {journal}
  {\bibinfo  {journal} {{Journal of Low Temperature Physics}}\ }\textbf
  {\bibinfo {volume} {{163}}},\ \bibinfo {pages} {{345}} (\bibinfo {year}
  {{2011}})}\BibitemShut {NoStop}%
\bibitem [{\citenamefont {Higashinaka}\ \emph {et~al.}({2002})\citenamefont
  {Higashinaka}, \citenamefont {Fukazawa}, \citenamefont {Yanagishima},\ and\
  \citenamefont {Maeno}}]{Higa02}%
  \BibitemOpen
  \bibfield  {author} {\bibinfo {author} {\bibfnamefont {R.}~\bibnamefont
  {Higashinaka}}, \bibinfo {author} {\bibfnamefont {H.}~\bibnamefont
  {Fukazawa}}, \bibinfo {author} {\bibfnamefont {D.}~\bibnamefont
  {Yanagishima}}, \ and\ \bibinfo {author} {\bibfnamefont {Y.}~\bibnamefont
  {Maeno}},\ }\bibfield  {title} {\enquote {\bibinfo {title} {{Specific heat of
  Dy$_2$Ti$_2$O$_7$ in magnetic fields: comparison between single-crystalline
  and polycrystalline}},}\ }\href@noop {} {\bibfield  {journal} {\bibinfo
  {journal} {{Journal of Physics and Chemistry of Solids}}\ }\textbf {\bibinfo
  {volume} {{63}}},\ \bibinfo {pages} {{1043}} (\bibinfo {year}
  {{2002}})}\BibitemShut {NoStop}%
\bibitem [{\citenamefont {Tabata}\ \emph {et~al.}(2006)\citenamefont {Tabata},
  \citenamefont {Kadowaki}, \citenamefont {Matsuhira}, \citenamefont {Hiroi},
  \citenamefont {Aso}, \citenamefont {Ressouche},\ and\ \citenamefont
  {F\aa{}k}}]{Taba06}%
  \BibitemOpen
  \bibfield  {author} {\bibinfo {author} {\bibfnamefont {Y.}~\bibnamefont
  {Tabata}}, \bibinfo {author} {\bibfnamefont {H.}~\bibnamefont {Kadowaki}},
  \bibinfo {author} {\bibfnamefont {K.}~\bibnamefont {Matsuhira}}, \bibinfo
  {author} {\bibfnamefont {Z.}~\bibnamefont {Hiroi}}, \bibinfo {author}
  {\bibfnamefont {N.}~\bibnamefont {Aso}}, \bibinfo {author} {\bibfnamefont
  {E.}~\bibnamefont {Ressouche}}, \ and\ \bibinfo {author} {\bibfnamefont
  {B.}~\bibnamefont {F\aa{}k}},\ }\bibfield  {title} {\enquote {\bibinfo
  {title} {{Kagome ice state in the dipolar spin ice Dy$_2$Ti$_2$O$_7$}},}\
  }\href@noop {} {\bibfield  {journal} {\bibinfo  {journal} {Phys. Rev. Lett.}\
  }\textbf {\bibinfo {volume} {97}},\ \bibinfo {pages} {257205} (\bibinfo
  {year} {2006})}\BibitemShut {NoStop}%
\bibitem [{\citenamefont {Yavors'kii}\ \emph {et~al.}(2008)\citenamefont
  {Yavors'kii}, \citenamefont {Fennell}, \citenamefont {Gingras},\ and\
  \citenamefont {Bramwell}}]{Yavo08}%
  \BibitemOpen
  \bibfield  {author} {\bibinfo {author} {\bibfnamefont {T.}~\bibnamefont
  {Yavors'kii}}, \bibinfo {author} {\bibfnamefont {T.}~\bibnamefont {Fennell}},
  \bibinfo {author} {\bibfnamefont {M.~J.~P.}\ \bibnamefont {Gingras}}, \ and\
  \bibinfo {author} {\bibfnamefont {S.~T.}\ \bibnamefont {Bramwell}},\
  }\bibfield  {title} {\enquote {\bibinfo {title} {{Dy$_2$Ti$_2$O$_7$ spin ice:
  a test case for emergent clusters in a frustrated magnet}},}\ }\href@noop {}
  {\bibfield  {journal} {\bibinfo  {journal} {Phys. Rev. Lett.}\ }\textbf
  {\bibinfo {volume} {101}},\ \bibinfo {pages} {037204} (\bibinfo {year}
  {2008})}\BibitemShut {NoStop}%
\bibitem [{\citenamefont {Ruff}\ \emph {et~al.}(2005)\citenamefont {Ruff},
  \citenamefont {Melko},\ and\ \citenamefont {Gingras}}]{Ruff}%
  \BibitemOpen
  \bibfield  {author} {\bibinfo {author} {\bibfnamefont {J.~P.~C.}\
  \bibnamefont {Ruff}}, \bibinfo {author} {\bibfnamefont {R.~G.}\ \bibnamefont
  {Melko}}, \ and\ \bibinfo {author} {\bibfnamefont {M.~J.~P.}\ \bibnamefont
  {Gingras}},\ }\bibfield  {title} {\enquote {\bibinfo {title}
  {Finite-temperature transitions in dipolar spin ice in a large magnetic
  field},}\ }\href@noop {} {\bibfield  {journal} {\bibinfo  {journal} {Phys.
  Rev. Lett.}\ }\textbf {\bibinfo {volume} {95}},\ \bibinfo {pages} {097202}
  (\bibinfo {year} {2005})}\BibitemShut {NoStop}%
\bibitem [{\citenamefont {Higashinaka}\ and\ \citenamefont
  {Maeno}(2005)}]{Higa_112}%
  \BibitemOpen
  \bibfield  {author} {\bibinfo {author} {\bibfnamefont {R.}~\bibnamefont
  {Higashinaka}}\ and\ \bibinfo {author} {\bibfnamefont {Y.}~\bibnamefont
  {Maeno}},\ }\bibfield  {title} {\enquote {\bibinfo {title} {{Field-induced
  transition on a triangular plane in the spin-ice compound
  Dy$_2$Ti$_2$O$_7$}},}\ }\href@noop {} {\bibfield  {journal} {\bibinfo
  {journal} {Phys. Rev. Lett.}\ }\textbf {\bibinfo {volume} {95}},\ \bibinfo
  {pages} {237208} (\bibinfo {year} {2005})}\BibitemShut {NoStop}%
\bibitem [{\citenamefont {Sato}\ \emph {et~al.}({2006})\citenamefont {Sato},
  \citenamefont {Matsuhira}, \citenamefont {Tayama}, \citenamefont {Hiroi},
  \citenamefont {Takagi},\ and\ \citenamefont {Sakakibara}}]{Sato06}%
  \BibitemOpen
  \bibfield  {author} {\bibinfo {author} {\bibfnamefont {H.}~\bibnamefont
  {Sato}}, \bibinfo {author} {\bibfnamefont {K.}~\bibnamefont {Matsuhira}},
  \bibinfo {author} {\bibfnamefont {T.}~\bibnamefont {Tayama}}, \bibinfo
  {author} {\bibfnamefont {Z.}~\bibnamefont {Hiroi}}, \bibinfo {author}
  {\bibfnamefont {S.}~\bibnamefont {Takagi}}, \ and\ \bibinfo {author}
  {\bibfnamefont {T.}~\bibnamefont {Sakakibara}},\ }\bibfield  {title}
  {\enquote {\bibinfo {title} {{Ferromagnetic ordering on the triangular
  lattice in the pyrochlore spin-ice compound Dy$_2$Ti$_2$O$_7$}},}\
  }\href@noop {} {\bibfield  {journal} {\bibinfo  {journal} {{Journal of
  Physics - Condensed Matter}}\ }\textbf {\bibinfo {volume} {{18}}},\ \bibinfo
  {pages} {{L297}} (\bibinfo {year} {{2006}})}\BibitemShut {NoStop}%
\bibitem [{\citenamefont {Fennell}\ \emph {et~al.}(2004)\citenamefont
  {Fennell}, \citenamefont {Petrenko}, \citenamefont {F\aa{}k}, \citenamefont
  {Bramwell}, \citenamefont {Enjalran}, \citenamefont {Yavors'kii},
  \citenamefont {Gingras}, \citenamefont {Melko},\ and\ \citenamefont
  {Balakrishnan}}]{Fennell04}%
  \BibitemOpen
  \bibfield  {author} {\bibinfo {author} {\bibfnamefont {T.}~\bibnamefont
  {Fennell}}, \bibinfo {author} {\bibfnamefont {O.~A.}\ \bibnamefont
  {Petrenko}}, \bibinfo {author} {\bibfnamefont {B.}~\bibnamefont {F\aa{}k}},
  \bibinfo {author} {\bibfnamefont {S.~T.}\ \bibnamefont {Bramwell}}, \bibinfo
  {author} {\bibfnamefont {M.}~\bibnamefont {Enjalran}}, \bibinfo {author}
  {\bibfnamefont {T.}~\bibnamefont {Yavors'kii}}, \bibinfo {author}
  {\bibfnamefont {M.~J.~P.}\ \bibnamefont {Gingras}}, \bibinfo {author}
  {\bibfnamefont {R.~G.}\ \bibnamefont {Melko}}, \ and\ \bibinfo {author}
  {\bibfnamefont {G.}~\bibnamefont {Balakrishnan}},\ }\bibfield  {title}
  {\enquote {\bibinfo {title} {{Neutron scattering investigation of the spin
  ice state in Dy$_2$Ti$_2$O$_7$}},}\ }\href@noop {} {\bibfield  {journal}
  {\bibinfo  {journal} {Phys. Rev. B}\ }\textbf {\bibinfo {volume} {70}},\
  \bibinfo {pages} {134408} (\bibinfo {year} {2004})}\BibitemShut {NoStop}%
\bibitem [{Note1()}]{Note1}%
  \BibitemOpen
  \bibinfo {note} {By ``refrustrate'' we mean that exchange interactions beyond
  nearest-neighbor compete with the meager tendency of the long-range part of
  the dipolar interactions to lift the degeneracy of the ice-rule states\cite
  {Gingras_CJP,Isakov_SS} with an associated transition at a temperature
  $T_c\sim 0.13D$\cite {Melko01,Melko04} and further depress the ordering
  transition to an even lower temperature.}\BibitemShut {Stop}%
\bibitem [{\citenamefont {McClarty}\ \emph {et~al.}(2015)\citenamefont
  {McClarty}, \citenamefont {Sikora}, \citenamefont {Moessner}, \citenamefont
  {Penc}, \citenamefont {Pollmann},\ and\ \citenamefont {Shannon}}]{McClarty}%
  \BibitemOpen
  \bibfield  {author} {\bibinfo {author} {\bibfnamefont {P.~A.}\ \bibnamefont
  {McClarty}}, \bibinfo {author} {\bibfnamefont {O.}~\bibnamefont {Sikora}},
  \bibinfo {author} {\bibfnamefont {R.}~\bibnamefont {Moessner}}, \bibinfo
  {author} {\bibfnamefont {K.}~\bibnamefont {Penc}}, \bibinfo {author}
  {\bibfnamefont {F.}~\bibnamefont {Pollmann}}, \ and\ \bibinfo {author}
  {\bibfnamefont {N.}~\bibnamefont {Shannon}},\ }\bibfield  {title} {\enquote
  {\bibinfo {title} {Chain-based order and quantum spin liquids in dipolar spin
  ice},}\ }\href@noop {} {\bibfield  {journal} {\bibinfo  {journal} {Phys. Rev.
  B}\ }\textbf {\bibinfo {volume} {92}},\ \bibinfo {pages} {094418} (\bibinfo
  {year} {2015})}\BibitemShut {NoStop}%
\bibitem [{\citenamefont {Ross}\ \emph {et~al.}(2012)\citenamefont {Ross},
  \citenamefont {Proffen}, \citenamefont {Dabkowska}, \citenamefont {Quilliam},
  \citenamefont {Yaraskavitch}, \citenamefont {Kycia},\ and\ \citenamefont
  {Gaulin}}]{Ross12}%
  \BibitemOpen
  \bibfield  {author} {\bibinfo {author} {\bibfnamefont {K.~A.}\ \bibnamefont
  {Ross}}, \bibinfo {author} {\bibfnamefont {Th.}\ \bibnamefont {Proffen}},
  \bibinfo {author} {\bibfnamefont {H.~A.}\ \bibnamefont {Dabkowska}}, \bibinfo
  {author} {\bibfnamefont {J.~A.}\ \bibnamefont {Quilliam}}, \bibinfo {author}
  {\bibfnamefont {L.~R.}\ \bibnamefont {Yaraskavitch}}, \bibinfo {author}
  {\bibfnamefont {J.~B.}\ \bibnamefont {Kycia}}, \ and\ \bibinfo {author}
  {\bibfnamefont {B.~D.}\ \bibnamefont {Gaulin}},\ }\bibfield  {title}
  {\enquote {\bibinfo {title} {{Lightly stuffed pyrochlore structure of
  single-crystalline Yb$_2$Ti$_2$O$_7$ grown by the optical floating zone
  technique}},}\ }\href@noop {} {\bibfield  {journal} {\bibinfo  {journal}
  {Phys. Rev. B}\ }\textbf {\bibinfo {volume} {86}},\ \bibinfo {pages} {174424}
  (\bibinfo {year} {2012})}\BibitemShut {NoStop}%
\bibitem [{\citenamefont {Baroudi}\ \emph {et~al.}(2015)\citenamefont
  {Baroudi}, \citenamefont {Gaulin}, \citenamefont {Lapidus}, \citenamefont
  {Gaudet},\ and\ \citenamefont {Cava}}]{Cava15}%
  \BibitemOpen
  \bibfield  {author} {\bibinfo {author} {\bibfnamefont {Kristen}\ \bibnamefont
  {Baroudi}}, \bibinfo {author} {\bibfnamefont {Bruce~D.}\ \bibnamefont
  {Gaulin}}, \bibinfo {author} {\bibfnamefont {Saul~H.}\ \bibnamefont
  {Lapidus}}, \bibinfo {author} {\bibfnamefont {Jonathan}\ \bibnamefont
  {Gaudet}}, \ and\ \bibinfo {author} {\bibfnamefont {R.~J.}\ \bibnamefont
  {Cava}},\ }\bibfield  {title} {\enquote {\bibinfo {title} {Symmetry and light
  stuffing of
  $\mathrm{H}{\mathrm{o}}_{2}\mathrm{T}{\mathrm{i}}_{2}{\mathrm{o}}_{7},
  \mathrm{E}{\mathrm{r}}_{2}\mathrm{T}{\mathrm{i}}_{2}{\mathrm{o}}_{7}$, and
  $\mathrm{Y}{\mathrm{b}}_{2}\mathrm{T}{\mathrm{i}}_{2}{\mathrm{o}}_{7}$
  characterized by synchrotron x-ray diffraction},}\ }\href@noop {} {\bibfield
  {journal} {\bibinfo  {journal} {Phys. Rev. B}\ }\textbf {\bibinfo {volume}
  {92}},\ \bibinfo {pages} {024110} (\bibinfo {year} {2015})}\BibitemShut
  {NoStop}%
\bibitem [{\citenamefont {Hao}\ \emph {et~al.}(2014)\citenamefont {Hao},
  \citenamefont {Day},\ and\ \citenamefont {Gingras}}]{Hao_hopping}%
  \BibitemOpen
  \bibfield  {author} {\bibinfo {author} {\bibfnamefont {Z.}~\bibnamefont
  {Hao}}, \bibinfo {author} {\bibfnamefont {A.~G.~R.}\ \bibnamefont {Day}}, \
  and\ \bibinfo {author} {\bibfnamefont {M.~J.~P.}\ \bibnamefont {Gingras}},\
  }\bibfield  {title} {\enquote {\bibinfo {title} {Bosonic many-body theory of
  quantum spin ice},}\ }\href@noop {} {\bibfield  {journal} {\bibinfo
  {journal} {Phys. Rev. B}\ }\textbf {\bibinfo {volume} {90}},\ \bibinfo
  {pages} {214430} (\bibinfo {year} {2014})}\BibitemShut {NoStop}%
\bibitem [{\citenamefont {{Kato}}\ and\ \citenamefont
  {{Onoda}}(2015)}]{Kato14}%
  \BibitemOpen
  \bibfield  {author} {\bibinfo {author} {\bibfnamefont {Y.}~\bibnamefont
  {{Kato}}}\ and\ \bibinfo {author} {\bibfnamefont {S.}~\bibnamefont
  {{Onoda}}},\ }\bibfield  {title} {\enquote {\bibinfo {title} {{Numerical
  evidence of quantum melting of spin ice: quantum-classical crossover}},}\
  }\href@noop {} {\bibfield  {journal} {\bibinfo  {journal} {Phys. Rev. Lett}\
  }\textbf {\bibinfo {volume} {115}},\ \bibinfo {pages} {077202} (\bibinfo
  {year} {2015})}\BibitemShut {NoStop}%
\bibitem [{\citenamefont {Santini}\ \emph {et~al.}(2009)\citenamefont
  {Santini}, \citenamefont {Carretta}, \citenamefont {Amoretti}, \citenamefont
  {Caciuffo}, \citenamefont {Magnani},\ and\ \citenamefont
  {Lander}}]{santini2009multipolar}%
  \BibitemOpen
  \bibfield  {author} {\bibinfo {author} {\bibfnamefont {P.}~\bibnamefont
  {Santini}}, \bibinfo {author} {\bibfnamefont {S.}~\bibnamefont {Carretta}},
  \bibinfo {author} {\bibfnamefont {G.}~\bibnamefont {Amoretti}}, \bibinfo
  {author} {\bibfnamefont {R.}~\bibnamefont {Caciuffo}}, \bibinfo {author}
  {\bibfnamefont {N.}~\bibnamefont {Magnani}}, \ and\ \bibinfo {author}
  {\bibfnamefont {G.~H.}\ \bibnamefont {Lander}},\ }\bibfield  {title}
  {\enquote {\bibinfo {title} {Multipolar interactions in f-electron systems:
  The paradigm of actinide dioxides},}\ }\href@noop {} {\bibfield  {journal}
  {\bibinfo  {journal} {Rev. Mod. Phys.}\ }\textbf {\bibinfo {volume} {81}},\
  \bibinfo {pages} {807} (\bibinfo {year} {2009})}\BibitemShut {NoStop}%
\bibitem [{\citenamefont {Rau}\ and\ \citenamefont {Gingras}(2015)}]{Rau15}%
  \BibitemOpen
  \bibfield  {author} {\bibinfo {author} {\bibfnamefont {J.~G.}\ \bibnamefont
  {Rau}}\ and\ \bibinfo {author} {\bibfnamefont {M.~J.~P.}\ \bibnamefont
  {Gingras}},\ }\bibfield  {title} {\enquote {\bibinfo {title} {{Magnitude of
  quantum effects in classical spin ices}},}\ }\href@noop {} {\bibfield
  {journal} {\bibinfo  {journal} {Phys. Rev. B}\ }\textbf {\bibinfo {volume}
  {92}},\ \bibinfo {pages} {144417} (\bibinfo {year} {2015})}\BibitemShut
  {NoStop}%
\bibitem [{Note2()}]{Note2}%
  \BibitemOpen
  \bibinfo {note} {Here, we have chosen to write the expected effective Ising
  exchange interactions for Dy$^{3+}$ ions\cite {Rau15} as ${\protect \bm
  {S}}_i \cdot {\protect \bm {S}}_j$ in order to keep with the notation used in
  several previous works.}\BibitemShut {Stop}%
\bibitem [{\citenamefont {Enjalran}\ and\ \citenamefont
  {Gingras}(2004)}]{Enjalran_mft}%
  \BibitemOpen
  \bibfield  {author} {\bibinfo {author} {\bibfnamefont {M.}~\bibnamefont
  {Enjalran}}\ and\ \bibinfo {author} {\bibfnamefont {M.~J.~P.}\ \bibnamefont
  {Gingras}},\ }\bibfield  {title} {\enquote {\bibinfo {title} {{Theory of
  paramagnetic scattering in highly frustrated magnets with long-range
  dipole-dipole interactions: The case of the Tb$_2$Ti$_2$O$_7$ pyrochlore
  antiferromagnet}},}\ }\href@noop {} {\bibfield  {journal} {\bibinfo
  {journal} {Phys. Rev. B}\ }\textbf {\bibinfo {volume} {70}},\ \bibinfo
  {pages} {174426} (\bibinfo {year} {2004})}\BibitemShut {NoStop}%
\bibitem [{\citenamefont {Marinari}\ and\ \citenamefont
  {Parisi}(1992)}]{Marinari92}%
  \BibitemOpen
  \bibfield  {author} {\bibinfo {author} {\bibfnamefont {E.}~\bibnamefont
  {Marinari}}\ and\ \bibinfo {author} {\bibfnamefont {G.}~\bibnamefont
  {Parisi}},\ }\href@noop {} {\bibfield  {journal} {\bibinfo  {journal}
  {Europhys. Lett.}\ }\textbf {\bibinfo {volume} {19}},\ \bibinfo {pages} {451}
  (\bibinfo {year} {1992})}\BibitemShut {NoStop}%
\bibitem [{\citenamefont {Brown}(1983-1993)}]{Brown93}%
  \BibitemOpen
  \bibfield  {author} {\bibinfo {author} {\bibfnamefont {P.~J.}\ \bibnamefont
  {Brown}},\ }\bibfield  {title} {\enquote {\bibinfo {title} {International
  tables for crystallography},}\ }in\ \href@noop {} {\emph {\bibinfo
  {booktitle} {International Tables for Crystallography}}},\ Vol.~\bibinfo
  {volume} {C},\ \bibinfo {editor} {edited by\ \bibinfo {editor} {\bibfnamefont
  {A.~J.~C.}\ \bibnamefont {Wilson}}}\ (\bibinfo  {publisher} {D. Reidel Pub.
  Co, Dordrecht, Holland},\ \bibinfo {year} {1983-1993})\ Chap.\ \bibinfo
  {chapter} {4.4.5}, p.\ \bibinfo {pages} {391}\BibitemShut {NoStop}%
\bibitem [{Note3()}]{Note3}%
  \BibitemOpen
  \bibinfo {note} {Recent x-ray synchrotron work\cite {Cava15} has shown that a
  small percentage of rare-earth (RE) ions occupying (``stuffing'') the B-site
  nominally occupied by the tetravalent transition metal ion (e.g. Ti$^{4+}$)
  in the RE$_2$Ti$_2$O$_7$ pyrochlore oxides is quite common, and found to
  occur, for example, in Ho$_2$Ti$_2$O$_7$, Er$_2$Ti$_2$O$_7$ and
  Yb$_2$Ti$_2$O$_7$.}\BibitemShut {Stop}%
\bibitem [{\citenamefont {Selke}({1988})}]{Selke}%
  \BibitemOpen
  \bibfield  {author} {\bibinfo {author} {\bibfnamefont {W.}~\bibnamefont
  {Selke}},\ }\bibfield  {title} {\enquote {\bibinfo {title} {{The ANNNI model
  -- theoretical analysis and experimental application}},}\ }\href@noop {}
  {\bibfield  {journal} {\bibinfo  {journal} {{Physics Reports -- Review
  Section of Physics Letters}}\ }\textbf {\bibinfo {volume} {{170}}},\ \bibinfo
  {pages} {{213}} (\bibinfo {year} {{1988}})}\BibitemShut {NoStop}%
\bibitem [{\citenamefont {Henley}(2010)}]{Henley_AnnRev}%
  \BibitemOpen
  \bibfield  {author} {\bibinfo {author} {\bibfnamefont {C.~L.}\ \bibnamefont
  {Henley}},\ }\bibfield  {title} {\enquote {\bibinfo {title} {The coulomb
  phase in frustrated systems},}\ }\href@noop {} {\bibfield  {journal}
  {\bibinfo  {journal} {Annual Review of Condensed Matter Physics}\ }\textbf
  {\bibinfo {volume} {1}},\ \bibinfo {pages} {179} (\bibinfo {year}
  {2010})}\BibitemShut {NoStop}%
\bibitem [{\citenamefont {Sen}\ \emph {et~al.}(2013)\citenamefont {Sen},
  \citenamefont {Moessner},\ and\ \citenamefont {Sondhi}}]{Sen_dip}%
  \BibitemOpen
  \bibfield  {author} {\bibinfo {author} {\bibfnamefont {A.}~\bibnamefont
  {Sen}}, \bibinfo {author} {\bibfnamefont {R.}~\bibnamefont {Moessner}}, \
  and\ \bibinfo {author} {\bibfnamefont {S.~L.}\ \bibnamefont {Sondhi}},\
  }\bibfield  {title} {\enquote {\bibinfo {title} {{Coulomb phase diagnostics
  as a function of temperature, interaction range, and disorder}},}\
  }\href@noop {} {\bibfield  {journal} {\bibinfo  {journal} {Phys. Rev. Lett.}\
  }\textbf {\bibinfo {volume} {110}},\ \bibinfo {pages} {107202} (\bibinfo
  {year} {2013})}\BibitemShut {NoStop}%
\bibitem [{\citenamefont {Fennell}\ \emph {et~al.}(2009)\citenamefont
  {Fennell}, \citenamefont {Deen}, \citenamefont {Wildes}, \citenamefont
  {Schmalzl}, \citenamefont {Prabhakaran}, \citenamefont {Boothroyd},
  \citenamefont {Aldus}, \citenamefont {McMorrow},\ and\ \citenamefont
  {Bramwell}}]{Fennell_Science}%
  \BibitemOpen
  \bibfield  {author} {\bibinfo {author} {\bibfnamefont {T.}~\bibnamefont
  {Fennell}}, \bibinfo {author} {\bibfnamefont {P.~P.}\ \bibnamefont {Deen}},
  \bibinfo {author} {\bibfnamefont {A.~R.}\ \bibnamefont {Wildes}}, \bibinfo
  {author} {\bibfnamefont {K.}~\bibnamefont {Schmalzl}}, \bibinfo {author}
  {\bibfnamefont {D.}~\bibnamefont {Prabhakaran}}, \bibinfo {author}
  {\bibfnamefont {A.~T.}\ \bibnamefont {Boothroyd}}, \bibinfo {author}
  {\bibfnamefont {R.~J.}\ \bibnamefont {Aldus}}, \bibinfo {author}
  {\bibfnamefont {D.~F.}\ \bibnamefont {McMorrow}}, \ and\ \bibinfo {author}
  {\bibfnamefont {S.~T.}\ \bibnamefont {Bramwell}},\ }\bibfield  {title}
  {\enquote {\bibinfo {title} {{Magnetic Coulomb phase in the spin ice
  Ho$_2$Ti$_2$O$_7$}},}\ }\href@noop {} {\bibfield  {journal} {\bibinfo
  {journal} {Science}\ }\textbf {\bibinfo {volume} {326}},\ \bibinfo {pages}
  {415} (\bibinfo {year} {2009})}\BibitemShut {NoStop}%
\bibitem [{\citenamefont {Morris}\ \emph {et~al.}(2009)\citenamefont {Morris},
  \citenamefont {Tennant}, \citenamefont {Grigera}, \citenamefont {Klemke},
  \citenamefont {Castelnovo}, \citenamefont {Moessner}, \citenamefont
  {Czternasty}, \citenamefont {Meissner}, \citenamefont {Rule}, \citenamefont
  {Hoffmann}, \citenamefont {Kiefer}, \citenamefont {Gerischer}, \citenamefont
  {Slobinsky},\ and\ \citenamefont {Perry}}]{Morris2009}%
  \BibitemOpen
  \bibfield  {author} {\bibinfo {author} {\bibfnamefont {D.~J.~P.}\
  \bibnamefont {Morris}}, \bibinfo {author} {\bibfnamefont {D.~A.}\
  \bibnamefont {Tennant}}, \bibinfo {author} {\bibfnamefont {S.~A.}\
  \bibnamefont {Grigera}}, \bibinfo {author} {\bibfnamefont {B.}~\bibnamefont
  {Klemke}}, \bibinfo {author} {\bibfnamefont {C.}~\bibnamefont {Castelnovo}},
  \bibinfo {author} {\bibfnamefont {R.}~\bibnamefont {Moessner}}, \bibinfo
  {author} {\bibfnamefont {C.}~\bibnamefont {Czternasty}}, \bibinfo {author}
  {\bibfnamefont {M.}~\bibnamefont {Meissner}}, \bibinfo {author}
  {\bibfnamefont {K.~C.}\ \bibnamefont {Rule}}, \bibinfo {author}
  {\bibfnamefont {J.-U.}\ \bibnamefont {Hoffmann}}, \bibinfo {author}
  {\bibfnamefont {K.}~\bibnamefont {Kiefer}}, \bibinfo {author} {\bibfnamefont
  {S.}~\bibnamefont {Gerischer}}, \bibinfo {author} {\bibfnamefont
  {D.}~\bibnamefont {Slobinsky}}, \ and\ \bibinfo {author} {\bibfnamefont
  {R.~S.}\ \bibnamefont {Perry}},\ }\bibfield  {title} {\enquote {\bibinfo
  {title} {{Dirac strings and magnetic monopoles in the spin sce
  Dy$_2$Ti$_2$O$_7$}},}\ }\href@noop {} {\bibfield  {journal} {\bibinfo
  {journal} {Science}\ }\textbf {\bibinfo {volume} {326}},\ \bibinfo {pages}
  {411} (\bibinfo {year} {2009})}\BibitemShut {NoStop}%
\bibitem [{\citenamefont {Clancy}\ \emph {et~al.}(2009)\citenamefont {Clancy},
  \citenamefont {Ruff}, \citenamefont {Dunsiger}, \citenamefont {Zhao},
  \citenamefont {Dabkowska}, \citenamefont {Gardner}, \citenamefont {Qiu},
  \citenamefont {Copley}, \citenamefont {Jenkins},\ and\ \citenamefont
  {Gaulin}}]{Clancy}%
  \BibitemOpen
  \bibfield  {author} {\bibinfo {author} {\bibfnamefont {J.~P.}\ \bibnamefont
  {Clancy}}, \bibinfo {author} {\bibfnamefont {J.~P.~C.}\ \bibnamefont {Ruff}},
  \bibinfo {author} {\bibfnamefont {S.~R.}\ \bibnamefont {Dunsiger}}, \bibinfo
  {author} {\bibfnamefont {Y.}~\bibnamefont {Zhao}}, \bibinfo {author}
  {\bibfnamefont {H.~A.}\ \bibnamefont {Dabkowska}}, \bibinfo {author}
  {\bibfnamefont {J.~S.}\ \bibnamefont {Gardner}}, \bibinfo {author}
  {\bibfnamefont {Y.}~\bibnamefont {Qiu}}, \bibinfo {author} {\bibfnamefont
  {J.~R.~D.}\ \bibnamefont {Copley}}, \bibinfo {author} {\bibfnamefont
  {T.}~\bibnamefont {Jenkins}}, \ and\ \bibinfo {author} {\bibfnamefont
  {B.~D.}\ \bibnamefont {Gaulin}},\ }\bibfield  {title} {\enquote {\bibinfo
  {title} {{Revisiting static and dynamic spin-ice correlations in
  ${\text{Ho}}_{2}{\text{Ti}}_{2}{\text{O}}_{7}$ with neutron scattering}},}\
  }\href@noop {} {\bibfield  {journal} {\bibinfo  {journal} {Phys. Rev. B}\
  }\textbf {\bibinfo {volume} {79}},\ \bibinfo {pages} {014408} (\bibinfo
  {year} {2009})}\BibitemShut {NoStop}%
\bibitem [{\citenamefont {Conlon}\ and\ \citenamefont
  {Chalker}(2010)}]{Conlon_ZBS}%
  \BibitemOpen
  \bibfield  {author} {\bibinfo {author} {\bibfnamefont {P.~H.}\ \bibnamefont
  {Conlon}}\ and\ \bibinfo {author} {\bibfnamefont {J.~T.}\ \bibnamefont
  {Chalker}},\ }\bibfield  {title} {\enquote {\bibinfo {title} {Absent pinch
  points and emergent clusters: further neighbor interactions in the pyrochlore
  heisenberg antiferromagnet},}\ }\href {\doibase 10.1103/PhysRevB.81.224413}
  {\bibfield  {journal} {\bibinfo  {journal} {Phys. Rev. B}\ }\textbf {\bibinfo
  {volume} {81}},\ \bibinfo {pages} {224413} (\bibinfo {year}
  {2010})}\BibitemShut {NoStop}%
\bibitem [{\citenamefont {Bramwell}\ \emph {et~al.}(2001)\citenamefont
  {Bramwell}, \citenamefont {Harris}, \citenamefont {den Hertog}, \citenamefont
  {Gingras}, \citenamefont {Gardner}, \citenamefont {McMorrow}, \citenamefont
  {Wildes}, \citenamefont {Cornelius}, \citenamefont {Champion}, \citenamefont
  {Melko},\ and\ \citenamefont {Fennell}}]{BramwellHo2Ti2O72001}%
  \BibitemOpen
  \bibfield  {author} {\bibinfo {author} {\bibfnamefont {S.~T.}\ \bibnamefont
  {Bramwell}}, \bibinfo {author} {\bibfnamefont {M.~J.}\ \bibnamefont
  {Harris}}, \bibinfo {author} {\bibfnamefont {B.~C.}\ \bibnamefont {den
  Hertog}}, \bibinfo {author} {\bibfnamefont {M.~J.~P.}\ \bibnamefont
  {Gingras}}, \bibinfo {author} {\bibfnamefont {J.~S.}\ \bibnamefont
  {Gardner}}, \bibinfo {author} {\bibfnamefont {D.~F.}\ \bibnamefont
  {McMorrow}}, \bibinfo {author} {\bibfnamefont {A.~R.}\ \bibnamefont
  {Wildes}}, \bibinfo {author} {\bibfnamefont {A.~L.}\ \bibnamefont
  {Cornelius}}, \bibinfo {author} {\bibfnamefont {J.~D.~M.}\ \bibnamefont
  {Champion}}, \bibinfo {author} {\bibfnamefont {R.~G.}\ \bibnamefont {Melko}},
  \ and\ \bibinfo {author} {\bibfnamefont {T.}~\bibnamefont {Fennell}},\
  }\bibfield  {title} {\enquote {\bibinfo {title} {{Spin correlations in
  ${\mathrm{Ho}}_{2}{\mathrm{Ti}}_{2}{O}_{7}$: A dipolar spin ice system}},}\
  }\href@noop {} {\bibfield  {journal} {\bibinfo  {journal} {Phys. Rev. Lett.}\
  }\textbf {\bibinfo {volume} {87}},\ \bibinfo {pages} {047205} (\bibinfo
  {year} {2001})}\BibitemShut {NoStop}%
\bibitem [{\citenamefont {den Hertog}\ \emph {et~al.}(1999)\citenamefont {den
  Hertog}, \citenamefont {Gingras}, \citenamefont {Bramwell},\ and\
  \citenamefont {Harris}}]{arXivHTO}%
  \BibitemOpen
  \bibfield  {author} {\bibinfo {author} {\bibfnamefont {B.~C.}\ \bibnamefont
  {den Hertog}}, \bibinfo {author} {\bibfnamefont {M.~J.~P.}\ \bibnamefont
  {Gingras}}, \bibinfo {author} {\bibfnamefont {S.~T.}\ \bibnamefont
  {Bramwell}}, \ and\ \bibinfo {author} {\bibfnamefont {M.~J.}\ \bibnamefont
  {Harris}},\ }\bibfield  {title} {\enquote {\bibinfo {title} {{ Comment on
  `Ising Pyrochlore Magnets: Low temperature troperties, ``ice rules,'' and
  beyond' by R. Siddharthan et al., Phys. Rev. Lett. 83, 1854 (1999)}},}\
  }\href@noop {} {\bibfield  {journal} {\bibinfo  {journal} {ArXiv e-prints}\ }
  (\bibinfo {year} {1999})},\ \Eprint
  {http://arxiv.org/abs/arXiv:cond-mat/9912220} {arXiv:arXiv:cond-mat/9912220}
  \BibitemShut {NoStop}%
\bibitem [{\citenamefont {Cornelius}\ and\ \citenamefont
  {Gardner}(2001)}]{Cornelius}%
  \BibitemOpen
  \bibfield  {author} {\bibinfo {author} {\bibfnamefont {A.~L.}\ \bibnamefont
  {Cornelius}}\ and\ \bibinfo {author} {\bibfnamefont {J.~S.}\ \bibnamefont
  {Gardner}},\ }\bibfield  {title} {\enquote {\bibinfo {title} {{Short-range
  magnetic interactions in the spin-ice compound
  ${\mathrm{Ho}}_{2}{\mathrm{Ti}}_{2}{\mathrm{O}}_{7}$ }},}\ }\href@noop {}
  {\bibfield  {journal} {\bibinfo  {journal} {Phys. Rev. B}\ }\textbf {\bibinfo
  {volume} {64}},\ \bibinfo {pages} {060406} (\bibinfo {year}
  {2001})}\BibitemShut {NoStop}%
\bibitem [{\citenamefont {Filippi}\ \emph {et~al.}(1977)\citenamefont
  {Filippi}, \citenamefont {Lasjaunias}, \citenamefont {Ravex}, \citenamefont
  {Tch\'eou},\ and\ \citenamefont {Rossat-Mignod}}]{Filippi}%
  \BibitemOpen
  \bibfield  {author} {\bibinfo {author} {\bibfnamefont {J.}~\bibnamefont
  {Filippi}}, \bibinfo {author} {\bibfnamefont {J.~C.}\ \bibnamefont
  {Lasjaunias}}, \bibinfo {author} {\bibfnamefont {A.}~\bibnamefont {Ravex}},
  \bibinfo {author} {\bibfnamefont {F.}~\bibnamefont {Tch\'eou}}, \ and\
  \bibinfo {author} {\bibfnamefont {J.}~\bibnamefont {Rossat-Mignod}},\
  }\bibfield  {title} {\enquote {\bibinfo {title} {{Specific heat of dysprosium
  gallium garnet between 37 mk and 2 K}},}\ }\href@noop {} {\bibfield
  {journal} {\bibinfo  {journal} {Sol. St. Comm.}\ }\textbf {\bibinfo {volume}
  {23}},\ \bibinfo {pages} {613} (\bibinfo {year} {1977})}\BibitemShut
  {NoStop}%
\bibitem [{\citenamefont {Mattis}\ and\ \citenamefont {Wolf}(1966)}]{Mattis}%
  \BibitemOpen
  \bibfield  {author} {\bibinfo {author} {\bibfnamefont {D.~C.}\ \bibnamefont
  {Mattis}}\ and\ \bibinfo {author} {\bibfnamefont {W.~P.}\ \bibnamefont
  {Wolf}},\ }\bibfield  {title} {\enquote {\bibinfo {title} {{Soluble extension
  of the Ising model}},}\ }\href@noop {} {\bibfield  {journal} {\bibinfo
  {journal} {Phys. Rev. Lett.}\ }\textbf {\bibinfo {volume} {16}},\ \bibinfo
  {pages} {899} (\bibinfo {year} {1966})}\BibitemShut {NoStop}%
\bibitem [{\citenamefont {Lin}\ \emph {et~al.}(2014)\citenamefont {Lin},
  \citenamefont {Ke}, \citenamefont {Thesberg}, \citenamefont {Schiffer},
  \citenamefont {Melko},\ and\ \citenamefont {Gingras}}]{Lin14}%
  \BibitemOpen
  \bibfield  {author} {\bibinfo {author} {\bibfnamefont {T.}~\bibnamefont
  {Lin}}, \bibinfo {author} {\bibfnamefont {X.}~\bibnamefont {Ke}}, \bibinfo
  {author} {\bibfnamefont {M.}~\bibnamefont {Thesberg}}, \bibinfo {author}
  {\bibfnamefont {P.}~\bibnamefont {Schiffer}}, \bibinfo {author}
  {\bibfnamefont {R.~G.}\ \bibnamefont {Melko}}, \ and\ \bibinfo {author}
  {\bibfnamefont {M.~J.~P.}\ \bibnamefont {Gingras}},\ }\bibfield  {title}
  {\enquote {\bibinfo {title} {{Nonmonotonic residual entropy in diluted spin
  ice: A comparison between Monte Carlo simulations of diluted dipolar spin ice
  models and experimental results}},}\ }\href@noop {} {\bibfield  {journal}
  {\bibinfo  {journal} {Phys. Rev. B}\ }\textbf {\bibinfo {volume} {90}},\
  \bibinfo {pages} {214433} (\bibinfo {year} {2014})}\BibitemShut {NoStop}%
\bibitem [{\citenamefont {Savary}\ and\ \citenamefont
  {Balents}(2012)}]{Savary_QSI}%
  \BibitemOpen
  \bibfield  {author} {\bibinfo {author} {\bibfnamefont {L.}~\bibnamefont
  {Savary}}\ and\ \bibinfo {author} {\bibfnamefont {L.}~\bibnamefont
  {Balents}},\ }\bibfield  {title} {\enquote {\bibinfo {title} {Coulombic
  quantum liquids in spin-$1/2$ pyrochlores},}\ }\href@noop {} {\bibfield
  {journal} {\bibinfo  {journal} {Phys. Rev. Lett.}\ }\textbf {\bibinfo
  {volume} {108}},\ \bibinfo {pages} {037202} (\bibinfo {year}
  {2012})}\BibitemShut {NoStop}%
\bibitem [{\citenamefont {Molavian}\ \emph {et~al.}(2007)\citenamefont
  {Molavian}, \citenamefont {Gingras},\ and\ \citenamefont
  {Canals}}]{Molavian07}%
  \BibitemOpen
  \bibfield  {author} {\bibinfo {author} {\bibfnamefont {H.~R.}\ \bibnamefont
  {Molavian}}, \bibinfo {author} {\bibfnamefont {M.~J.~P.}\ \bibnamefont
  {Gingras}}, \ and\ \bibinfo {author} {\bibfnamefont {B.}~\bibnamefont
  {Canals}},\ }\bibfield  {title} {\enquote {\bibinfo {title} {{Dynamically
  induced frustration as a route to a quantum spin ice state in
  Tb${}_{2}$Ti${}_{2}$O${}_{7}$ via virtual crystal field excitations and
  quantum many-body effects}},}\ }\href@noop {} {\bibfield  {journal} {\bibinfo
   {journal} {Phys. Rev. Lett.}\ }\textbf {\bibinfo {volume} {98}},\ \bibinfo
  {pages} {157204} (\bibinfo {year} {2007})}\BibitemShut {NoStop}%
\bibitem [{\citenamefont {Onoda}\ and\ \citenamefont
  {Tanaka}(2011)}]{onoda2011quantum}%
  \BibitemOpen
  \bibfield  {author} {\bibinfo {author} {\bibfnamefont {S.}~\bibnamefont
  {Onoda}}\ and\ \bibinfo {author} {\bibfnamefont {Y.}~\bibnamefont {Tanaka}},\
  }\bibfield  {title} {\enquote {\bibinfo {title} {Quantum fluctuations in the
  effective pseudospin-1/2 model for magnetic pyrochlore oxides},}\ }\href@noop
  {} {\bibfield  {journal} {\bibinfo  {journal} {Phys. Rev. B}\ }\textbf
  {\bibinfo {volume} {83}},\ \bibinfo {pages} {094411} (\bibinfo {year}
  {2011})}\BibitemShut {NoStop}%
\bibitem [{\citenamefont {Scharffe}\ \emph {et~al.}(2015)\citenamefont
  {Scharffe}, \citenamefont {Breunig}, \citenamefont {Cho}, \citenamefont
  {Laschitzky}, \citenamefont {Valldor}, \citenamefont {Welter},\ and\
  \citenamefont {Lorenz}}]{Scharffe}%
  \BibitemOpen
  \bibfield  {author} {\bibinfo {author} {\bibfnamefont {S.}~\bibnamefont
  {Scharffe}}, \bibinfo {author} {\bibfnamefont {O.}~\bibnamefont {Breunig}},
  \bibinfo {author} {\bibfnamefont {V.}~\bibnamefont {Cho}}, \bibinfo {author}
  {\bibfnamefont {P.}~\bibnamefont {Laschitzky}}, \bibinfo {author}
  {\bibfnamefont {M.}~\bibnamefont {Valldor}}, \bibinfo {author} {\bibfnamefont
  {J.~F.}\ \bibnamefont {Welter}}, \ and\ \bibinfo {author} {\bibfnamefont
  {T.}~\bibnamefont {Lorenz}},\ }\bibfield  {title} {\enquote {\bibinfo {title}
  {Suppression of pauling's residual entropy in the dilute spin ice
  ${({\mathrm{Dy}}_{1\ensuremath{-}x}{\mathrm{Y}}_{x})}_{2}{\mathrm{ti}}_{2}{\mathrm{o}}_{7}$},}\
  }\href@noop {} {\bibfield  {journal} {\bibinfo  {journal} {Phys. Rev. B}\
  }\textbf {\bibinfo {volume} {92}},\ \bibinfo {pages} {180405} (\bibinfo
  {year} {2015})}\BibitemShut {NoStop}%
\bibitem [{\citenamefont {Shannon}(1976)}]{Shannon_radii}%
  \BibitemOpen
  \bibfield  {author} {\bibinfo {author} {\bibfnamefont {R.~D.}\ \bibnamefont
  {Shannon}},\ }\bibfield  {title} {\enquote {\bibinfo {title} {{Revised
  effective ionic radii and systematic studies of interatomic distances in
  halides and chalcogenides}},}\ }\href@noop {} {\bibfield  {journal} {\bibinfo
   {journal} {Acta Crys.}\ }\textbf {\bibinfo {volume} {A32}},\ \bibinfo
  {pages} {751} (\bibinfo {year} {1976})}\BibitemShut {NoStop}%
\bibitem [{tra()}]{transverse}%
  \BibitemOpen
  \href@noop {} {}\bibinfo {note} {{Because of the large crystal-field gap
  $\Delta$ and the strictly Ising nature of the ground doublet of Dy$^{3+}$ in
  Dy$_2$Ti$_2$O$_7$ there are in this material negligible quantum transverse
  field effects, unlike in the LiHoF$_4$ dipolar Ising ferromagnet. See M. J.
  P. Gingras and P. Henelius, ``Collective Phenomena in the
  {LiHo$_x$Y$_{1-x}$F$_4$} Quantum {I}sing magnet: recent progress and open
  questions'', J. Phys.: Conf. Ser. {\bf 320}, 012001 (2011).}}\BibitemShut
  {Stop}%
\bibitem [{hor()}]{horizvert_foot}%
  \BibitemOpen
  \href@noop {} {}\bibinfo {note} {Equivalently, this experiment can be done
  with a so-called ``vector magnet'' in which one can tune independently a
  ``horizontal'' and a ``vertical'' magnetic field. Such an apparatus was used
  in the experiment exploring the physics of magnetic ordering in a strong
  $[112]$ field and reported in Ref.~[\onlinecite{Higa_112}].}\BibitemShut
  {Stop}%
\bibitem [{\citenamefont {Savary}\ \emph {et~al.}(2012)\citenamefont {Savary},
  \citenamefont {Ross}, \citenamefont {Gaulin}, \citenamefont {Ruff},\ and\
  \citenamefont {Balents}}]{Ross_ETO}%
  \BibitemOpen
  \bibfield  {author} {\bibinfo {author} {\bibfnamefont {L.}~\bibnamefont
  {Savary}}, \bibinfo {author} {\bibfnamefont {K.~A.}\ \bibnamefont {Ross}},
  \bibinfo {author} {\bibfnamefont {B.~D.}\ \bibnamefont {Gaulin}}, \bibinfo
  {author} {\bibfnamefont {J.~P.~C.}\ \bibnamefont {Ruff}}, \ and\ \bibinfo
  {author} {\bibfnamefont {L.}~\bibnamefont {Balents}},\ }\bibfield  {title}
  {\enquote {\bibinfo {title} {{Order by quantum disorder in
  Er$_2$Ti$_2$O$_7$}},}\ }\href@noop {} {\bibfield  {journal} {\bibinfo
  {journal} {Phys. Rev. Lett.}\ }\textbf {\bibinfo {volume} {109}},\ \bibinfo
  {pages} {167201} (\bibinfo {year} {2012})}\BibitemShut {NoStop}%
\bibitem [{\citenamefont {Ross}\ \emph {et~al.}(2011)\citenamefont {Ross},
  \citenamefont {Savary}, \citenamefont {Gaulin},\ and\ \citenamefont
  {Balents}}]{Ross_YbTO}%
  \BibitemOpen
  \bibfield  {author} {\bibinfo {author} {\bibfnamefont {K.~A.}\ \bibnamefont
  {Ross}}, \bibinfo {author} {\bibfnamefont {L.}~\bibnamefont {Savary}},
  \bibinfo {author} {\bibfnamefont {B.~D.}\ \bibnamefont {Gaulin}}, \ and\
  \bibinfo {author} {\bibfnamefont {L.}~\bibnamefont {Balents}},\ }\bibfield
  {title} {\enquote {\bibinfo {title} {Quantum excitations in quantum spin
  ice},}\ }\href@noop {} {\bibfield  {journal} {\bibinfo  {journal} {Phys. Rev.
  X}\ }\textbf {\bibinfo {volume} {1}},\ \bibinfo {pages} {021002} (\bibinfo
  {year} {2011})}\BibitemShut {NoStop}%
\bibitem [{\citenamefont {Applegate}\ \emph {et~al.}(2012)\citenamefont
  {Applegate}, \citenamefont {Hayre}, \citenamefont {Singh}, \citenamefont
  {Lin}, \citenamefont {Day},\ and\ \citenamefont {Gingras}}]{Applegate}%
  \BibitemOpen
  \bibfield  {author} {\bibinfo {author} {\bibfnamefont {R.}~\bibnamefont
  {Applegate}}, \bibinfo {author} {\bibfnamefont {N.~R.}\ \bibnamefont
  {Hayre}}, \bibinfo {author} {\bibfnamefont {R.~R.~P.}\ \bibnamefont {Singh}},
  \bibinfo {author} {\bibfnamefont {T.}~\bibnamefont {Lin}}, \bibinfo {author}
  {\bibfnamefont {A.~G.~R.}\ \bibnamefont {Day}}, \ and\ \bibinfo {author}
  {\bibfnamefont {M.~J.~P.}\ \bibnamefont {Gingras}},\ }\bibfield  {title}
  {\enquote {\bibinfo {title} {{Vindication of Yb${}_{2}$Ti${}_{2}$O${}_{7}$ as
  a model exchange quantum spin ice}},}\ }\href@noop {} {\bibfield  {journal}
  {\bibinfo  {journal} {Phys. Rev. Lett.}\ }\textbf {\bibinfo {volume} {109}},\
  \bibinfo {pages} {097205} (\bibinfo {year} {2012})}\BibitemShut {NoStop}%
\bibitem [{\citenamefont {Hayre}\ \emph {et~al.}(2013)\citenamefont {Hayre},
  \citenamefont {Ross}, \citenamefont {Applegate}, \citenamefont {Lin},
  \citenamefont {Singh}, \citenamefont {Gaulin},\ and\ \citenamefont
  {Gingras}}]{Hayre}%
  \BibitemOpen
  \bibfield  {author} {\bibinfo {author} {\bibfnamefont {N.~R.}\ \bibnamefont
  {Hayre}}, \bibinfo {author} {\bibfnamefont {K.~A.}\ \bibnamefont {Ross}},
  \bibinfo {author} {\bibfnamefont {R.}~\bibnamefont {Applegate}}, \bibinfo
  {author} {\bibfnamefont {T.}~\bibnamefont {Lin}}, \bibinfo {author}
  {\bibfnamefont {R.~R.~P.}\ \bibnamefont {Singh}}, \bibinfo {author}
  {\bibfnamefont {B.~D.}\ \bibnamefont {Gaulin}}, \ and\ \bibinfo {author}
  {\bibfnamefont {M.~J.~P.}\ \bibnamefont {Gingras}},\ }\bibfield  {title}
  {\enquote {\bibinfo {title} {{Thermodynamic properties of
  Yb${}_{2}$Ti${}_{2}$O${}_{7}$ pyrochlore as a function of temperature and
  magnetic field: Validation of a quantum spin ice exchange Hamiltonian}},}\
  }\href@noop {} {\bibfield  {journal} {\bibinfo  {journal} {Phys. Rev. B}\
  }\textbf {\bibinfo {volume} {87}},\ \bibinfo {pages} {184423} (\bibinfo
  {year} {2013})}\BibitemShut {NoStop}%
\bibitem [{\citenamefont {Oitmaa}\ \emph {et~al.}(2013)\citenamefont {Oitmaa},
  \citenamefont {Singh}, \citenamefont {Javanparast}, \citenamefont {Day},
  \citenamefont {Bagheri},\ and\ \citenamefont {Gingras}}]{Oitmaa_ETO}%
  \BibitemOpen
  \bibfield  {author} {\bibinfo {author} {\bibfnamefont {J.}~\bibnamefont
  {Oitmaa}}, \bibinfo {author} {\bibfnamefont {R.~R.~P.}\ \bibnamefont
  {Singh}}, \bibinfo {author} {\bibfnamefont {B.}~\bibnamefont {Javanparast}},
  \bibinfo {author} {\bibfnamefont {A.~G.~R.}\ \bibnamefont {Day}}, \bibinfo
  {author} {\bibfnamefont {B.~V.}\ \bibnamefont {Bagheri}}, \ and\ \bibinfo
  {author} {\bibfnamefont {M.~J.~P.}\ \bibnamefont {Gingras}},\ }\bibfield
  {title} {\enquote {\bibinfo {title} {{Phase transition and thermal
  order-by-disorder in the pyrochlore antiferromagnet Er$_2$Ti$_2$O$_7$:a
  high-temperature series expansion study}},}\ }\href@noop {} {\bibfield
  {journal} {\bibinfo  {journal} {Phys. Rev. B}\ }\textbf {\bibinfo {volume}
  {88}},\ \bibinfo {pages} {220404} (\bibinfo {year} {2013})}\BibitemShut
  {NoStop}%
\bibitem [{\citenamefont {Vierti\"o}\ and\ \citenamefont
  {Oja}(1993)}]{Viertio93}%
  \BibitemOpen
  \bibfield  {author} {\bibinfo {author} {\bibfnamefont {H.~E.}\ \bibnamefont
  {Vierti\"o}}\ and\ \bibinfo {author} {\bibfnamefont {A.~S.}\ \bibnamefont
  {Oja}},\ }\bibfield  {title} {\enquote {\bibinfo {title} {Interplay of three
  antiferromagnetic modulations in the nuclear-spin system of copper},}\
  }\href@noop {} {\bibfield  {journal} {\bibinfo  {journal} {Phys. Rev. B}\
  }\textbf {\bibinfo {volume} {48}},\ \bibinfo {pages} {1062} (\bibinfo {year}
  {1993})}\BibitemShut {NoStop}%
\bibitem [{j3()}]{j3}%
  \BibitemOpen
  \href@noop {} {}\bibinfo {note} {For fixed $J_1$, tuning $J_{3a}$, and
  $J_{3b}$ through the constraints of Eq.~(\ref{eq:j3}), drives a transition
  between the two ground states.}\BibitemShut {Stop}%
\bibitem [{\citenamefont {Anderson}\ \emph {et~al.}(1969)\citenamefont
  {Anderson}, \citenamefont {Holmstr{\"o}m}, \citenamefont {Krusius},\ and\
  \citenamefont {Pickett}}]{Anderson69}%
  \BibitemOpen
  \bibfield  {author} {\bibinfo {author} {\bibfnamefont {A.~C.}\ \bibnamefont
  {Anderson}}, \bibinfo {author} {\bibfnamefont {B.}~\bibnamefont
  {Holmstr{\"o}m}}, \bibinfo {author} {\bibfnamefont {M.}~\bibnamefont
  {Krusius}}, \ and\ \bibinfo {author} {\bibfnamefont {G.~R.}\ \bibnamefont
  {Pickett}},\ }\bibfield  {title} {\enquote {\bibinfo {title} {{Calorimetric
  investigation of the hyperfine interactions in metallic Nd, Sm, and Dy}},}\
  }\href@noop {} {\bibfield  {journal} {\bibinfo  {journal} {Phys. Rev.}\
  }\textbf {\bibinfo {volume} {183}},\ \bibinfo {pages} {546} (\bibinfo {year}
  {1969})}\BibitemShut {NoStop}%
\bibitem [{\citenamefont {Cooke}\ and\ \citenamefont {Park}(1956)}]{Dy_salt_1}%
  \BibitemOpen
  \bibfield  {author} {\bibinfo {author} {\bibfnamefont {A.~H.}\ \bibnamefont
  {Cooke}}\ and\ \bibinfo {author} {\bibfnamefont {J.G.}\ \bibnamefont
  {Park}},\ }\bibfield  {title} {\enquote {\bibinfo {title} {{Nuclear spins and
  magnetic moments of 161 Dy, 163 Dy, 171 Yb and 173 Yb}},}\ }\href@noop {}
  {\bibfield  {journal} {\bibinfo  {journal} {Proceedings of the Physical
  Society. Section A}\ }\textbf {\bibinfo {volume} {69}},\ \bibinfo {pages}
  {282} (\bibinfo {year} {1956})}\BibitemShut {NoStop}%
\bibitem [{\citenamefont {Wickman}\ and\ \citenamefont
  {Nowik}(1967)}]{Dy_salt_2}%
  \BibitemOpen
  \bibfield  {author} {\bibinfo {author} {\bibfnamefont {H.~H.}\ \bibnamefont
  {Wickman}}\ and\ \bibinfo {author} {\bibfnamefont {I.}~\bibnamefont
  {Nowik}},\ }\bibfield  {title} {\enquote {\bibinfo {title} {{The hyperfine
  structure of $^{161}$Dy in Dysprosium salts}},}\ }\href@noop {} {\bibfield
  {journal} {\bibinfo  {journal} {J. Phys. Chem. Solids}\ }\textbf {\bibinfo
  {volume} {28}},\ \bibinfo {pages} {2099} (\bibinfo {year}
  {1967})}\BibitemShut {NoStop}%
\bibitem [{\citenamefont {Brunhart}\ \emph {et~al.}(1971)\citenamefont
  {Brunhart}, \citenamefont {Postma}, \citenamefont {Rorer}, \citenamefont
  {Sailor},\ and\ \citenamefont {Vanneste}}]{Dy_salt_3}%
  \BibitemOpen
  \bibfield  {author} {\bibinfo {author} {\bibfnamefont {G.}~\bibnamefont
  {Brunhart}}, \bibinfo {author} {\bibfnamefont {H.}~\bibnamefont {Postma}},
  \bibinfo {author} {\bibfnamefont {D.~C.}\ \bibnamefont {Rorer}}, \bibinfo
  {author} {\bibfnamefont {V.~L.}\ \bibnamefont {Sailor}}, \ and\ \bibinfo
  {author} {\bibfnamefont {L.}~\bibnamefont {Vanneste}},\ }\bibfield  {title}
  {\enquote {\bibinfo {title} {{Absolute spin assignments of Dy-161 and Dy-163
  neutron resonances and hyperfine coupling constants in Dy-163}},}\
  }\href@noop {} {\bibfield  {journal} {\bibinfo  {journal} {Zeitschrift F\"ur
  Naturforschung Part A -- -Astrophysik Physik Und Physikalische Chemie}\
  }\textbf {\bibinfo {volume} {A26}},\ \bibinfo {pages} {334} (\bibinfo {year}
  {1971})}\BibitemShut {NoStop}%
\bibitem [{\citenamefont {Catanese}\ and\ \citenamefont
  {Meissner}(1973)}]{Dy_salt_4}%
  \BibitemOpen
  \bibfield  {author} {\bibinfo {author} {\bibfnamefont {C.~A.}\ \bibnamefont
  {Catanese}}\ and\ \bibinfo {author} {\bibfnamefont {H.~E.}\ \bibnamefont
  {Meissner}},\ }\bibfield  {title} {\enquote {\bibinfo {title} {{Magnetic
  Ordering in Dy${(\mathrm{OH})}_{3}$ and Ho${(\mathrm{OH})}_{3}$}},}\
  }\href@noop {} {\bibfield  {journal} {\bibinfo  {journal} {Phys. Rev. B}\
  }\textbf {\bibinfo {volume} {8}},\ \bibinfo {pages} {2060} (\bibinfo {year}
  {1973})}\BibitemShut {NoStop}%
\end{thebibliography}%

\end{document}